\documentclass[aps,singlecolumn]{revtex4-1}

    \usepackage{psfrag,graphicx}

    \usepackage[]{tcolorbox}
    \usepackage{parskip} % Stop auto-indenting (to mimic markdown behaviour)
    
    \usepackage{iftex}
    \ifPDFTeX
    	\usepackage[T1]{fontenc}
    	\usepackage{mathpazo}
    \else
    	\usepackage{fontspec}
    \fi

    \usepackage{graphicx}
    \usepackage{caption}

    \usepackage[Export]{adjustbox} % Used to constrain images to a maximum size
    \adjustboxset{max size={0.9\linewidth}{0.9\paperheight}}
    \usepackage{float}
    \usepackage{xcolor} % Allow colors to be defined
    \usepackage{enumerate} % Needed for markdown enumerations to work
    \usepackage{geometry} % Used to adjust the document margins
    \usepackage{amsmath} % Equations
    \usepackage{amssymb} % Equations
    \usepackage{textcomp} % defines textquotesingle
    \AtBeginDocument{%
        \def\PYZsq{\textquotesingle}% Upright quotes in Pygmentized code
    }
    \usepackage{upquote} % Upright quotes for verbatim code
    \usepackage{eurosym} % defines \euro
    \usepackage[mathletters]{ucs} % Extended unicode (utf-8) support
    \usepackage{fancyvrb} % verbatim replacement that allows latex
    \usepackage{grffile} % extends the file name processing of package graphics 
    \makeatletter % fix for grffile with XeLaTeX
    \def\Gread@@xetex#1{%
      \IfFileExists{"\Gin@base".bb}%
      {\Gread@eps{\Gin@base.bb}}%
      {\Gread@@xetex@aux#1}%
    }
    \makeatother

    \usepackage{hyperref}

    \definecolor{urlcolor}{rgb}{0,.145,.698}
    \definecolor{linkcolor}{rgb}{.71,0.21,0.01}
    \definecolor{citecolor}{rgb}{.12,.54,.11}

    \definecolor{ansi-black}{HTML}{3E424D}
    \definecolor{ansi-black-intense}{HTML}{282C36}
    \definecolor{ansi-red}{HTML}{E75C58}
    \definecolor{ansi-red-intense}{HTML}{B22B31}
    \definecolor{ansi-green}{HTML}{00A250}
    \definecolor{ansi-green-intense}{HTML}{007427}
    \definecolor{ansi-yellow}{HTML}{DDB62B}
    \definecolor{ansi-yellow-intense}{HTML}{B27D12}
    \definecolor{ansi-blue}{HTML}{208FFB}
    \definecolor{ansi-blue-intense}{HTML}{0065CA}
    \definecolor{ansi-magenta}{HTML}{D160C4}
    \definecolor{ansi-magenta-intense}{HTML}{A03196}
    \definecolor{ansi-cyan}{HTML}{60C6C8}
    \definecolor{ansi-cyan-intense}{HTML}{258F8F}
    \definecolor{ansi-white}{HTML}{C5C1B4}
    \definecolor{ansi-white-intense}{HTML}{A1A6B2}
    \definecolor{ansi-default-inverse-fg}{HTML}{FFFFFF}
    \definecolor{ansi-default-inverse-bg}{HTML}{000000}

    \providecommand{\tightlist}{%
      \setlength{\itemsep}{0pt}\setlength{\parskip}{0pt}}
    \DefineVerbatimEnvironment{Highlighting}{Verbatim}{commandchars=\\\{\}}
    \let\Oldtex\TeX
    \let\Oldlatex\LaTeX
    \renewcommand{\TeX}{\textrm{\Oldtex}}
    \renewcommand{\LaTeX}{\textrm{\Oldlatex}}
\makeatletter
\def\PY@reset{\let\PY@it=\relax \let\PY@bf=\relax%
    \let\PY@ul=\relax \let\PY@tc=\relax%
    \let\PY@bc=\relax \let\PY@ff=\relax}
\def\PY@tok#1{\csname PY@tok@#1\endcsname}
\def\PY@toks#1+{\ifx\relax#1\empty\else%
    \PY@tok{#1}\expandafter\PY@toks\fi}
\def\PY@do#1{\PY@bc{\PY@tc{\PY@ul{%
    \PY@it{\PY@bf{\PY@ff{#1}}}}}}}
\def\PY#1#2{\PY@reset\PY@toks#1+\relax+\PY@do{#2}}

\expandafter\def\csname PY@tok@w\endcsname{\def\PY@tc##1{\textcolor[rgb]{0.73,0.73,0.73}{##1}}}
\expandafter\def\csname PY@tok@c\endcsname{\let\PY@it=\textit\def\PY@tc##1{\textcolor[rgb]{0.25,0.50,0.50}{##1}}}
\expandafter\def\csname PY@tok@cp\endcsname{\def\PY@tc##1{\textcolor[rgb]{0.74,0.48,0.00}{##1}}}
\expandafter\def\csname PY@tok@k\endcsname{\let\PY@bf=\textbf\def\PY@tc##1{\textcolor[rgb]{0.00,0.50,0.00}{##1}}}
\expandafter\def\csname PY@tok@kp\endcsname{\def\PY@tc##1{\textcolor[rgb]{0.00,0.50,0.00}{##1}}}
\expandafter\def\csname PY@tok@kt\endcsname{\def\PY@tc##1{\textcolor[rgb]{0.69,0.00,0.25}{##1}}}
\expandafter\def\csname PY@tok@o\endcsname{\def\PY@tc##1{\textcolor[rgb]{0.40,0.40,0.40}{##1}}}
\expandafter\def\csname PY@tok@ow\endcsname{\let\PY@bf=\textbf\def\PY@tc##1{\textcolor[rgb]{0.67,0.13,1.00}{##1}}}
\expandafter\def\csname PY@tok@nb\endcsname{\def\PY@tc##1{\textcolor[rgb]{0.00,0.50,0.00}{##1}}}
\expandafter\def\csname PY@tok@nf\endcsname{\def\PY@tc##1{\textcolor[rgb]{0.00,0.00,1.00}{##1}}}
\expandafter\def\csname PY@tok@nc\endcsname{\let\PY@bf=\textbf\def\PY@tc##1{\textcolor[rgb]{0.00,0.00,1.00}{##1}}}
\expandafter\def\csname PY@tok@nn\endcsname{\let\PY@bf=\textbf\def\PY@tc##1{\textcolor[rgb]{0.00,0.00,1.00}{##1}}}
\expandafter\def\csname PY@tok@ne\endcsname{\let\PY@bf=\textbf\def\PY@tc##1{\textcolor[rgb]{0.82,0.25,0.23}{##1}}}
\expandafter\def\csname PY@tok@nv\endcsname{\def\PY@tc##1{\textcolor[rgb]{0.10,0.09,0.49}{##1}}}
\expandafter\def\csname PY@tok@no\endcsname{\def\PY@tc##1{\textcolor[rgb]{0.53,0.00,0.00}{##1}}}
\expandafter\def\csname PY@tok@nl\endcsname{\def\PY@tc##1{\textcolor[rgb]{0.63,0.63,0.00}{##1}}}
\expandafter\def\csname PY@tok@ni\endcsname{\let\PY@bf=\textbf\def\PY@tc##1{\textcolor[rgb]{0.60,0.60,0.60}{##1}}}
\expandafter\def\csname PY@tok@na\endcsname{\def\PY@tc##1{\textcolor[rgb]{0.49,0.56,0.16}{##1}}}
\expandafter\def\csname PY@tok@nt\endcsname{\let\PY@bf=\textbf\def\PY@tc##1{\textcolor[rgb]{0.00,0.50,0.00}{##1}}}
\expandafter\def\csname PY@tok@nd\endcsname{\def\PY@tc##1{\textcolor[rgb]{0.67,0.13,1.00}{##1}}}
\expandafter\def\csname PY@tok@s\endcsname{\def\PY@tc##1{\textcolor[rgb]{0.73,0.13,0.13}{##1}}}
\expandafter\def\csname PY@tok@sd\endcsname{\let\PY@it=\textit\def\PY@tc##1{\textcolor[rgb]{0.73,0.13,0.13}{##1}}}
\expandafter\def\csname PY@tok@si\endcsname{\let\PY@bf=\textbf\def\PY@tc##1{\textcolor[rgb]{0.73,0.40,0.53}{##1}}}
\expandafter\def\csname PY@tok@se\endcsname{\let\PY@bf=\textbf\def\PY@tc##1{\textcolor[rgb]{0.73,0.40,0.13}{##1}}}
\expandafter\def\csname PY@tok@sr\endcsname{\def\PY@tc##1{\textcolor[rgb]{0.73,0.40,0.53}{##1}}}
\expandafter\def\csname PY@tok@ss\endcsname{\def\PY@tc##1{\textcolor[rgb]{0.10,0.09,0.49}{##1}}}
\expandafter\def\csname PY@tok@sx\endcsname{\def\PY@tc##1{\textcolor[rgb]{0.00,0.50,0.00}{##1}}}
\expandafter\def\csname PY@tok@m\endcsname{\def\PY@tc##1{\textcolor[rgb]{0.40,0.40,0.40}{##1}}}
\expandafter\def\csname PY@tok@gh\endcsname{\let\PY@bf=\textbf\def\PY@tc##1{\textcolor[rgb]{0.00,0.00,0.50}{##1}}}
\expandafter\def\csname PY@tok@gu\endcsname{\let\PY@bf=\textbf\def\PY@tc##1{\textcolor[rgb]{0.50,0.00,0.50}{##1}}}
\expandafter\def\csname PY@tok@gd\endcsname{\def\PY@tc##1{\textcolor[rgb]{0.63,0.00,0.00}{##1}}}
\expandafter\def\csname PY@tok@gi\endcsname{\def\PY@tc##1{\textcolor[rgb]{0.00,0.63,0.00}{##1}}}
\expandafter\def\csname PY@tok@gr\endcsname{\def\PY@tc##1{\textcolor[rgb]{1.00,0.00,0.00}{##1}}}
\expandafter\def\csname PY@tok@ge\endcsname{\let\PY@it=\textit}
\expandafter\def\csname PY@tok@gs\endcsname{\let\PY@bf=\textbf}
\expandafter\def\csname PY@tok@gp\endcsname{\let\PY@bf=\textbf\def\PY@tc##1{\textcolor[rgb]{0.00,0.00,0.50}{##1}}}
\expandafter\def\csname PY@tok@go\endcsname{\def\PY@tc##1{\textcolor[rgb]{0.53,0.53,0.53}{##1}}}
\expandafter\def\csname PY@tok@gt\endcsname{\def\PY@tc##1{\textcolor[rgb]{0.00,0.27,0.87}{##1}}}
\expandafter\def\csname PY@tok@err\endcsname{\def\PY@bc##1{\setlength{\fboxsep}{0pt}\fcolorbox[rgb]{1.00,0.00,0.00}{1,1,1}{\strut ##1}}}
\expandafter\def\csname PY@tok@kc\endcsname{\let\PY@bf=\textbf\def\PY@tc##1{\textcolor[rgb]{0.00,0.50,0.00}{##1}}}
\expandafter\def\csname PY@tok@kd\endcsname{\let\PY@bf=\textbf\def\PY@tc##1{\textcolor[rgb]{0.00,0.50,0.00}{##1}}}
\expandafter\def\csname PY@tok@kn\endcsname{\let\PY@bf=\textbf\def\PY@tc##1{\textcolor[rgb]{0.00,0.50,0.00}{##1}}}
\expandafter\def\csname PY@tok@kr\endcsname{\let\PY@bf=\textbf\def\PY@tc##1{\textcolor[rgb]{0.00,0.50,0.00}{##1}}}
\expandafter\def\csname PY@tok@bp\endcsname{\def\PY@tc##1{\textcolor[rgb]{0.00,0.50,0.00}{##1}}}
\expandafter\def\csname PY@tok@fm\endcsname{\def\PY@tc##1{\textcolor[rgb]{0.00,0.00,1.00}{##1}}}
\expandafter\def\csname PY@tok@vc\endcsname{\def\PY@tc##1{\textcolor[rgb]{0.10,0.09,0.49}{##1}}}
\expandafter\def\csname PY@tok@vg\endcsname{\def\PY@tc##1{\textcolor[rgb]{0.10,0.09,0.49}{##1}}}
\expandafter\def\csname PY@tok@vi\endcsname{\def\PY@tc##1{\textcolor[rgb]{0.10,0.09,0.49}{##1}}}
\expandafter\def\csname PY@tok@vm\endcsname{\def\PY@tc##1{\textcolor[rgb]{0.10,0.09,0.49}{##1}}}
\expandafter\def\csname PY@tok@sa\endcsname{\def\PY@tc##1{\textcolor[rgb]{0.73,0.13,0.13}{##1}}}
\expandafter\def\csname PY@tok@sb\endcsname{\def\PY@tc##1{\textcolor[rgb]{0.73,0.13,0.13}{##1}}}
\expandafter\def\csname PY@tok@sc\endcsname{\def\PY@tc##1{\textcolor[rgb]{0.73,0.13,0.13}{##1}}}
\expandafter\def\csname PY@tok@dl\endcsname{\def\PY@tc##1{\textcolor[rgb]{0.73,0.13,0.13}{##1}}}
\expandafter\def\csname PY@tok@s2\endcsname{\def\PY@tc##1{\textcolor[rgb]{0.73,0.13,0.13}{##1}}}
\expandafter\def\csname PY@tok@sh\endcsname{\def\PY@tc##1{\textcolor[rgb]{0.73,0.13,0.13}{##1}}}
\expandafter\def\csname PY@tok@s1\endcsname{\def\PY@tc##1{\textcolor[rgb]{0.73,0.13,0.13}{##1}}}
\expandafter\def\csname PY@tok@mb\endcsname{\def\PY@tc##1{\textcolor[rgb]{0.40,0.40,0.40}{##1}}}
\expandafter\def\csname PY@tok@mf\endcsname{\def\PY@tc##1{\textcolor[rgb]{0.40,0.40,0.40}{##1}}}
\expandafter\def\csname PY@tok@mh\endcsname{\def\PY@tc##1{\textcolor[rgb]{0.40,0.40,0.40}{##1}}}
\expandafter\def\csname PY@tok@mi\endcsname{\def\PY@tc##1{\textcolor[rgb]{0.40,0.40,0.40}{##1}}}
\expandafter\def\csname PY@tok@il\endcsname{\def\PY@tc##1{\textcolor[rgb]{0.40,0.40,0.40}{##1}}}
\expandafter\def\csname PY@tok@mo\endcsname{\def\PY@tc##1{\textcolor[rgb]{0.40,0.40,0.40}{##1}}}
\expandafter\def\csname PY@tok@ch\endcsname{\let\PY@it=\textit\def\PY@tc##1{\textcolor[rgb]{0.25,0.50,0.50}{##1}}}
\expandafter\def\csname PY@tok@cm\endcsname{\let\PY@it=\textit\def\PY@tc##1{\textcolor[rgb]{0.25,0.50,0.50}{##1}}}
\expandafter\def\csname PY@tok@cpf\endcsname{\let\PY@it=\textit\def\PY@tc##1{\textcolor[rgb]{0.25,0.50,0.50}{##1}}}
\expandafter\def\csname PY@tok@c1\endcsname{\let\PY@it=\textit\def\PY@tc##1{\textcolor[rgb]{0.25,0.50,0.50}{##1}}}
\expandafter\def\csname PY@tok@cs\endcsname{\let\PY@it=\textit\def\PY@tc##1{\textcolor[rgb]{0.25,0.50,0.50}{##1}}}

\def\PYZbs{\char`\\}
\def\PYZus{\char`\_}
\def\PYZob{\char`\{}
\def\PYZcb{\char`\}}

\def\PYZlt{\char`\<}
\def\PYZgt{\char`\>}
\def\PYZsh{\char`\#}

\def\PYZhy{\char`\-}
\def\PYZsq{\char`\'}
\def\PYZdq{\char`\"}

\makeatother

    \makeatletter
        \newbox\Wrappedcontinuationbox 
        \newbox\Wrappedvisiblespacebox 
        \newcommand*\Wrappedvisiblespace {\textcolor{red}{\textvisiblespace}} 
        \newcommand*\Wrappedcontinuationsymbol {\textcolor{red}{\llap{\tiny$\m@th\hookrightarrow$}}} 
        \newcommand*\Wrappedcontinuationindent {3ex } 
        \newcommand*\Wrappedafterbreak {\kern\Wrappedcontinuationindent\copy\Wrappedcontinuationbox} 
        \newcommand*\Wrappedbreaksatspecials {% 
            \def\PYGZus{\discretionary{\char`\_}{\Wrappedafterbreak}{\char`\_}}% 
            \def\PYGZob{\discretionary{}{\Wrappedafterbreak\char`\{}{\char`\{}}% 
            \def\PYGZcb{\discretionary{\char`\}}{\Wrappedafterbreak}{\char`\}}}% 
            \def\PYGZca{\discretionary{\char`\^}{\Wrappedafterbreak}{\char`\^}}% 
            \def\PYGZam{\discretionary{\char`\&}{\Wrappedafterbreak}{\char`\&}}% 
            \def\PYGZlt{\discretionary{}{\Wrappedafterbreak\char`\<}{\char`\<}}% 
            \def\PYGZgt{\discretionary{\char`\>}{\Wrappedafterbreak}{\char`\>}}% 
            \def\PYGZsh{\discretionary{}{\Wrappedafterbreak\char`\#}{\char`\#}}% 
            \def\PYGZpc{\discretionary{}{\Wrappedafterbreak\char`\%}{\char`\%}}% 
            \def\PYGZdl{\discretionary{}{\Wrappedafterbreak\char`\$}{\char`\$}}% 
            \def\PYGZhy{\discretionary{\char`\-}{\Wrappedafterbreak}{\char`\-}}% 
            \def\PYGZsq{\discretionary{}{\Wrappedafterbreak\textquotesingle}{\textquotesingle}}% 
            \def\PYGZdq{\discretionary{}{\Wrappedafterbreak\char`\"}{\char`\"}}% 
            \def\PYGZti{\discretionary{\char`\~}{\Wrappedafterbreak}{\char`\~}}% 
        } 
        \newcommand*\Wrappedbreaksatpunct {% 
            \lccode`\~`\.\lowercase{\def~}{\discretionary{\hbox{\char`\.}}{\Wrappedafterbreak}{\hbox{\char`\.}}}% 
            \lccode`\~`\,\lowercase{\def~}{\discretionary{\hbox{\char`\,}}{\Wrappedafterbreak}{\hbox{\char`\,}}}% 
            \lccode`\~`\;\lowercase{\def~}{\discretionary{\hbox{\char`\;}}{\Wrappedafterbreak}{\hbox{\char`\;}}}% 
            \lccode`\~`\:\lowercase{\def~}{\discretionary{\hbox{\char`\:}}{\Wrappedafterbreak}{\hbox{\char`\:}}}% 
            \lccode`\~`\?\lowercase{\def~}{\discretionary{\hbox{\char`\?}}{\Wrappedafterbreak}{\hbox{\char`\?}}}% 
            \lccode`\~`\!\lowercase{\def~}{\discretionary{\hbox{\char`\!}}{\Wrappedafterbreak}{\hbox{\char`\!}}}% 
            \lccode`\~`\/\lowercase{\def~}{\discretionary{\hbox{\char`\/}}{\Wrappedafterbreak}{\hbox{\char`\/}}}% 
            \catcode`\.\active
            \catcode`\,\active 
            \catcode`\;\active
            \catcode`\:\active
            \catcode`\?\active
            \catcode`\!\active
            \catcode`\/\active 
            \lccode`\~`\~ 	
        }
    \makeatother

    \let\OriginalVerbatim=\Verbatim
    \makeatletter
    \renewcommand{\Verbatim}[1][1]{%
        \sbox\Wrappedcontinuationbox {\Wrappedcontinuationsymbol}%
        \sbox\Wrappedvisiblespacebox {\FV@SetupFont\Wrappedvisiblespace}%
        \def\FancyVerbFormatLine ##1{\hsize\linewidth
            \vtop{\raggedright\hyphenpenalty\z@\exhyphenpenalty\z@
                \doublehyphendemerits\z@\finalhyphendemerits\z@
                \strut ##1\strut}%
        }%
        \def\FV@Space {%
            \nobreak\hskip\z@ plus\fontdimen3\font minus\fontdimen4\font
            \discretionary{\copy\Wrappedvisiblespacebox}{\Wrappedafterbreak}
            {\kern\fontdimen2\font}%
        }%
        
        \Wrappedbreaksatspecials
        \OriginalVerbatim[#1,codes*=\Wrappedbreaksatpunct]%
    }
    \makeatother

    \definecolor{incolor}{HTML}{303F9F}
    \definecolor{outcolor}{HTML}{D84315}
    \definecolor{cellborder}{HTML}{CFCFCF}
    \definecolor{cellbackground}{HTML}{F7F7F7}
    
    \makeatletter
    \newcommand{\boxspacing}{\kern\kvtcb@left@rule\kern\kvtcb@boxsep}
    \makeatother
    \newcommand{\prompt}[4]{
    }

    \hypersetup{
      breaklinks=true,  % so long urls are correctly broken across lines
      colorlinks=true,
      urlcolor=urlcolor,
      linkcolor=linkcolor,
      citecolor=citecolor,
      }
\begin{document}

\title{Benchmarking near-term devices with quantum error correction}
\author{James R. Wootton}
\affiliation{IBM Research - Zurich}

\begin{abstract}

Now that ever more sophisticated devices for quantum computing are being developed, we require ever more sophisticated benchmarks. This includes a need to determine how well these devices support the techniques required for quantum error correction. In this paper we introduce the \texttt{topological\_codes} module of Qiskit-Ignis, which is designed to provide the tools necessary to perform such tests. Specifically, we use the \texttt{RepetitionCode} and \texttt{GraphDecoder} classes to run tests based on the repetition code and process the results. As an example, data from a 43 qubit code running on IBM's \emph{Rochester} device is presented.

\end{abstract}
    
\maketitle

\section{Introduction}

Software comes in many forms. The most prominent forms of software for classical computers are dedicated to applications, in which the device performs a useful task for the end user. Though this is also the goal of quantum software, there will instead be a heavy focus on benchmarking, testing and validation of quantum devices in the near-term. The \texttt{topological\_codes} module of Qiskit Ignis is one means by which this can be done. In this paper we introduce this new module, and describe its implementation and the methodology behind it.

Quantum software is based on the idea of encoding information in qubits. Most
quantum algorithms developed over the past few decades have assumed that
these qubits are perfect: they can be prepared in any state we desire,
and be manipulated with complete precision. Qubits that obey these
assumptions are often known as \emph{logical qubits}.

The last few decades have also seen great advances in finding physical
systems that behave as qubits with ever greater fidelity. However, the imperfections can never be removed
entirely. These qubits will always be much too imprecise to serve
directly as logical qubits. Instead, we refer to them as \emph{physical
qubits}.

In the current era of quantum computing, we seek to use physical qubits
despite their imperfections, by designing custom algorithms and using
error mitigation\cite{temme:16,endo:18,murali:19}. For the future era of fault-tolerance,
however, we must find ways to build logical qubits from physical qubits.
This will be done through the process of quantum error correction~\cite{lidar:13}, in
which logical qubits are encoded in a large numbers of physical
qubits. The encoding is maintained by constantly putting the physical
qubits through a highly entangling circuit. Auxiliary degrees of
freedom are then constantly measured, to detect signs of errors and
allow their effects to be removed.

Because of the vast amount effort required for this process, most
operations performed in fault-tolerant quantum computers will be done to
serve the purpose of error detection and correction. The logical  operations required for 
quantum computation are essentially just small perturbations to the error correction procedure. As such, as we
benchmark our progress towards fault-tolerant quantum computation, we
must keep track of how well our devices perform error correction.

Various experiments testing the ideas behind quantum error correction have already been performed ~\cite{kelly:14,riste:15,corcoles:15,takita:16,wootton:majorana,takita:17,linke:17,vuillot:18,wootton:18,naveh:18,gong:19,andersen:19}. These include several experiments based on repetition codes ~\cite{kelly:14,wootton:18,naveh:18}. This is the simplest example of error detection and correction that can be done using the standard techniques of quantum stabilizer codes~\cite{gottesman:96}. Though not a true example of quantum error correction - it uses physical qubits to encode a logical \emph{bit}, rather than a qubit - it serves as a simple guide to all the basic concepts in any quantum error correcting code.  Its requirements in terms of qubit number and connectivity are very flexible, allowing it to be straightforwardly implemented on almost any device. This makes it an excellent general-purpose benchmark.

In this paper we will provide a simple introduction to the code, and show how to run instances of it on current prototype devices using the open-source \emph{Qiskit} framework~\cite{qiskit}. Specifically, we will use the \texttt{topological\_codes} module of Qiskit-Ignis, which provides tools to create the quantum circuits required for simple quantum error correcting codes, as well as process the results.

\section{Introduction to the repetition code}

\subsection{The basics of error correction}
The basic ideas behind error correction are the same for quantum
information as for classical information. This allows us to begin by
considering a very straightforward example: speaking on the phone. If
someone asks you a question to which the answer is `yes' or `no', the
way you give your response will depend on two factors:

\begin{itemize}
\tightlist
\item
  How important is it that you are understood correctly?
\item
  How good is your connection?
\end{itemize}

Both of these can be paramaterized with probabilities. For the first, we
can use \(P_a\), the maximum acceptable probability of being
misunderstood. If you are being asked to confirm a preference for ice
cream flavours, and don't mind too much if you get vanilla rather than
chocolate, \(P_a\) might be quite high. If you are being asked a
question on which someone's life depends, however, \(P_a\) will be much
lower.

For the second we can use \(\rho\), the probability that your answer is
garbled by a bad connection. For simplicity, let's imagine a case where
a garbled `yes' doesn't simply sound like nonsense, but sounds like a
`no'. And similarly a `no' is transformed into `yes'. Then \(\rho\) is the
probability that you are completely misunderstood.

A good connection or a relatively unimportant question will result in
\(\rho<P_a\). In this case it is fine to simply answer in the most direct
way possible: you just say `yes' or `no'.

If, however, your connection is poor and your answer is important, we
will have \(\rho>P_a\). A single `yes' or `no' is not enough in this case.
The probability of being misunderstood would be too high. Instead we
must encode our answer in a more complex structure, allowing the
receiver to decode our meaning despite the possibility of the message
being disrupted. The simplest method is the one that many would do
without thinking: simply repeat the answer many times. For example say
`yes, yes, yes' instead of `yes' or `no, no no' instead of `no'.

If the receiver hears `yes, yes, yes' in this case, they will of course
conclude that the sender meant `yes'. If they hear `no, yes, yes', `yes,
no, yes' or `yes, yes, no', they will probably conclude the same thing,
since there is more positivity than negativity in the answer. To be
misunderstood in this case, at least two of the replies need to be
garbled. The probability for this, \(P\), will be less than \(\rho\). When
encoded in this way, the message therefore becomes more likely to be
understood. The code cell below shows an example of this.

    \begin{tcolorbox}[ size=fbox, boxrule=1pt, colback=cellbackground, colframe=cellborder]
\prompt{In}{incolor}{1}{\boxspacing}
\begin{Verbatim}[commandchars=\\\{\}]
\PY{n}{p} \PY{o}{=} \PY{l+m+mf}{0.01}
\PY{n}{P} \PY{o}{=} \PY{l+m+mi}{3} \PY{o}{*} \PY{n}{p}\PY{o}{*}\PY{o}{*}\PY{l+m+mi}{2} \PY{o}{*} \PY{p}{(}\PY{l+m+mi}{1}\PY{o}{\PYZhy{}}\PY{n}{p}\PY{p}{)} \PY{o}{+} \PY{n}{p}\PY{o}{*}\PY{o}{*}\PY{l+m+mi}{3} \PY{c+c1}{\PYZsh{} probability of 2 or 3 errors}
\PY{n+nb}{print}\PY{p}{(}\PY{l+s+s1}{\PYZsq{}}\PY{l+s+s1}{Probability of a single reply being garbled:}\PY{l+s+s1}{\PYZsq{}}\PY{p}{,}\PY{n}{p}\PY{p}{)}
\PY{n+nb}{print}\PY{p}{(}\PY{l+s+s1}{\PYZsq{}}\PY{l+s+s1}{Probability of a the majority of three replies being garbled:}\PY{l+s+s1}{\PYZsq{}}\PY{p}{,}\PY{n}{P}\PY{p}{)}
\end{Verbatim}
\end{tcolorbox}

The output obtained from running the above program snippet is as follows (from henceforth, any such output is displayed directly beneath the cell that it pertains to).

\vspace{0.25cm}
\begin{Verbatim}[commandchars=\\\{\}]
Probability of a single reply being garbled: 0.01
Probability of a the majority of three replies being garbled:
0.00029800000000000003
\end{Verbatim}
\vspace{0.25cm}
\vspace{0.25cm}

    If \(P<P_a\), this technique solves our problem. If not, we can simply
add more repetitions. The fact that \(P<\rho\) above comes from the fact
that we need at least two replies to be garbled to flip the majority,
and so even the most likely possibilities have a probability of
\(\sim \rho^2\). For five repetitions we'd need at least three replies to
be garbled to flip the majority, which happens with probability
\(\sim \rho^3\). The value for \(P\) in this case would then be even lower.
Indeed, as we increase the number of repetitions, \(P\) will decrease
exponentially. No matter how bad the connection, or how certain we need
to be of our message getting through correctly, we can achieve it by
just repeating our answer enough times.

Though this is a simple example, it contains all the aspects of error
correction.

\begin{itemize}
\tightlist
\item
There is some information to be sent or stored: In this
case, a `yes' or `no'.
\item
The information is encoded in a larger system to
protect it against noise: In this case, by repeating the message.
\item
The
information is finally decoded, mitigating for the effects of noise: In
this case, by trusting the majority of the transmitted messages.
\end{itemize}

This same encoding scheme can also be used for binary, by simply
substituting \texttt{0} and \texttt{1} for `yes' and 'no. It can
therefore also be easily generalised to qubits by using the states
\(\left|0\right\rangle\) and \(\left|1\right\rangle\). In each case it
is known as the \emph{repetition code}. Many other forms of encoding are
also possible in both the classical and quantum cases, which outperform
the repetition code in many ways. However, its status as the simplest
encoding does lend it to certain applications. One is exactly what it is
used for in Qiskit: as the first and simplest test of implementing the
ideas behind quantum error correction.

    \hypertarget{correcting-errors-in-qubits}{%
\subsection{Correcting errors in
qubits}\label{correcting-errors-in-qubits}}

We will now implement these ideas explicitly using Qiskit. To see the
effects of imperfect qubits, we can simply use the qubits of the
prototype devices. We can also reproduce the effects in simulations. The
function below creates a simple noise models in order to do the latter. The noise models it creates
go beyond the simple case discussed earlier, of a single noise event
which happens with a probability \(\rho\). Instead we consider two forms of
error that can occur. One is a gate error: an imperfection in any
operation we perform. We model this here in a simple way, using
so-called depolarizing noise. The effect of this will be, with
probability \(\rho_{gate}\) ,to replace the state of any qubit with a
completely random state. For two qubit gates, it is applied
independently to each qubit. The other form of noise is that for
measurement. This simply flips between a \texttt{0} to a \texttt{1} and
vice-versa immediately before measurement with probability \(\rho_{meas}\).

    \begin{tcolorbox}[ size=fbox, boxrule=1pt, colback=cellbackground, colframe=cellborder]
\prompt{In}{incolor}{2}{\boxspacing}
\begin{Verbatim}[commandchars=\\\{\}]
\PY{k+kn}{from} \PY{n+nn}{qiskit}\PY{n+nn}{.}\PY{n+nn}{providers}\PY{n+nn}{.}\PY{n+nn}{aer}\PY{n+nn}{.}\PY{n+nn}{noise} \PY{k+kn}{import} \PY{n}{NoiseModel}
\PY{k+kn}{from} \PY{n+nn}{qiskit}\PY{n+nn}{.}\PY{n+nn}{providers}\PY{n+nn}{.}\PY{n+nn}{aer}\PY{n+nn}{.}\PY{n+nn}{noise}\PY{n+nn}{.}\PY{n+nn}{errors} \PY{k+kn}{import} \PY{n}{pauli\PYZus{}error}\PY{p}{,} \PY{n}{depolarizing\PYZus{}error}

\PY{k}{def} \PY{n+nf}{get\PYZus{}noise}\PY{p}{(}\PY{n}{rho\PYZus{}meas}\PY{p}{,}\PY{n}{rho\PYZus{}gate}\PY{p}{)}\PY{p}{:}

    \PY{n}{error\PYZus{}meas} \PY{o}{=} \PY{n}{pauli\PYZus{}error}\PY{p}{(}\PY{p}{[}\PY{p}{(}\PY{l+s+s1}{\PYZsq{}}\PY{l+s+s1}{X}\PY{l+s+s1}{\PYZsq{}}\PY{p}{,}\PY{n}{rho\PYZus{}meas}\PY{p}{)}\PY{p}{,} \PY{p}{(}\PY{l+s+s1}{\PYZsq{}}\PY{l+s+s1}{I}\PY{l+s+s1}{\PYZsq{}}\PY{p}{,} \PY{l+m+mi}{1} \PY{o}{\PYZhy{}} \PY{n}{rho\PYZus{}meas}\PY{p}{)}\PY{p}{]}\PY{p}{)}
    \PY{n}{error\PYZus{}gate1} \PY{o}{=} \PY{n}{depolarizing\PYZus{}error}\PY{p}{(}\PY{n}{rho\PYZus{}gate}\PY{p}{,} \PY{l+m+mi}{1}\PY{p}{)}
    \PY{n}{error\PYZus{}gate2} \PY{o}{=} \PY{n}{error\PYZus{}gate1}\PY{o}{.}\PY{n}{tensor}\PY{p}{(}\PY{n}{error\PYZus{}gate1}\PY{p}{)}

    \PY{n}{noise\PYZus{}model} \PY{o}{=} \PY{n}{NoiseModel}\PY{p}{(}\PY{p}{)}
    \PY{c+c1}{\PYZsh{} measurement error is applied to measurements}
    \PY{n}{noise\PYZus{}model}\PY{o}{.}\PY{n}{add\PYZus{}all\PYZus{}qubit\PYZus{}quantum\PYZus{}error}\PY{p}{(}\PY{n}{error\PYZus{}meas}\PY{p}{,} \PY{l+s+s2}{\PYZdq{}}\PY{l+s+s2}{measure}\PY{l+s+s2}{\PYZdq{}}\PY{p}{)}
    \PY{c+c1}{\PYZsh{} single qubit gate error is applied to x gates}
    \PY{n}{noise\PYZus{}model}\PY{o}{.}\PY{n}{add\PYZus{}all\PYZus{}qubit\PYZus{}quantum\PYZus{}error}\PY{p}{(}\PY{n}{error\PYZus{}gate1}\PY{p}{,} \PY{p}{[}\PY{l+s+s2}{\PYZdq{}}\PY{l+s+s2}{x}\PY{l+s+s2}{\PYZdq{}}\PY{p}{]}\PY{p}{)}
    \PY{c+c1}{\PYZsh{} two qubit gate error is applied to cx gates}
    \PY{n}{noise\PYZus{}model}\PY{o}{.}\PY{n}{add\PYZus{}all\PYZus{}qubit\PYZus{}quantum\PYZus{}error}\PY{p}{(}\PY{n}{error\PYZus{}gate2}\PY{p}{,} \PY{p}{[}\PY{l+s+s2}{\PYZdq{}}\PY{l+s+s2}{cx}\PY{l+s+s2}{\PYZdq{}}\PY{p}{]}\PY{p}{)} 
        
    \PY{k}{return} \PY{n}{noise\PYZus{}model}
\end{Verbatim}
\end{tcolorbox}

    With this we'll now create such a noise model with a probability of
\(1\%\) for each type of error.

    \begin{tcolorbox}[ size=fbox, boxrule=1pt, colback=cellbackground, colframe=cellborder]
\prompt{In}{incolor}{3}{\boxspacing}
\begin{Verbatim}[commandchars=\\\{\}]
\PY{n}{noise\PYZus{}model} \PY{o}{=} \PY{n}{get\PYZus{}noise}\PY{p}{(}\PY{l+m+mf}{0.01}\PY{p}{,}\PY{l+m+mf}{0.01}\PY{p}{)}
\end{Verbatim}
\end{tcolorbox}

    Let's see what affect this has when try to store a \texttt{0} using
three qubits in state \(\left|0\right\rangle\). We'll repeat the process
\texttt{shots=1024} times to see how likely different results are.

    \begin{tcolorbox}[ size=fbox, boxrule=1pt, colback=cellbackground, colframe=cellborder]
\prompt{In}{incolor}{4}{\boxspacing}
\begin{Verbatim}[commandchars=\\\{\}]
\PY{k+kn}{from} \PY{n+nn}{qiskit} \PY{k+kn}{import} \PY{n}{QuantumCircuit}\PY{p}{,} \PY{n}{execute}\PY{p}{,} \PY{n}{Aer}

\PY{c+c1}{\PYZsh{} initialize circuit with three qubits in the 0 state}
\PY{n}{qc0} \PY{o}{=} \PY{n}{QuantumCircuit}\PY{p}{(}\PY{l+m+mi}{3}\PY{p}{,}\PY{l+m+mi}{3}\PY{p}{,}\PY{n}{name}\PY{o}{=}\PY{l+s+s1}{\PYZsq{}}\PY{l+s+s1}{0}\PY{l+s+s1}{\PYZsq{}}\PY{p}{)}

\PY{n}{qc0}\PY{o}{.}\PY{n}{measure}\PY{p}{(}\PY{n}{qc0}\PY{o}{.}\PY{n}{qregs}\PY{p}{[}\PY{l+m+mi}{0}\PY{p}{]}\PY{p}{,}\PY{n}{qc0}\PY{o}{.}\PY{n}{cregs}\PY{p}{[}\PY{l+m+mi}{0}\PY{p}{]}\PY{p}{)} \PY{c+c1}{\PYZsh{} measure the qubits}

\PY{c+c1}{\PYZsh{} run the circuit with th noise model and extract the counts}
\PY{n}{counts} \PY{o}{=} \PY{n}{execute}\PY{p}{(} \PY{n}{qc0}\PY{p}{,} \PY{n}{Aer}\PY{o}{.}\PY{n}{get\PYZus{}backend}\PY{p}{(}\PY{l+s+s1}{\PYZsq{}}\PY{l+s+s1}{qasm\PYZus{}simulator}\PY{l+s+s1}{\PYZsq{}}\PY{p}{)}\PY{p}{,}
                 \PY{n}{noise\PYZus{}model}\PY{o}{=}\PY{n}{noise\PYZus{}model}\PY{p}{,}
                 \PY{n}{shots}\PY{o}{=}\PY{n}{1024}\PY{p}{)}\PY{o}{.}\PY{n}{result}\PY{p}{(}\PY{p}{)}\PY{o}{.}\PY{n}{get\PYZus{}counts}\PY{p}{(}\PY{p}{)}

\PY{n+nb}{print}\PY{p}{(}\PY{n}{counts}\PY{p}{)}
\end{Verbatim}
\end{tcolorbox}

    \begin{Verbatim}[commandchars=\\\{\}]
\{'110':2, '001': 9, '100': 11, '010': 14, '000': 988\}
    \end{Verbatim}
\vspace{0.5cm}

Note that the \texttt{shots=1024} argument here is actually the default argument for the \texttt{execute} function, and so it nexed not be included (unless a different number of shots is required). As such, it will not be included in future code snippets.

    Here a set of typical results are shown. Results will vary for different runs, but will be qualitatively the same. Specifically, almost all results still come out
\texttt{\textquotesingle{}000\textquotesingle{}}, as they would if there
was no noise. Of the remaining possibilities, those with a majority of
\texttt{0}s are most likely. Much less than 10 of the 1024 samples will come
out with a majority of \texttt{1}s. When using this circuit to encode a
\texttt{0}, this means that \(P<1\%\)

Now let's try the same for storing a \texttt{1} using three qubits in
state \(\left|1\right\rangle\).

    \begin{tcolorbox}[ size=fbox, boxrule=1pt, colback=cellbackground, colframe=cellborder]
\prompt{In}{incolor}{5}{\boxspacing}
\begin{Verbatim}[commandchars=\\\{\}]
\PY{c+c1}{\PYZsh{} initialize circuit with three qubits in the 0 state}
\PY{n}{qc1} \PY{o}{=} \PY{n}{QuantumCircuit}\PY{p}{(}\PY{l+m+mi}{3}\PY{p}{,}\PY{l+m+mi}{3}\PY{p}{,}\PY{n}{name}\PY{o}{=}\PY{l+s+s1}{\PYZsq{}}\PY{l+s+s1}{0}\PY{l+s+s1}{\PYZsq{}}\PY{p}{)}
\PY{n}{qc1}\PY{o}{.}\PY{n}{x}\PY{p}{(}\PY{n}{qc1}\PY{o}{.}\PY{n}{qregs}\PY{p}{[}\PY{l+m+mi}{0}\PY{p}{]}\PY{p}{)} \PY{c+c1}{\PYZsh{} flip each 0 to 1}

\PY{n}{qc1}\PY{o}{.}\PY{n}{measure}\PY{p}{(}\PY{n}{qc1}\PY{o}{.}\PY{n}{qregs}\PY{p}{[}\PY{l+m+mi}{0}\PY{p}{]}\PY{p}{,}\PY{n}{qc1}\PY{o}{.}\PY{n}{cregs}\PY{p}{[}\PY{l+m+mi}{0}\PY{p}{]}\PY{p}{)} \PY{c+c1}{\PYZsh{} measure the qubits}

\PY{c+c1}{\PYZsh{} run the circuit with th noise model and extract the counts}
\PY{n}{counts} \PY{o}{=} \PY{n}{execute}\PY{p}{(} \PY{n}{qc1}\PY{p}{,} \PY{n}{Aer}\PY{o}{.}\PY{n}{get\PYZus{}backend}\PY{p}{(}\PY{l+s+s1}{\PYZsq{}}\PY{l+s+s1}{qasm\PYZus{}simulator}\PY{l+s+s1}{\PYZsq{}}\PY{p}{)}\PY{p}{,}
                 \PY{n}{noise\PYZus{}model}\PY{o}{=}\PY{n}{noise\PYZus{}model}\PY{p}{)}\PY{o}{.}\PY{n}{result}\PY{p}{(}\PY{p}{)}\PY{o}{.}\PY{n}{get\PYZus{}counts}\PY{p}{(}\PY{p}{)}

\PY{n+nb}{print}\PY{p}{(}\PY{n}{counts}\PY{p}{)}
\end{Verbatim}
\end{tcolorbox}

    \begin{Verbatim}[commandchars=\\\{\}]
\{'110': 15, '011': 17, '111': 974, '101': 18\}
    \end{Verbatim}
\vspace{0.5cm}

    The number of samples that come out with a majority in the wrong state
(\texttt{0} in this case) is again much less than 100, so \(P<1\%\).
Whether we store a \texttt{0} or a \texttt{1}, we can retrieve the
information with a smaller probability of error than either of our
sources of noise.

This was possible because the noise we considered was relatively weak.
As we increase \(\rho_{meas}\) and \(\rho_{gate}\), the higher the probability
\(P\) will be. The extreme case of this is for either of them to have a
\(50/50\) chance of applying the bit flip error, \texttt{x}. For
example, let's run the same circuit as before but with \(\rho_{meas}=0.5\)
and \(\rho_{gate}=0\).

    \begin{tcolorbox}[ size=fbox, boxrule=1pt, colback=cellbackground, colframe=cellborder]
\prompt{In}{incolor}{6}{\boxspacing}
\begin{Verbatim}[commandchars=\\\{\}]
\PY{n}{noise\PYZus{}model} \PY{o}{=} \PY{n}{get\PYZus{}noise}\PY{p}{(}\PY{l+m+mf}{0.5}\PY{p}{,}\PY{l+m+mf}{0.0}\PY{p}{)}
\PY{n}{counts} \PY{o}{=} \PY{n}{execute}\PY{p}{(} \PY{n}{qc1}\PY{p}{,} \PY{n}{Aer}\PY{o}{.}\PY{n}{get\PYZus{}backend}\PY{p}{(}\PY{l+s+s1}{\PYZsq{}}\PY{l+s+s1}{qasm\PYZus{}simulator}\PY{l+s+s1}{\PYZsq{}}\PY{p}{)}\PY{p}{,}
                 \PY{n}{noise\PYZus{}model}\PY{o}{=}\PY{n}{noise\PYZus{}model}\PY{p}{)}\PY{o}{.}\PY{n}{result}\PY{p}{(}\PY{p}{)}\PY{o}{.}\PY{n}{get\PYZus{}counts}\PY{p}{(}\PY{p}{)}
\PY{n+nb}{print}\PY{p}{(}\PY{n}{counts}\PY{p}{)}
\end{Verbatim}
\end{tcolorbox}

    \begin{Verbatim}[commandchars=\\\{\}]
\{'000': 123, '001': 125, '011': 121, '100': 131, '010': 124, '110': 130, '111':
140, '101': 130\}
    \end{Verbatim}
    \vspace{0.5cm}

    With this noise, all outcomes occur with equal probability, with
differences in results being due only to statistical noise. No trace of
the encoded state remains. This is an important point to consider for
error correction: sometimes the noise is too strong to be corrected. The
optimal approach is to combine a good way of encoding the information
you require, with hardware whose noise is not too strong.

    \hypertarget{storing-qubits}{%
\subsection{Storing qubits}\label{storing-qubits}}

So far, we have considered cases where there is no delay between
encoding and decoding. For qubits, this means that there is no
significant amount of time that passes between initializing the circuit,
and making the final measurements.

However, there are many cases for which there will be a significant
delay. As an obvious example, one may wish to encode a quantum state and
store it for a long time, like a quantum hard drive. A less obvious but
much more important example is performing fault-tolerant quantum
computation itself. For this, we need to store quantum states and
preserve their integrity during the computation. This must also be done
in a way that allows us to manipulate the stored information in any way
we need, and which corrects any errors we may introduce when performing
the manipulations.

In all cases, we need account for the fact that errors do not only occur
when something happens (like a gate or measurement), they also occur
when the qubits are idle. Such noise is due to the fact that the qubits
interact with each other and their environment. The longer we leave our
qubits idle for, the greater the effects of this noise becomes. If we
leave them for long enough, we'll encounter a situation like the
\(\rho_{meas}=0.5\) case above, where the noise is too strong for errors to
be reliably corrected.

The solution is to keep measuring throughout. No qubit is left idle for
too long. Instead, information is constantly being extracted from the
system to keep track of the errors that have occurred.

For the case of classical information, where we simply wish to store a
\texttt{0} or \texttt{1}, this can be done by just constantly measuring
the value of each qubit. By keeping track of when the values change due
to noise, we can easily deduce a history of when errors occurred.

For quantum information, however, it is not so easy. For example,
consider the case that we wish to encode the logical state
\(\left|+\right\rangle\). Our encoding is such that

\[\left|0\right\rangle \rightarrow \left|000\right\rangle,~~~ \left|1\right\rangle \rightarrow \left|111\right\rangle.\]

To encode the logical \(\left|+\right\rangle\) state we therefore need

\[\left|+\right\rangle=\frac{1}{\sqrt{2}}\left(\left|0\right\rangle+\left|1\right\rangle\right)\rightarrow \frac{1}{\sqrt{2}}\left(\left|000\right\rangle+\left|111\right\rangle\right).\]

With the repetition encoding that we are using, a z measurement (which
distinguishes between the \(\left|0\right\rangle\) and
\(\left|1\right\rangle\) states) of the logical qubit is done using a z
measurement of each physical qubit. The final result for the logical
measurement is decoded from the physical qubit measurement results by
simply looking which output is in the majority.

As mentioned earlier, we can keep track of errors on logical qubits that
are stored for a long time by constantly performing z measurements of
the physical qubits. However, note that this effectively corresponds to
constantly performing z measurements of the logical qubit. This is fine
if we are simply storing a \texttt{0} or \texttt{1}, but it has
undesired effects if we are storing a superposition. Specifically: the
first time we do such a check for errors, we will collapse the
superposition.

This is not ideal. If we wanted to do some computation on our logical
qubit, or if we wish to perform a basis change before final measurement,
we need to preserve the superposition. Destroying it is an error. But
this is not an error caused by imperfections in our devices. It is an
error that we have introduced as part of our attempts to correct errors.
And since we cannot hope to recreate any arbitrary superposition stored
in our quantum computer, it is an error that cannot be corrected.

For this reason, we must find another way of keeping track of the errors
that occur when our logical qubit is stored for long times. This should
give us the information we need to detect and correct errors, and to
decode the final measurement result with high probability. However, it
should not cause uncorrectable errors to occur during the process by
collapsing superpositions that we need to preserve.

The way to do this is with the following circuit element.
    \begin{tcolorbox}[ size=fbox, boxrule=1pt, colback=cellbackground, colframe=cellborder]
\prompt{In}{incolor}{7}{\boxspacing}
\begin{Verbatim}[commandchars=\\\{\}]
\PY{k+kn}{from} \PY{n+nn}{qiskit} \PY{k+kn}{import} \PY{n}{QuantumRegister}\PY{p}{,} \PY{n}{ClassicalRegister}

\PY{n}{cq} \PY{o}{=} \PY{n}{QuantumRegister}\PY{p}{(}\PY{l+m+mi}{2}\PY{p}{,}\PY{l+s+s1}{\PYZsq{}}\PY{l+s+s1}{code}\PY{l+s+s1}{\PYZbs{}}\PY{l+s+s1}{ qubit}\PY{l+s+s1}{\PYZbs{}}\PY{l+s+s1}{ }\PY{l+s+s1}{\PYZsq{}}\PY{p}{)}
\PY{n}{lq} \PY{o}{=} \PY{n}{QuantumRegister}\PY{p}{(}\PY{l+m+mi}{1}\PY{p}{,}\PY{l+s+s1}{\PYZsq{}}\PY{l+s+s1}{ancilla}\PY{l+s+s1}{\PYZbs{}}\PY{l+s+s1}{ qubit}\PY{l+s+s1}{\PYZbs{}}\PY{l+s+s1}{ }\PY{l+s+s1}{\PYZsq{}}\PY{p}{)}
\PY{n}{sb} \PY{o}{=} \PY{n}{ClassicalRegister}\PY{p}{(}\PY{l+m+mi}{1}\PY{p}{,}\PY{l+s+s1}{\PYZsq{}}\PY{l+s+s1}{syndrome}\PY{l+s+s1}{\PYZbs{}}\PY{l+s+s1}{ bit}\PY{l+s+s1}{\PYZbs{}}\PY{l+s+s1}{ }\PY{l+s+s1}{\PYZsq{}}\PY{p}{)}
\PY{n}{qc} \PY{o}{=} \PY{n}{QuantumCircuit}\PY{p}{(}\PY{n}{cq}\PY{p}{,}\PY{n}{lq}\PY{p}{,}\PY{n}{sb}\PY{p}{)}
\PY{n}{qc}\PY{o}{.}\PY{n}{cx}\PY{p}{(}\PY{n}{cq}\PY{p}{[}\PY{l+m+mi}{0}\PY{p}{]}\PY{p}{,}\PY{n}{lq}\PY{p}{[}\PY{l+m+mi}{0}\PY{p}{]}\PY{p}{)}
\PY{n}{qc}\PY{o}{.}\PY{n}{cx}\PY{p}{(}\PY{n}{cq}\PY{p}{[}\PY{l+m+mi}{1}\PY{p}{]}\PY{p}{,}\PY{n}{lq}\PY{p}{[}\PY{l+m+mi}{0}\PY{p}{]}\PY{p}{)}
\PY{n}{qc}\PY{o}{.}\PY{n}{measure}\PY{p}{(}\PY{n}{lq}\PY{p}{,}\PY{n}{sb}\PY{p}{)}
\PY{n}{qc}\PY{o}{.}\PY{n}{draw}\PY{p}{(}\PY{n}{output}\PY{o}{=}\PY{l+s+s1}{\PYZsq{}}\PY{l+s+s1}{mpl}\PY{l+s+s1}{\PYZsq{}}\PY{p}{)}
\end{Verbatim}
\end{tcolorbox}

    \begin{center}
    \adjustimage{max size={0.667\linewidth}{0.9\paperheight}}{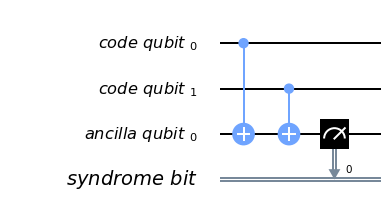}
    \end{center}
     \vspace{0.5cm}

    Here we have three physical qubits. Two are called `code qubits', and
the other is called an `ancilla qubit'. One bit of output is extracted,
called the syndrome bit. The ancilla qubit is always initialized in
state \(\left|0\right\rangle\). The code qubits, however, can be
initialized in different states. To see what affect different inputs
have on the output, we can create a circuit \texttt{qc\_init} that
prepares the code qubits in some state, and then run the circuit
\texttt{qc\_init+qc}.

First, the trivial case: \texttt{qc\_init} does nothing, and so the code
qubits are initially \(\left|00\right\rangle\).

    \begin{tcolorbox}[ size=fbox, boxrule=1pt, colback=cellbackground, colframe=cellborder]
\prompt{In}{incolor}{8}{\boxspacing}
\begin{Verbatim}[commandchars=\\\{\}]
\PY{n}{qc\PYZus{}init} \PY{o}{=} \PY{n}{QuantumCircuit}\PY{p}{(}\PY{n}{cq}\PY{p}{)}

\PY{p}{(}\PY{n}{qc\PYZus{}init}\PY{o}{+}\PY{n}{qc}\PY{p}{)}\PY{o}{.}\PY{n}{draw}\PY{p}{(}\PY{n}{output}\PY{o}{=}\PY{l+s+s1}{\PYZsq{}}\PY{l+s+s1}{mpl}\PY{l+s+s1}{\PYZsq{}}\PY{p}{)}
\end{Verbatim}
\end{tcolorbox}

    \begin{center}
    \adjustimage{max size={0.667\linewidth}{0.9\paperheight}}{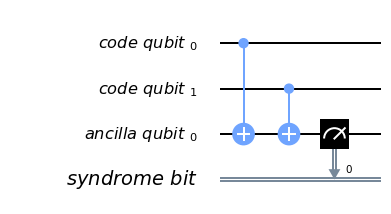}
    \end{center}
     \vspace{0.5cm}

    \begin{tcolorbox}[ size=fbox, boxrule=1pt, colback=cellbackground, colframe=cellborder]
\prompt{In}{incolor}{9}{\boxspacing}
\begin{Verbatim}[commandchars=\\\{\}]
\PY{n}{counts} \PY{o}{=} \PY{n}{execute}\PY{p}{(} \PY{n}{qc\PYZus{}init}\PY{o}{+}\PY{n}{qc}\PY{p}{,} \PY{n}{Aer}\PY{o}{.}\PY{n}{get\PYZus{}backend}\PY{p}{(}\PY{l+s+s1}{\PYZsq{}}\PY{l+s+s1}{qasm\PYZus{}simulator}\PY{l+s+s1}{\PYZsq{}}\PY{p}{)}
                \PY{p}{)}\PY{o}{.}\PY{n}{result}\PY{p}{(}\PY{p}{)}\PY{o}{.}\PY{n}{get\PYZus{}counts}\PY{p}{(}\PY{p}{)}
\PY{n+nb}{print}\PY{p}{(}\PY{l+s+s1}{\PYZsq{}}\PY{l+s+s1}{Results:}\PY{l+s+s1}{\PYZsq{}}\PY{p}{,}\PY{n}{counts}\PY{p}{)}
\end{Verbatim}
\end{tcolorbox}

    \begin{Verbatim}[commandchars=\\\{\}]
Results: \{'0': 1024\}
    \end{Verbatim}
    \vspace{0.5cm}

    The outcome, in all cases, is \texttt{0}.

Now let's try an initial state of \(\left|11\right\rangle\).

    \begin{tcolorbox}[ size=fbox, boxrule=1pt, colback=cellbackground, colframe=cellborder]
\prompt{In}{incolor}{10}{\boxspacing}
\begin{Verbatim}[commandchars=\\\{\}]
\PY{n}{qc\PYZus{}init} \PY{o}{=} \PY{n}{QuantumCircuit}\PY{p}{(}\PY{n}{cq}\PY{p}{)}
\PY{n}{qc\PYZus{}init}\PY{o}{.}\PY{n}{x}\PY{p}{(}\PY{n}{cq}\PY{p}{)}

\PY{p}{(}\PY{n}{qc\PYZus{}init}\PY{o}{+}\PY{n}{qc}\PY{p}{)}\PY{o}{.}\PY{n}{draw}\PY{p}{(}\PY{n}{output}\PY{o}{=}\PY{l+s+s1}{\PYZsq{}}\PY{l+s+s1}{mpl}\PY{l+s+s1}{\PYZsq{}}\PY{p}{)}
\end{Verbatim}
\end{tcolorbox}

    \begin{center}
    \adjustimage{max size={0.667\linewidth}{0.9\paperheight}}{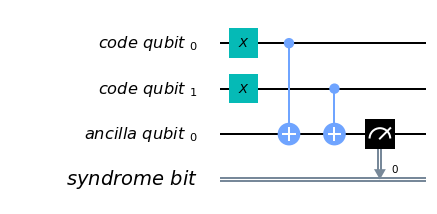}
    \end{center}
     \vspace{0.5cm}

    \begin{tcolorbox}[ size=fbox, boxrule=1pt, colback=cellbackground, colframe=cellborder]
\prompt{In}{incolor}{11}{\boxspacing}
\begin{Verbatim}[commandchars=\\\{\}]
\PY{n}{counts} \PY{o}{=} \PY{n}{execute}\PY{p}{(} \PY{n}{qc\PYZus{}init}\PY{o}{+}\PY{n}{qc}\PY{p}{,} \PY{n}{Aer}\PY{o}{.}\PY{n}{get\PYZus{}backend}\PY{p}{(}\PY{l+s+s1}{\PYZsq{}}\PY{l+s+s1}{qasm\PYZus{}simulator}\PY{l+s+s1}{\PYZsq{}}\PY{p}{)}
                \PY{p}{)}\PY{o}{.}\PY{n}{result}\PY{p}{(}\PY{p}{)}\PY{o}{.}\PY{n}{get\PYZus{}counts}\PY{p}{(}\PY{p}{)}
\PY{n+nb}{print}\PY{p}{(}\PY{l+s+s1}{\PYZsq{}}\PY{l+s+s1}{Results:}\PY{l+s+s1}{\PYZsq{}}\PY{p}{,}\PY{n}{counts}\PY{p}{)}
\end{Verbatim}
\end{tcolorbox}

    \begin{Verbatim}[commandchars=\\\{\}]
Results: \{'0': 1024\}
    \end{Verbatim}
    \vspace{0.5cm}

    The outcome in this case is also always \texttt{0}. Given the linearity
of quantum mechanics, we can expect the same to be true also for any
superposition of \(\left|00\right\rangle\) and
\(\left|11\right\rangle\), such as the example below.

    \begin{tcolorbox}[ size=fbox, boxrule=1pt, colback=cellbackground, colframe=cellborder]
\prompt{In}{incolor}{12}{\boxspacing}
\begin{Verbatim}[commandchars=\\\{\}]
\PY{n}{qc\PYZus{}init} \PY{o}{=} \PY{n}{QuantumCircuit}\PY{p}{(}\PY{n}{cq}\PY{p}{)}
\PY{n}{qc\PYZus{}init}\PY{o}{.}\PY{n}{h}\PY{p}{(}\PY{n}{cq}\PY{p}{[}\PY{l+m+mi}{0}\PY{p}{]}\PY{p}{)}
\PY{n}{qc\PYZus{}init}\PY{o}{.}\PY{n}{cx}\PY{p}{(}\PY{n}{cq}\PY{p}{[}\PY{l+m+mi}{0}\PY{p}{]}\PY{p}{,}\PY{n}{cq}\PY{p}{[}\PY{l+m+mi}{1}\PY{p}{]}\PY{p}{)}

\PY{p}{(}\PY{n}{qc\PYZus{}init}\PY{o}{+}\PY{n}{qc}\PY{p}{)}\PY{o}{.}\PY{n}{draw}\PY{p}{(}\PY{n}{output}\PY{o}{=}\PY{l+s+s1}{\PYZsq{}}\PY{l+s+s1}{mpl}\PY{l+s+s1}{\PYZsq{}}\PY{p}{)}
\end{Verbatim}
\end{tcolorbox}

    \begin{center}
    \adjustimage{max size={0.667\linewidth}{0.9\paperheight}}{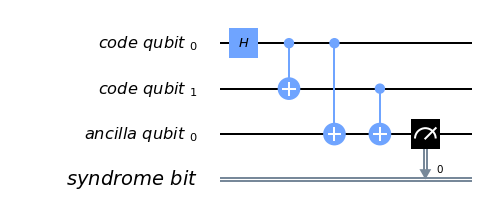}
    \end{center}
     \vspace{0.5cm}

    \begin{tcolorbox}[ size=fbox, boxrule=1pt, colback=cellbackground, colframe=cellborder]
\prompt{In}{incolor}{13}{\boxspacing}
\begin{Verbatim}[commandchars=\\\{\}]
\PY{n}{counts} \PY{o}{=} \PY{n}{execute}\PY{p}{(} \PY{n}{qc\PYZus{}init}\PY{o}{+}\PY{n}{qc}\PY{p}{,} \PY{n}{Aer}\PY{o}{.}\PY{n}{get\PYZus{}backend}\PY{p}{(}\PY{l+s+s1}{\PYZsq{}}\PY{l+s+s1}{qasm\PYZus{}simulator}\PY{l+s+s1}{\PYZsq{}}\PY{p}{)}
                \PY{p}{)}\PY{o}{.}\PY{n}{result}\PY{p}{(}\PY{p}{)}\PY{o}{.}\PY{n}{get\PYZus{}counts}\PY{p}{(}\PY{p}{)}
\PY{n+nb}{print}\PY{p}{(}\PY{l+s+s1}{\PYZsq{}}\PY{l+s+s1}{Results:}\PY{l+s+s1}{\PYZsq{}}\PY{p}{,}\PY{n}{counts}\PY{p}{)}
\end{Verbatim}
\end{tcolorbox}

    \begin{Verbatim}[commandchars=\\\{\}]
Results: \{'0': 1024\}
    \end{Verbatim}
    \vspace{0.5cm}

    The opposite outcome will be found for an initial state of
\(\left|01\right\rangle\), \(\left|10\right\rangle\) or any
superposition thereof.

    \begin{tcolorbox}[ size=fbox, boxrule=1pt, colback=cellbackground, colframe=cellborder]
\prompt{In}{incolor}{14}{\boxspacing}
\begin{Verbatim}[commandchars=\\\{\}]
\PY{n}{qc\PYZus{}init} \PY{o}{=} \PY{n}{QuantumCircuit}\PY{p}{(}\PY{n}{cq}\PY{p}{)}
\PY{n}{qc\PYZus{}init}\PY{o}{.}\PY{n}{h}\PY{p}{(}\PY{n}{cq}\PY{p}{[}\PY{l+m+mi}{0}\PY{p}{]}\PY{p}{)}
\PY{n}{qc\PYZus{}init}\PY{o}{.}\PY{n}{cx}\PY{p}{(}\PY{n}{cq}\PY{p}{[}\PY{l+m+mi}{0}\PY{p}{]}\PY{p}{,}\PY{n}{cq}\PY{p}{[}\PY{l+m+mi}{1}\PY{p}{]}\PY{p}{)}
\PY{n}{qc\PYZus{}init}\PY{o}{.}\PY{n}{x}\PY{p}{(}\PY{n}{cq}\PY{p}{[}\PY{l+m+mi}{0}\PY{p}{]}\PY{p}{)}

\PY{p}{(}\PY{n}{qc\PYZus{}init}\PY{o}{+}\PY{n}{qc}\PY{p}{)}\PY{o}{.}\PY{n}{draw}\PY{p}{(}\PY{n}{output}\PY{o}{=}\PY{l+s+s1}{\PYZsq{}}\PY{l+s+s1}{mpl}\PY{l+s+s1}{\PYZsq{}}\PY{p}{)}
\end{Verbatim}
\end{tcolorbox}

    \begin{center}
    \adjustimage{max size={0.667\linewidth}{0.9\paperheight}}{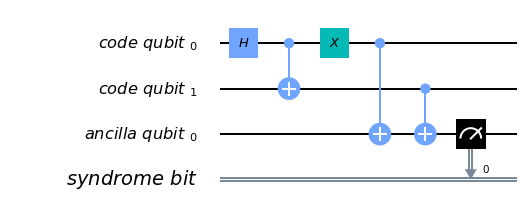}
    \end{center}
     \vspace{0.5cm}

    \begin{tcolorbox}[ size=fbox, boxrule=1pt, colback=cellbackground, colframe=cellborder]
\prompt{In}{incolor}{15}{\boxspacing}
\begin{Verbatim}[commandchars=\\\{\}]
\PY{n}{counts} \PY{o}{=} \PY{n}{execute}\PY{p}{(} \PY{n}{qc\PYZus{}init}\PY{o}{+}\PY{n}{qc}\PY{p}{,} \PY{n}{Aer}\PY{o}{.}\PY{n}{get\PYZus{}backend}\PY{p}{(}\PY{l+s+s1}{\PYZsq{}}\PY{l+s+s1}{qasm\PYZus{}simulator}\PY{l+s+s1}{\PYZsq{}}\PY{p}{)}
                \PY{p}{)}\PY{o}{.}\PY{n}{result}\PY{p}{(}\PY{p}{)}\PY{o}{.}\PY{n}{get\PYZus{}counts}\PY{p}{(}\PY{p}{)}
\PY{n+nb}{print}\PY{p}{(}\PY{l+s+s1}{\PYZsq{}}\PY{l+s+s1}{Results:}\PY{l+s+s1}{\PYZsq{}}\PY{p}{,}\PY{n}{counts}\PY{p}{)}
\end{Verbatim}
\end{tcolorbox}

    \begin{Verbatim}[commandchars=\\\{\}]
Results: \{'1': 1024\}
    \end{Verbatim}
    \vspace{0.5cm}

    In such cases the output is always
\texttt{\textquotesingle{}1\textquotesingle{}}.

This measurement is therefore telling us about a collective property of
multiple qubits. Specifically, it looks at the two code qubits and
determines whether their state is the same or different in the z basis.
For basis states that are the same in the z basis, like
\(\left|00\right\rangle\) and \(\left|11\right\rangle\), the measurement
simply returns \texttt{0}. It also does so for any superposition of
these. Since it does not distinguish between these states in any way, it
also does not collapse such a superposition.

Similarly, For basis states that are different in the z basis it returns
a \texttt{1}. This occurs for \(\left|01\right\rangle\),
\(\left|10\right\rangle\) or any superposition thereof.

Now suppose we apply such a `syndrome measurement' on all pairs of
physical qubits in our repetition code. If their state is described by a
repeated \(\left|0\right\rangle\), a repeated \(\left|1\right\rangle\),
or any superposition thereof, all the syndrome measurements will return
\texttt{0}. Given this result, we will know that our states are indeed
encoded in the repeated states that we want them to be, and can deduce
that no errors have occurred. If some syndrome measurements return
\texttt{1}, however, it is a signature of an error. We can therefore use
these measurement results to determine how to decode the result.

    \hypertarget{quantum-repetition-code}{%
\subsection{Quantum repetition code}\label{quantum-repetition-code}}

We now know enough to understand exactly how the quantum version of the
repetition code is implemented

We can use it in Qiskit by importing the required tools from Ignis.

    \begin{tcolorbox}[ size=fbox, boxrule=1pt, colback=cellbackground, colframe=cellborder]
\prompt{In}{incolor}{16}{\boxspacing}
\begin{Verbatim}[commandchars=\\\{\}]
\PY{k+kn}{from} \PY{n+nn}{qiskit}\PY{n+nn}{.}\PY{n+nn}{ignis}\PY{n+nn}{.}\PY{n+nn}{verification}\PY{n+nn}{.}\PY{n+nn}{topological\PYZus{}codes} \PY{k+kn}{import} \PY{n}{RepetitionCode}
\PY{k+kn}{from} \PY{n+nn}{qiskit}\PY{n+nn}{.}\PY{n+nn}{ignis}\PY{n+nn}{.}\PY{n+nn}{verification}\PY{n+nn}{.}\PY{n+nn}{topological\PYZus{}codes} \PY{k+kn}{import} \PY{n}{lookuptable\PYZus{}decoding}
\PY{k+kn}{from} \PY{n+nn}{qiskit}\PY{n+nn}{.}\PY{n+nn}{ignis}\PY{n+nn}{.}\PY{n+nn}{verification}\PY{n+nn}{.}\PY{n+nn}{topological\PYZus{}codes} \PY{k+kn}{import} \PY{n}{GraphDecoder}
\end{Verbatim}
\end{tcolorbox}

    We are free to choose how many physical qubits we want the logical qubit
to be encoded in. We can also choose how many times the syndrome
measurements will be applied while we store our logical qubit, before
the final readout measurement. Let us start with the smallest
non-trivial case: three repetitions and one syndrome measurement round.
The circuits for the repetition code can then be created automatically
from the using the \texttt{RepetitionCode} object from Qiskit-Ignis.

    \begin{tcolorbox}[ size=fbox, boxrule=1pt, colback=cellbackground, colframe=cellborder]
\prompt{In}{incolor}{17}{\boxspacing}
\begin{Verbatim}[commandchars=\\\{\}]
\PY{n}{n} \PY{o}{=} \PY{l+m+mi}{3}
\PY{n}{T} \PY{o}{=} \PY{l+m+mi}{1}

\PY{n}{code} \PY{o}{=} \PY{n}{RepetitionCode}\PY{p}{(}\PY{n}{n}\PY{p}{,}\PY{n}{T}\PY{p}{)}
\end{Verbatim}
\end{tcolorbox}

    With this we can inspect various properties of the code, such as the
names of the qubit registers used for the code and ancilla qubits.

    The \texttt{RepetitionCode} contains two quantum circuits that implement
the code: One for each of the two possible logical bit values. Here are
those for logical \texttt{0} and \texttt{1}, respectively.

    \begin{tcolorbox}[ size=fbox, boxrule=1pt, colback=cellbackground, colframe=cellborder]
\prompt{In}{incolor}{18}{\boxspacing}
\begin{Verbatim}[commandchars=\\\{\}]
\PY{c+c1}{\PYZsh{} this bit is just needed to make the labels look nice}
\PY{k}{for} \PY{n}{reg} \PY{o+ow}{in} \PY{n}{code}\PY{o}{.}\PY{n}{circuit}\PY{p}{[}\PY{l+s+s1}{\PYZsq{}}\PY{l+s+s1}{0}\PY{l+s+s1}{\PYZsq{}}\PY{p}{]}\PY{o}{.}\PY{n}{qregs}\PY{o}{+}\PY{n}{code}\PY{o}{.}\PY{n}{circuit}\PY{p}{[}\PY{l+s+s1}{\PYZsq{}}\PY{l+s+s1}{1}\PY{l+s+s1}{\PYZsq{}}\PY{p}{]}\PY{o}{.}\PY{n}{cregs}\PY{p}{:}
    \PY{n}{reg}\PY{o}{.}\PY{n}{name} \PY{o}{=} \PY{n}{reg}\PY{o}{.}\PY{n}{name}\PY{o}{.}\PY{n}{replace}\PY{p}{(}\PY{l+s+s1}{\PYZsq{}}\PY{l+s+s1}{\PYZus{}}\PY{l+s+s1}{\PYZsq{}}\PY{p}{,}\PY{l+s+s1}{\PYZsq{}}\PY{l+s+s1}{\PYZbs{}}\PY{l+s+s1}{ }\PY{l+s+s1}{\PYZsq{}}\PY{p}{)} \PY{o}{+} \PY{l+s+s1}{\PYZsq{}}\PY{l+s+s1}{\PYZbs{}}\PY{l+s+s1}{ }\PY{l+s+s1}{\PYZsq{}}

\PY{n}{code}\PY{o}{.}\PY{n}{circuit}\PY{p}{[}\PY{l+s+s1}{\PYZsq{}}\PY{l+s+s1}{0}\PY{l+s+s1}{\PYZsq{}}\PY{p}{]}\PY{o}{.}\PY{n}{draw}\PY{p}{(}\PY{n}{output}\PY{o}{=}\PY{l+s+s1}{\PYZsq{}}\PY{l+s+s1}{mpl}\PY{l+s+s1}{\PYZsq{}}\PY{p}{)}
\end{Verbatim}
\end{tcolorbox}

    \begin{center}
    \adjustimage{max size={0.667\linewidth}{0.9\paperheight}}{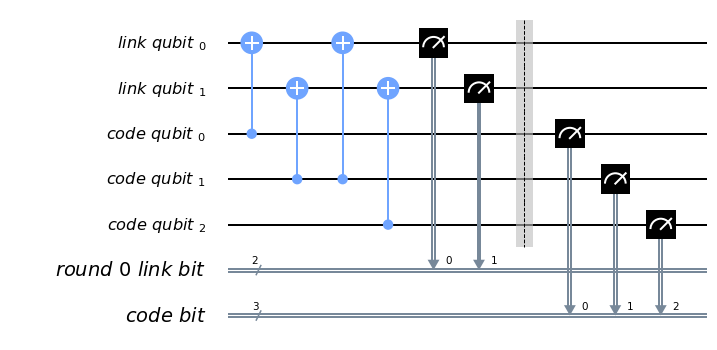}
    \end{center}
     \vspace{0.5cm}

    \begin{tcolorbox}[ size=fbox, boxrule=1pt, colback=cellbackground, colframe=cellborder]
\prompt{In}{incolor}{19}{\boxspacing}
\begin{Verbatim}[commandchars=\\\{\}]
\PY{n}{code}\PY{o}{.}\PY{n}{circuit}\PY{p}{[}\PY{l+s+s1}{\PYZsq{}}\PY{l+s+s1}{1}\PY{l+s+s1}{\PYZsq{}}\PY{p}{]}\PY{o}{.}\PY{n}{draw}\PY{p}{(}\PY{n}{output}\PY{o}{=}\PY{l+s+s1}{\PYZsq{}}\PY{l+s+s1}{mpl}\PY{l+s+s1}{\PYZsq{}}\PY{p}{)}
\end{Verbatim}
\end{tcolorbox}

    \begin{center}
    \adjustimage{max size={0.667\linewidth}{0.9\paperheight}}{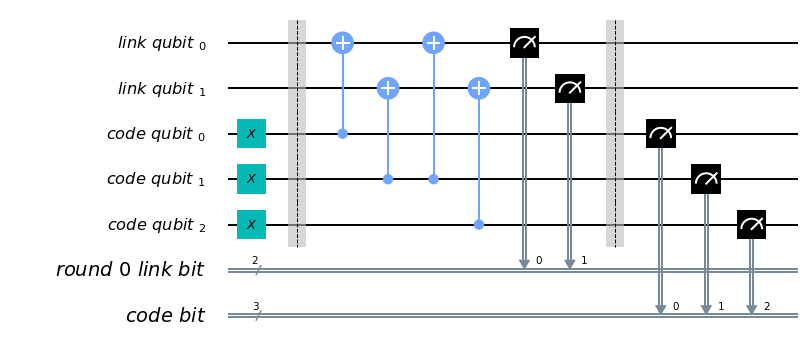}
    \end{center}
     \vspace{0.5cm}

    In these circuits, we have two types of physical qubits. There are the
`code qubits', which are the three physical qubits across which the
logical state is encoded. There are also the `link qubits', which serve
as the ancilla qubits for the syndrome measurements.

Our single round of syndrome measurements in these circuits consist of
just two syndrome measurements. One compares code qubits 0 and 1, and
the other compares code qubits 1 and 2. One might expect that a further
measurement, comparing code qubits 0 and 2, should be required to create
a full set. However, these two are sufficient. This is because the
information on whether 0 and 2 have the same z basis state can be
inferred from the same information about 0 and 1 with that for 1 and 2.
Indeed, for \(n\) qubits, we can get the required information from just
\(n-1\) syndrome measurements of neighbouring pairs of qubits.

Running these circuits on a simulator without any noise leads to very
simple results.

    \begin{tcolorbox}[ size=fbox, boxrule=1pt, colback=cellbackground, colframe=cellborder]
\prompt{In}{incolor}{20}{\boxspacing}
\begin{Verbatim}[commandchars=\\\{\}]
\PY{k}{def} \PY{n+nf}{get\PYZus{}raw\PYZus{}results}\PY{p}{(}\PY{n}{code}\PY{p}{,}\PY{n}{noise\PYZus{}model}\PY{o}{=}\PY{k+kc}{None}\PY{p}{)}\PY{p}{:}

    \PY{n}{circuits} \PY{o}{=} \PY{n}{code}\PY{o}{.}\PY{n}{get\PYZus{}circuit\PYZus{}list}\PY{p}{(}\PY{p}{)}
    \PY{n}{job} \PY{o}{=} \PY{n}{execute}\PY{p}{(} \PY{n}{circuits}\PY{p}{,} \PY{n}{Aer}\PY{o}{.}\PY{n}{get\PYZus{}backend}\PY{p}{(}\PY{l+s+s1}{\PYZsq{}}\PY{l+s+s1}{qasm\PYZus{}simulator}\PY{l+s+s1}{\PYZsq{}}\PY{p}{)}\PY{p}{,}
                  \PY{n}{noise\PYZus{}model}\PY{o}{=}\PY{n}{noise\PYZus{}model} \PY{p}{)}
    \PY{n}{raw\PYZus{}results} \PY{o}{=} \PY{p}{\PYZob{}}\PY{p}{\PYZcb{}}
    \PY{k}{for} \PY{n}{log} \PY{o+ow}{in} \PY{p}{[}\PY{l+s+s1}{\PYZsq{}}\PY{l+s+s1}{0}\PY{l+s+s1}{\PYZsq{}}\PY{p}{,}\PY{l+s+s1}{\PYZsq{}}\PY{l+s+s1}{1}\PY{l+s+s1}{\PYZsq{}}\PY{p}{]}\PY{p}{:}
        \PY{n}{raw\PYZus{}results}\PY{p}{[}\PY{n}{log}\PY{p}{]} \PY{o}{=} \PY{n}{job}\PY{o}{.}\PY{n}{result}\PY{p}{(}\PY{p}{)}\PY{o}{.}\PY{n}{get\PYZus{}counts}\PY{p}{(}\PY{n}{log}\PY{p}{)}
    \PY{k}{return} \PY{n}{raw\PYZus{}results}

\PY{n}{raw\PYZus{}results} \PY{o}{=} \PY{n}{get\PYZus{}raw\PYZus{}results}\PY{p}{(}\PY{n}{code}\PY{p}{)}
\PY{k}{for} \PY{n}{log} \PY{o+ow}{in} \PY{n}{raw\PYZus{}results}\PY{p}{:}
    \PY{n+nb}{print}\PY{p}{(}\PY{l+s+s1}{\PYZsq{}}\PY{l+s+s1}{Logical}\PY{l+s+s1}{\PYZsq{}}\PY{p}{,}\PY{n}{log}\PY{p}{,}\PY{l+s+s1}{\PYZsq{}}\PY{l+s+s1}{:}\PY{l+s+s1}{\PYZsq{}}\PY{p}{,}\PY{n}{raw\PYZus{}results}\PY{p}{[}\PY{n}{log}\PY{p}{]}\PY{p}{,}\PY{l+s+s1}{\PYZsq{}}\PY{l+s+se}{\PYZbs{}n}\PY{l+s+s1}{\PYZsq{}}\PY{p}{)}
\end{Verbatim}
\end{tcolorbox}

    \begin{Verbatim}[commandchars=\\\{\}]
Logical 0 : \{'000 00': 1024\}

Logical 1 : \{'111 00': 1024\}

    \end{Verbatim}
    \vspace{0.5cm}

    Here we see that the output comes in two parts. The part on the right
holds the outcomes of the two syndrome measurements. That on the left
holds the outcomes of the three final measurements of the code qubits.

For more measurement rounds, \(T=4\) for example, we would have the
results of more syndrome measurements on the right.

    \begin{tcolorbox}[ size=fbox, boxrule=1pt, colback=cellbackground, colframe=cellborder]
\prompt{In}{incolor}{21}{\boxspacing}
\begin{Verbatim}[commandchars=\\\{\}]
\PY{n}{code} \PY{o}{=} \PY{n}{RepetitionCode}\PY{p}{(}\PY{n}{n}\PY{p}{,}\PY{l+m+mi}{4}\PY{p}{)}

\PY{n}{raw\PYZus{}results} \PY{o}{=} \PY{n}{get\PYZus{}raw\PYZus{}results}\PY{p}{(}\PY{n}{code}\PY{p}{)}
\PY{k}{for} \PY{n}{log} \PY{o+ow}{in} \PY{n}{raw\PYZus{}results}\PY{p}{:}
    \PY{n+nb}{print}\PY{p}{(}\PY{l+s+s1}{\PYZsq{}}\PY{l+s+s1}{Logical}\PY{l+s+s1}{\PYZsq{}}\PY{p}{,}\PY{n}{log}\PY{p}{,}\PY{l+s+s1}{\PYZsq{}}\PY{l+s+s1}{:}\PY{l+s+s1}{\PYZsq{}}\PY{p}{,}\PY{n}{raw\PYZus{}results}\PY{p}{[}\PY{n}{log}\PY{p}{]}\PY{p}{,}\PY{l+s+s1}{\PYZsq{}}\PY{l+s+se}{\PYZbs{}n}\PY{l+s+s1}{\PYZsq{}}\PY{p}{)}
\end{Verbatim}
\end{tcolorbox}

    \begin{Verbatim}[commandchars=\\\{\}]
Logical 0 : \{'000 00 00 00 00': 1024\}

Logical 1 : \{'111 00 00 00 00': 1024\}

    \end{Verbatim}
    \vspace{0.5cm}

    For more repetitions, \(n=5\) for example, each set of measurements
would be larger. The final measurement on the left would be of \(n\)
qubits. The \(T\) syndrome measurements would each be of the \(n-1\)
possible neighbouring pairs.

    \begin{tcolorbox}[ size=fbox, boxrule=1pt, colback=cellbackground, colframe=cellborder]
\prompt{In}{incolor}{22}{\boxspacing}
\begin{Verbatim}[commandchars=\\\{\}]
\PY{n}{code} \PY{o}{=} \PY{n}{RepetitionCode}\PY{p}{(}\PY{l+m+mi}{5}\PY{p}{,}\PY{l+m+mi}{4}\PY{p}{)}

\PY{n}{raw\PYZus{}results} \PY{o}{=} \PY{n}{get\PYZus{}raw\PYZus{}results}\PY{p}{(}\PY{n}{code}\PY{p}{)}
\PY{k}{for} \PY{n}{log} \PY{o+ow}{in} \PY{n}{raw\PYZus{}results}\PY{p}{:}
    \PY{n+nb}{print}\PY{p}{(}\PY{l+s+s1}{\PYZsq{}}\PY{l+s+s1}{Logical}\PY{l+s+s1}{\PYZsq{}}\PY{p}{,}\PY{n}{log}\PY{p}{,}\PY{l+s+s1}{\PYZsq{}}\PY{l+s+s1}{:}\PY{l+s+s1}{\PYZsq{}}\PY{p}{,}\PY{n}{raw\PYZus{}results}\PY{p}{[}\PY{n}{log}\PY{p}{]}\PY{p}{,}\PY{l+s+s1}{\PYZsq{}}\PY{l+s+se}{\PYZbs{}n}\PY{l+s+s1}{\PYZsq{}}\PY{p}{)}
\end{Verbatim}
\end{tcolorbox}

    \begin{Verbatim}[commandchars=\\\{\}]
Logical 0 : \{'00000 0000 0000 0000 0000': 1024\}

Logical 1 : \{'11111 0000 0000 0000 0000': 1024\}

    \end{Verbatim}
    \vspace{0.5cm}

    \hypertarget{lookup-table-decoding}{%
\subsection{Lookup table decoding}\label{lookup-table-decoding}}

Now let's return to the \(n=3\), \(T=1\) example and look at a case with
some noise.

    \begin{tcolorbox}[ size=fbox, boxrule=1pt, colback=cellbackground, colframe=cellborder]
\prompt{In}{incolor}{23}{\boxspacing}
\begin{Verbatim}[commandchars=\\\{\}]
\PY{n}{code} \PY{o}{=} \PY{n}{RepetitionCode}\PY{p}{(}\PY{l+m+mi}{3}\PY{p}{,}\PY{l+m+mi}{1}\PY{p}{)}

\PY{n}{noise\PYZus{}model} \PY{o}{=} \PY{n}{get\PYZus{}noise}\PY{p}{(}\PY{l+m+mf}{0.05}\PY{p}{,}\PY{l+m+mf}{0.05}\PY{p}{)}

\PY{n}{raw\PYZus{}results} \PY{o}{=} \PY{n}{get\PYZus{}raw\PYZus{}results}\PY{p}{(}\PY{n}{code}\PY{p}{,}\PY{n}{noise\PYZus{}model}\PY{p}{)}
\PY{k}{for} \PY{n}{log} \PY{o+ow}{in} \PY{n}{raw\PYZus{}results}\PY{p}{:}
    \PY{n+nb}{print}\PY{p}{(}\PY{l+s+s1}{\PYZsq{}}\PY{l+s+s1}{Logical}\PY{l+s+s1}{\PYZsq{}}\PY{p}{,}\PY{n}{log}\PY{p}{,}\PY{l+s+s1}{\PYZsq{}}\PY{l+s+s1}{:}\PY{l+s+s1}{\PYZsq{}}\PY{p}{,}\PY{n}{raw\PYZus{}results}\PY{p}{[}\PY{n}{log}\PY{p}{]}\PY{p}{,}\PY{l+s+s1}{\PYZsq{}}\PY{l+s+se}{\PYZbs{}n}\PY{l+s+s1}{\PYZsq{}}\PY{p}{)}
\end{Verbatim}
\end{tcolorbox}

    \begin{Verbatim}[commandchars=\\\{\}]
Logical 0 : \{'011 10': 1, '001 10': 5, '111 00': 2, '010 01': 24, '000 11': 6,
'101 00': 2, '100 01': 2, '100 11': 1, '100 10': 7, '001 11': 2, '010 00': 52,
'001 01': 2, '110 01': 4, '000 00': 642, '000 01': 64, '001 00': 49, '010 10':
5, '000 10': 73, '011 00': 5, '111 01': 1, '011 01': 1, '101 01': 1, '100 00':
67, '110 00': 5, '010 11': 1\}

Logical 1 : \{'011 10': 13, '001 10': 3, '111 00': 594, '010 01': 2, '011 11': 2,
'111 10': 65, '101 00': 55, '100 11': 3, '101 10': 11, '001 11': 1, '101 11':
16, '001 01': 1, '010 00': 9, '110 01': 22, '111 11': 9, '000 00': 1, '110 11':
4, '001 00': 7, '010 10': 6, '110 10': 2, '011 00': 44, '111 01': 61, '011 01':
9, '101 01': 17, '100 00': 4, '110 00': 63\}

    \end{Verbatim}
    \vspace{0.5cm}

    Here we have created \texttt{raw\_results}, a dictionary that holds both
the results for a circuit encoding a logical \texttt{0} and \texttt{1}
encoded for a logical \texttt{1}.

Our task when confronted with any of the possible outcomes we see here
is to determine what the outcome should have been, if there was no
noise. For an outcome of
\texttt{\textquotesingle{}000\ 00\textquotesingle{}} or
\texttt{\textquotesingle{}111\ 00\textquotesingle{}}, the answer is
obvious. These are the results we just saw for a logical \texttt{0} and
logical \texttt{1}, respectively, when no errors occur. The former is
the most common outcome for the logical \texttt{0} even with noise, and
the latter is the most common for the logical \texttt{1}. We will
therefore conclude that the outcome was indeed that for logical
\texttt{0} whenever we encounter
\texttt{\textquotesingle{}000\ 00\textquotesingle{}}, and the same for
logical \texttt{1} when we encounter
\texttt{\textquotesingle{}111\ 00\textquotesingle{}}.

Though this tactic is optimal, it can nevertheless fail. Note that
\texttt{\textquotesingle{}111\ 00\textquotesingle{}} typically occurs in
a handful of cases for an encoded \texttt{0}, and
\texttt{\textquotesingle{}000\ 00\textquotesingle{}} similarly occurs for
an encoded \texttt{1}. In this case, through no fault of our own, we
will incorrectly decode the output. In these cases, a large number of
errors conspired to make it look like we had a noiseless case of the
opposite logical value, and so correction becomes impossible.

We can employ a similar tactic to decode all other outcomes. The outcome
\texttt{\textquotesingle{}001\ 00\textquotesingle{}}, for example,
occurs far more for a logical \texttt{0} than a logical \texttt{1}. This
is because it could be caused by just a single measurement error in the
former case (which incorrectly reports a single \texttt{0} to be
\texttt{1}), but would require at least two errors in the latter. So
whenever we see \texttt{\textquotesingle{}001\ 00\textquotesingle{}}, we
can decode it as a logical \texttt{0}.

Applying this tactic over all the strings is a form of so-called `lookup
table decoding'. Whenever an output string is obtained, it is compared to a large body of results for known logical values. Then most likely logical value can then be inferred. For many qubits,
this quickly becomes intractable, as the number of possible outcomes
becomes so large. In these cases, more algorithmic decoders are needed.
However, lookup table decoding works well for testing out small codes.

We can use tools in Qiskit to implement lookup table decoding for any
code. For this we need two sets of results. One is the set of results
that we actually want to decode, and for which we want to calculate the
probability of incorrect decoding, \(P\). We will use the
\texttt{raw\_results} we already have for this.

The other set of results is one to be used as the lookup table. This
will need to be run for a large number of samples, to ensure that it
gets good statistics for each possible outcome. We'll use
\texttt{shots=10000}.

    \begin{tcolorbox}[ size=fbox, boxrule=1pt, colback=cellbackground, colframe=cellborder]
\prompt{In}{incolor}{24}{\boxspacing}
\begin{Verbatim}[commandchars=\\\{\}]
\PY{n}{circuits} \PY{o}{=} \PY{n}{code}\PY{o}{.}\PY{n}{get\PYZus{}circuit\PYZus{}list}\PY{p}{(}\PY{p}{)}
\PY{n}{job} \PY{o}{=} \PY{n}{execute}\PY{p}{(} \PY{n}{circuits}\PY{p}{,} \PY{n}{Aer}\PY{o}{.}\PY{n}{get\PYZus{}backend}\PY{p}{(}\PY{l+s+s1}{\PYZsq{}}\PY{l+s+s1}{qasm\PYZus{}simulator}\PY{l+s+s1}{\PYZsq{}}\PY{p}{)}\PY{p}{,}
              \PY{n}{noise\PYZus{}model}\PY{o}{=}\PY{n}{noise\PYZus{}model}\PY{p}{,} \PY{n}{shots}\PY{o}{=}\PY{l+m+mi}{10000} \PY{p}{)}
\PY{n}{table\PYZus{}results} \PY{o}{=} \PY{p}{\PYZob{}}\PY{p}{\PYZcb{}}
\PY{k}{for} \PY{n}{log} \PY{o+ow}{in} \PY{p}{[}\PY{l+s+s1}{\PYZsq{}}\PY{l+s+s1}{0}\PY{l+s+s1}{\PYZsq{}}\PY{p}{,}\PY{l+s+s1}{\PYZsq{}}\PY{l+s+s1}{1}\PY{l+s+s1}{\PYZsq{}}\PY{p}{]}\PY{p}{:}
    \PY{n}{table\PYZus{}results}\PY{p}{[}\PY{n}{log}\PY{p}{]} \PY{o}{=} \PY{n}{job}\PY{o}{.}\PY{n}{result}\PY{p}{(}\PY{p}{)}\PY{o}{.}\PY{n}{get\PYZus{}counts}\PY{p}{(}\PY{n}{log}\PY{p}{)}
\end{Verbatim}
\end{tcolorbox}

    With this data, which we call \texttt{table\_results}, we can now use
the \texttt{lookuptable\_decoding} function from Qiskit. This takes each
outcome from \texttt{raw\_results} and decodes it with the information
in \texttt{table\_results}. Then it checks if the decoding was correct,
and uses this information to calculate \(P\).

    \begin{tcolorbox}[ size=fbox, boxrule=1pt, colback=cellbackground, colframe=cellborder]
\prompt{In}{incolor}{25}{\boxspacing}
\begin{Verbatim}[commandchars=\\\{\}]
\PY{n}{P} \PY{o}{=} \PY{n}{lookuptable\PYZus{}decoding}\PY{p}{(}\PY{n}{raw\PYZus{}results}\PY{p}{,}\PY{n}{table\PYZus{}results}\PY{p}{)}
\PY{n+nb}{print}\PY{p}{(}\PY{l+s+s1}{\PYZsq{}}\PY{l+s+s1}{P =}\PY{l+s+s1}{\PYZsq{}}\PY{p}{,}\PY{n}{P}\PY{p}{)}
\end{Verbatim}
\end{tcolorbox}

    \begin{Verbatim}[commandchars=\\\{\}]
P = \{'0': 0.0238, '1': 0.0237\}
    \end{Verbatim}
    \vspace{0.5cm}

    Here we see that the values for \(P\) are lower than those for
\(\rho_{meas}\) and \(\rho_{gate}\), so we get an improvement in the
reliability for storing the bit value. Note also that the value of \(P\)
for an encoded \texttt{1} is higher than that for \texttt{0}. This is
because the encoding of \texttt{1} requires the application of
\texttt{x} gates, which are an additional source of noise.

    \hypertarget{graph-theoretic-decoding}{%
\subsection{Graph theoretic
decoding}\label{graph-theoretic-decoding}}

The decoding considered above produces the best possible results, and
does so without needing to use any details of the code. However, it has
a major drawback that counters these advantages: the lookup table grows
exponentially large as code size increases. For this reason, decoding is
typically done in a more algorithmic manner that takes into account the
structure of the code and its resulting syndromes.

The \texttt{topological\_codes} module is designed to support multiple codes
that share the same structure, and therefore can be decoded using the same
methods. These methods are all based on similar graph theoretic minimization
problems, where the graph in question is one that can be derived from the
syndrome. The repetition code is one example that can be decoding in this way,
and it is with this example that we will explain the graph-theoretic decoding in
this section. Other examples are the toric and surface codes\cite{double,dennis},
 2D color codes\cite{bombin:06,delfosse:14} and matching codes\cite{wootton:15}.
 All of these are examples of so-called topological quantum error
correcting codes, which led to the name of the module. However, note
that not all topological codes are compatible with such a decoder. Also,
some non-topological codes will be compatible (such as the repetition
code).

To find the the graph that will be used in the decoding, some post-processing of
the syndromes is required. Instead of using the form shown
above, with the final measurement of the code qubits on the left and the
outputs of the syndrome measurement rounds on the right, we use the
\texttt{process\_results} method of the code object to rewrite them in a
different form.

For example, below is the processed form of a \texttt{raw\_results}
dictionary, in this case for \(n=3\) and \(T=2\). Only results with 50
or more samples are shown for clarity.

    \begin{tcolorbox}[ size=fbox, boxrule=1pt, colback=cellbackground, colframe=cellborder]
\prompt{In}{incolor}{26}{\boxspacing}
\begin{Verbatim}[commandchars=\\\{\}]
\PY{n}{code} \PY{o}{=} \PY{n}{RepetitionCode}\PY{p}{(}\PY{l+m+mi}{3}\PY{p}{,}\PY{l+m+mi}{2}\PY{p}{)}

\PY{n}{raw\PYZus{}results} \PY{o}{=} \PY{n}{get\PYZus{}raw\PYZus{}results}\PY{p}{(}\PY{n}{code}\PY{p}{,}\PY{n}{noise\PYZus{}model}\PY{p}{)}

\PY{n}{results} \PY{o}{=} \PY{n}{code}\PY{o}{.}\PY{n}{process\PYZus{}results}\PY{p}{(} \PY{n}{raw\PYZus{}results} \PY{p}{)}

\PY{k}{for} \PY{n}{log} \PY{o+ow}{in} \PY{p}{[}\PY{l+s+s1}{\PYZsq{}}\PY{l+s+s1}{0}\PY{l+s+s1}{\PYZsq{}}\PY{p}{,}\PY{l+s+s1}{\PYZsq{}}\PY{l+s+s1}{1}\PY{l+s+s1}{\PYZsq{}}\PY{p}{]}\PY{p}{:}
    \PY{n+nb}{print}\PY{p}{(}\PY{l+s+s1}{\PYZsq{}}\PY{l+s+se}{\PYZbs{}n}\PY{l+s+s1}{Logical }\PY{l+s+s1}{\PYZsq{}} \PY{o}{+} \PY{n}{log} \PY{o}{+} \PY{l+s+s1}{\PYZsq{}}\PY{l+s+s1}{:}\PY{l+s+s1}{\PYZsq{}}\PY{p}{)}
    \PY{n+nb}{print}\PY{p}{(}\PY{l+s+s1}{\PYZsq{}}\PY{l+s+s1}{raw results       }\PY{l+s+s1}{\PYZsq{}}\PY{p}{,} \PY{p}{\PYZob{}}\PY{n}{string}\PY{p}{:}\PY{n}{raw\PYZus{}results}\PY{p}{[}\PY{n}{log}\PY{p}{]}\PY{p}{[}\PY{n}{string}\PY{p}{]} \PY{k}{for} \PY{n}{string} \PY{o+ow}{in} \PY{n}{raw\PYZus{}results}\PY{p}{[}\PY{n}{log}\PY{p}{]} \PY{k}{if} \PY{n}{raw\PYZus{}results}\PY{p}{[}\PY{n}{log}\PY{p}{]}\PY{p}{[}\PY{n}{string}\PY{p}{]}\PY{o}{\PYZgt{}}\PY{o}{=}\PY{l+m+mi}{50} \PY{p}{\PYZcb{}}\PY{p}{)}
    \PY{n+nb}{print}\PY{p}{(}\PY{l+s+s1}{\PYZsq{}}\PY{l+s+s1}{processed results }\PY{l+s+s1}{\PYZsq{}}\PY{p}{,} \PY{p}{\PYZob{}}\PY{n}{string}\PY{p}{:}\PY{n}{results}\PY{p}{[}\PY{n}{log}\PY{p}{]}\PY{p}{[}\PY{n}{string}\PY{p}{]} \PY{k}{for} \PY{n}{string} \PY{o+ow}{in} \PY{n}{results}\PY{p}{[}\PY{n}{log}\PY{p}{]} \PY{k}{if} \PY{n}{results}\PY{p}{[}\PY{n}{log}\PY{p}{]}\PY{p}{[}\PY{n}{string}\PY{p}{]}\PY{o}{\PYZgt{}}\PY{o}{=}\PY{l+m+mi}{50} \PY{p}{\PYZcb{}}\PY{p}{)}
\end{Verbatim}
\end{tcolorbox}

    \begin{Verbatim}[commandchars=\\\{\}]

Logical 0:
raw results        \{'000 00 00': 485, '000 00 01': 55\}
processed results  \{'0 0  00 00 00': 485, '0 0  01 01 00': 55\}

Logical 1:
raw results        \{'111 10 00': 51, '111 01 00': 57, '111 00 00': 455, '111 00
10': 51\}
processed results  \{'1 1  00 10 10': 51, '1 1  00 01 01': 57, '1 1  00 00 00':
455, '1 1  10 10 00': 51\}
    \end{Verbatim}
    \vspace{0.5cm}

    Here we can see that
\texttt{\textquotesingle{}000\ 00\ 00\textquotesingle{}} has been
transformed to
\texttt{\textquotesingle{}0\ 0\ \ 00\ 00\ 00\textquotesingle{}}, and
\texttt{\textquotesingle{}111\ 00\ 00\textquotesingle{}} to
\texttt{\textquotesingle{}1\ 1\ \ 00\ 00\ 00\textquotesingle{}}, and so
on.

In these new strings, the \texttt{0\ 0} to the far left for the logical
\texttt{0} results and the \texttt{1\ 1} to the far left of the logical
\texttt{1} results are the logical readout. Any code qubit could be used
for this readout, since they should (without errors) all be equal. It
would therefore be possible in principle to just have a single
\texttt{0} or \texttt{1} at this position. We could also do as in the
original form of the result and have \(n\), one for each qubit. Instead
we use two, from the two qubits at either end of the line. The reason
for this will be shown later. In the absence of errors, these two values
will always be equal, since they represent the same encoded bit value.

After the logical values follow the \(n-1\) results of the syndrome
measurements for the first round. A \texttt{0} implies that the
corresponding pair of qubits have the same value, and \texttt{1} implies
they they are different from each other. There are \(n-1\) results
because the line of \(d\) code qubits has \(n-1\) possible neighbouring
pairs. In the absence of errors, they will all be \texttt{0}. This is
exactly the same as the first such set of syndrome results from the
original form of the result.

The next block is the next round of syndrome results. However, rather
than presenting these results directly, it instead gives us the syndrome
change between the first and second rounds. It is therefore the bitwise
\texttt{OR} of the syndrome measurement results from the second round
with those from the first. In the absence of errors, they will all be
\texttt{0}.

Any subsequent blocks follow the same formula, though the last of all
requires some comment. This is not measured using the standard method
(with a link qubit). Instead it is calculated from the final readout
measurement of all code qubits. Again it is presented as a syndrome
change, and will be all \texttt{0} in the absence of errors. This is the
\(T+1\)-th block of syndrome measurements since, as it is not done in
the same way as the others, it is not counted among the \(T\) syndrome
measurement rounds.

The following examples further illustrate this convention.

\textbf{Example 1:} \texttt{0\ 0\ \ 0110\ 0000\ 0000} represents a
\(d=5\), \(T=2\) repetition code with encoded \texttt{0}. The syndrome
shows that (most likely) the middle code qubit was flipped by an error
before the first measurement round. This causes it to disagree with both
neighboring code qubits for the rest of the circuit. This is shown by
the syndrome in the first round, but the blocks for subsequent rounds do
not report it as it no longer represents a change. Other sets of errors
could also have caused this syndrome, but they would need to be more
complex and so presumably less likely.

\textbf{Example 2:} \texttt{0\ 0\ \ 0010\ 0010\ 0000} represents a
\(d=5\), \(T=2\) repetition code with encoded \texttt{0}. Here one of
the syndrome measurements reported a difference between two code qubits
in the first round, leading to a \texttt{1}. The next round did not see
the same effect, and so resulted in a \texttt{0}. However, since this
disagreed with the previous result for the same syndrome measurement,
and since we track syndrome changes, this change results in another
\texttt{1}. Subsequent rounds also do not detect anything, but this no
longer represents a change and hence results in a \texttt{0} in the same
position. Most likely the measurement result leading to the first
\texttt{1} was an error.

\textbf{Example 3:} \texttt{0\ 1\ \ 0000\ 0001\ 0000} represents a
\(d=5\), \(T=2\) repetition code with encoded \texttt{1}. A code qubit
on the end of the line is flipped before the second round of syndrome
measurements. This is detected by only a single syndrome measurement,
because it is on the end of the line. For the same reason, it also
disturbs one of the logical readouts.

Note that in all these examples, a single error causes exactly two
characters in the string to change from the value they would have with
no errors. This is the defining feature of the convention used to
represent the syndrome in \texttt{topological\_codes}. It is used to
define the graph on which the decoding problem is defined.

Specifically, the graph is constructed by first taking the circuit
encoding logical \texttt{0}, for which all bit values in the output
string should be \texttt{0}. Many copies of this are then created and
run on a simulator, with a different single Pauli operator inserted into
each. This is done for each of the three types of Pauli operator on each
of the qubits and at every circuit depth. The output from each of these
circuits can be used to determine the effects of each possible single
error. Since the circuit contains only Clifford operations, the
simulation can be performed efficiently.

In each case, the error will change exactly two of the characters
(unless it has no effect). A graph is then constructed for which each
bit of the output string corresponds to a node, and the pairs of bits
affected by the same error correspond to an edge.

The process of decoding a particular output string typically requires
the algorithm to deduce which set of errors occurred, given the syndrome
found in the output string. This can be done by constructing a second
graph, containing only nodes that correspond to non-trivial syndrome
bits in the output. An edge is then placed between each pair of nodes,
with an corresponding weight equal to the length of the minimal path
between those nodes in the original graph. A set of errors consistent
with the syndrome then corresponds then to finding a perfect matching of
this graph. To deduce the most likely set of errors to have occurred, a
good tactic would be to find one with the least possible number of
errors that is consistent with the observed syndrome. This corresponds
to a minimum weight perfect matching of the graph~\cite{dennis}.

Using minimal weight perfect matching is a standard decoding technique
for the repetition code and surface codes \cite{dennis,fowler:12}, and is implemented in Qiskit
Ignis. It can also be used in other cases, such as Color codes, but it
does not find the best approximation of the most likely set of errors
for every code and noise model. For that reason, other decoding
techniques based on the same graph can be used. The \texttt{GraphDecoder}
of Qiskit Ignis calculates these graphs for a given code, and will
provide a range of methods to analyze it. At time of writing, only
minimum weight perfect matching is implemented.

Note that, for codes such as the surface code, it is not strictly true
than each single error will change the value of only two bits in the
output string. A \(\sigma^y\) error, for example would flip a pair of
values corresponding to two different types of stabilizer, which are
typically decoded independently. Output for these codes will therefore
be presented in a way that acknowledges this, and analysis of such
syndromes will correspondingly create multiple independent graphs to
represent the different syndrome types.

    \hypertarget{running-a-repetition-code-benchmarking-procedure}{%
\section{Running a repetition code benchmarking
procedure}\label{running-a-repetition-code-benchmarking-procedure}}

We will now run examples of repetition codes on real devices, and use
the results as a benchmark. First, we will briefly summarize the
process. This applies to this example of the repetition code, but also
for other benchmarking procedures in \texttt{topological\_codes}, and
indeed for Qiskit Ignis in general. In each case, the following
three-step process is used.

\begin{enumerate}
\def\labelenumi{\arabic{enumi}.}
\tightlist
\item
  A task is defined. Qiskit Ignis determines the set of circuits that
  must be run and creates them.
\item
  The circuits are run. This is typically done using Qiskit. However, in
  principle any service or experimental equipment could be interfaced.
\item
  Qiskit Ignis is used to process the results from the circuits, to
  create the output required for the given task.
\end{enumerate}

For \texttt{topological\_codes}, step 1 requires the type and size of
quantum error correction code to be chosen. Each type of code has a
dedicated Python class. A corresponding object is initialized by
providing the parameters required, such as \texttt{n} and \texttt{T} for
a \texttt{RepetitionCode} object. The resulting object then contains the
circuits corresponding to the given code encoding simple logical qubit
states (such as \(\left|0\right\rangle\) and \(\left|1\right\rangle\)),
and then running the procedure of error detection for a specified number
of rounds, before final readout in a straightforward logical basis
(typically a standard \(\left|0\right\rangle\)/\(\left|1\right\rangle\)
measurement).

For \texttt{topological\_codes}, the main processing of step 3 is the
decoding, which aims to mitigate for any errors in the final readout by
using the information obtained from error detection. The optimal
algorithm for decoding typically varies between codes.

The decoding is done by the \texttt{GraphDecoder} class. A corresponding
object is initialized by providing the code object for which the
decoding will be performed. This is then used to determine the graph on
which the decoding problem will be defined. The results can then be
processed using the various methods of the decoder object.

In the following we will see the above ideas put into practice for the
repetition code. In doing this we will employ two Boolean variables,
\texttt{step\_2} and \texttt{step\_3}. The variable \texttt{step\_2} is
used to show which parts of the program need to be run when taking data
from a device, and \texttt{step\_3} is used to show the parts which
process the resulting data.

Both are set to false by default, to ensure that all the program
snippets below can be run using only previously collected and processed
data. However, to obtain new data one only needs to use
\texttt{step\_2\ =\ True}, and perform decoding on any data one only
needs to use \texttt{step\_3\ =\ True}.

    \begin{tcolorbox}[ size=fbox, boxrule=1pt, colback=cellbackground, colframe=cellborder]
\prompt{In}{incolor}{27}{\boxspacing}
\begin{Verbatim}[commandchars=\\\{\}]
\PY{n}{step\PYZus{}2} \PY{o}{=} \PY{k+kc}{False}
\PY{n}{step\PYZus{}3} \PY{o}{=} \PY{k+kc}{False}
\end{Verbatim}
\end{tcolorbox}

    To benchmark a real device we need the tools required to access that
device over the cloud, and compile circuits suitable to run on it. These
are imported as follows.

    \begin{tcolorbox}[ size=fbox, boxrule=1pt, colback=cellbackground, colframe=cellborder]
\prompt{In}{incolor}{28}{\boxspacing}
\begin{Verbatim}[commandchars=\\\{\}]
\PY{k+kn}{from} \PY{n+nn}{qiskit} \PY{k+kn}{import} \PY{n}{IBMQ}
\PY{k+kn}{from} \PY{n+nn}{qiskit}\PY{n+nn}{.}\PY{n+nn}{compiler} \PY{k+kn}{import} \PY{n}{transpile}
\PY{k+kn}{from} \PY{n+nn}{qiskit}\PY{n+nn}{.}\PY{n+nn}{transpiler} \PY{k+kn}{import} \PY{n}{PassManager}
\end{Verbatim}
\end{tcolorbox}

    We can now create the backend object, which is used to run the circuits.
This is done by supplying the string used to specify the device. Here
the 53 qubit
\texttt{\textquotesingle{}ibmq\_rochester\textquotesingle{}} device is
used. Of the publicly accessible devices,
\texttt{\textquotesingle{}ibmq\_16\_melbourne\textquotesingle{}} can also be used. This
has 15 active qubits at time of writing. To gather new from this device, the device name below is the only thing that would need to be changed.

    \begin{tcolorbox}[ size=fbox, boxrule=1pt, colback=cellbackground, colframe=cellborder]
\prompt{In}{incolor}{29}{\boxspacing}
\begin{Verbatim}[commandchars=\\\{\}]
\PY{n}{device\PYZus{}name} \PY{o}{=} \PY{l+s+s1}{\PYZsq{}}\PY{l+s+s1}{ibmq\PYZus{}rochester}\PY{l+s+s1}{\PYZsq{}}

\PY{k}{if} \PY{n}{step\PYZus{}2}\PY{p}{:}
    
    \PY{n}{IBMQ}\PY{o}{.}\PY{n}{load\PYZus{}account}\PY{p}{(}\PY{p}{)}
    
    \PY{k}{for} \PY{n}{provider} \PY{o+ow}{in} \PY{n}{IBMQ}\PY{o}{.}\PY{n}{providers}\PY{p}{(}\PY{p}{)}\PY{p}{:}
        \PY{k}{for} \PY{n}{potential\PYZus{}backend} \PY{o+ow}{in} \PY{n}{provider}\PY{o}{.}\PY{n}{backends}\PY{p}{(}\PY{p}{)}\PY{p}{:}
            \PY{k}{if} \PY{n}{potential\PYZus{}backend}\PY{o}{.}\PY{n}{name}\PY{p}{(}\PY{p}{)}\PY{o}{==}\PY{n}{device\PYZus{}name}\PY{p}{:}
                \PY{n}{backend} \PY{o}{=} \PY{n}{potential\PYZus{}backend}

    \PY{n}{coupling\PYZus{}map} \PY{o}{=} \PY{n}{backend}\PY{o}{.}\PY{n}{configuration}\PY{p}{(}\PY{p}{)}\PY{o}{.}\PY{n}{coupling\PYZus{}map}
\end{Verbatim}
\end{tcolorbox}

    When running a circuit on a real device, a transpilation process is
first implemented. This changes the gates of the circuit into the native
gate set implement by the device. In some cases these changes are fairly
trivial, such as expressing each Hadamard as a single qubit rotation by
the corresponding Euler angles. However, the changes can be more major
if the circuit does not respect the connectivity of the device. For
example, suppose the circuit requires a controlled-NOT that is not
directly implemented by the device. The effect must be then be
reproduced with techniques such as using additional controlled-NOT gates
to move the qubit states around. As well as introducing additional
noise, this also delocalizes any noise already present. A single qubit
error in the original circuit could become a multiqubit monstrosity
under the action of the additional transpilation. Such non-trivial
transpilation must therefore be prevented when running quantum error
correction circuits.

\begin{figure}[t]
\begin{center}
{\includegraphics[width=0.667\columnwidth]{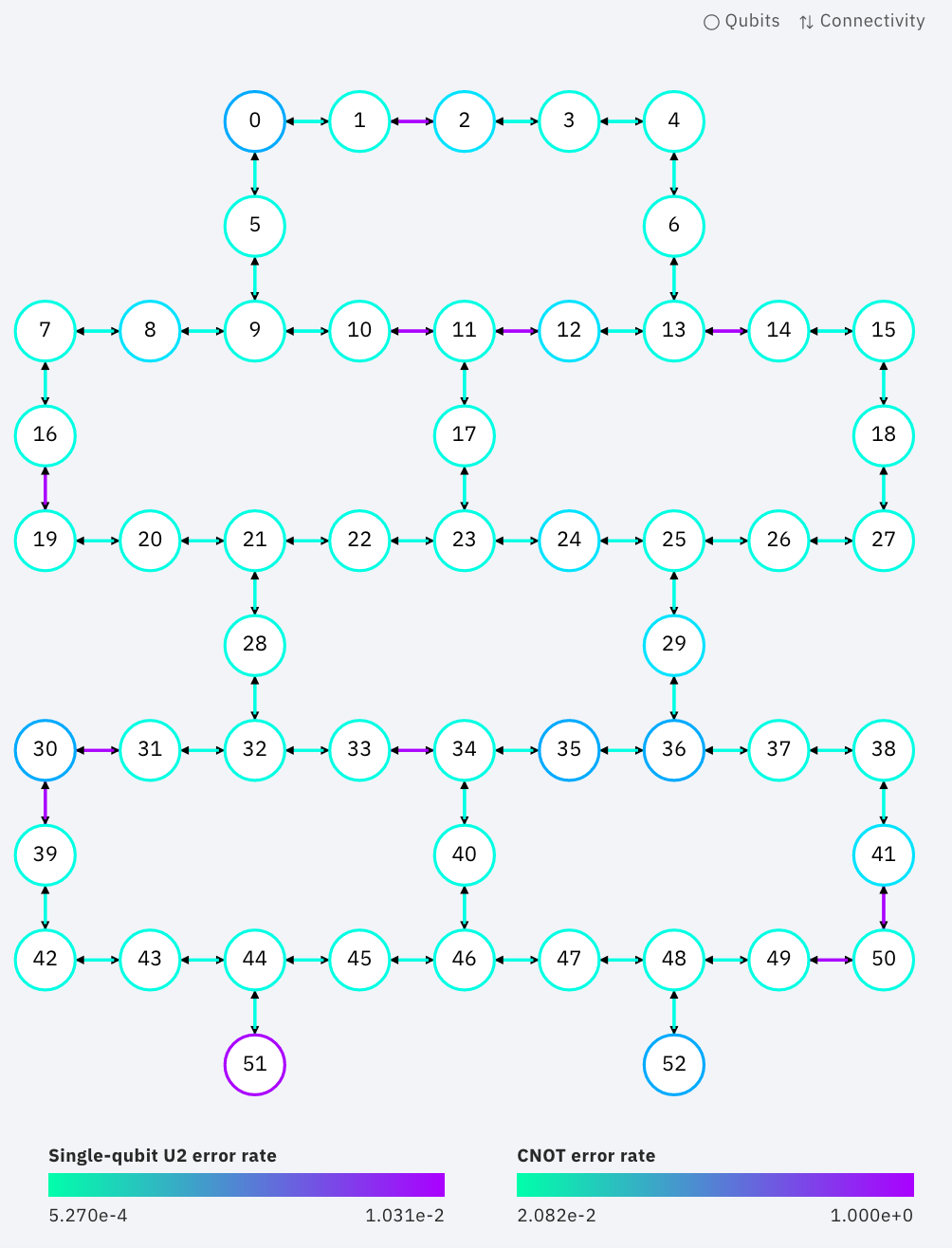}}
\caption{\label{device} The layout of the \emph{Rochester} device. Colours represent error probabilities for controlled-NOTs and readout on qubits.
}
\end{center}
\end{figure}

Tests of the repetition code require qubits to be effectively ordered
along a line. The only controlled-NOT gates required are between
neighbours along that line. Our first job is therefore to study the
coupling map of the device, shown in Fig.~\ref{device}, and find a line.

For Melbourne it is possible to find a line that covers all 15 qubits.
The choice one specified in the list \texttt{line} below is designed to
avoid the most error prone \texttt{cx} gates. For the 53 qubit
\emph{Rochester} device, there is no single line that covers all 53
qubits. Instead we can use the following choice, which covers 43.

    \begin{tcolorbox}[ size=fbox, boxrule=1pt, colback=cellbackground, colframe=cellborder]
\prompt{In}{incolor}{30}{\boxspacing}
\begin{Verbatim}[commandchars=\\\{\}]
\PY{k}{if} \PY{n}{device\PYZus{}name}\PY{o}{==}\PY{l+s+s1}{\PYZsq{}}\PY{l+s+s1}{ibmq\PYZus{}16\PYZus{}melbourne}\PY{l+s+s1}{\PYZsq{}}\PY{p}{:}
    \PY{n}{line} \PY{o}{=} \PY{p}{[}\PY{l+m+mi}{13}\PY{p}{,}\PY{l+m+mi}{14}\PY{p}{,}\PY{l+m+mi}{0}\PY{p}{,}\PY{l+m+mi}{1}\PY{p}{,}\PY{l+m+mi}{2}\PY{p}{,}\PY{l+m+mi}{12}\PY{p}{,}\PY{l+m+mi}{11}\PY{p}{,}\PY{l+m+mi}{3}\PY{p}{,}\PY{l+m+mi}{4}\PY{p}{,}\PY{l+m+mi}{10}\PY{p}{,}\PY{l+m+mi}{9}\PY{p}{,}\PY{l+m+mi}{5}\PY{p}{,}\PY{l+m+mi}{6}\PY{p}{,}\PY{l+m+mi}{8}\PY{p}{,}\PY{l+m+mi}{7}\PY{p}{]}
\PY{k}{elif} \PY{n}{device\PYZus{}name}\PY{o}{==}\PY{l+s+s1}{\PYZsq{}}\PY{l+s+s1}{ibmq\PYZus{}rochester}\PY{l+s+s1}{\PYZsq{}}\PY{p}{:}
    \PY{n}{line} \PY{o}{=} \PY{p}{[}\PY{l+m+mi}{10}\PY{p}{,}\PY{l+m+mi}{11}\PY{p}{,}\PY{l+m+mi}{17}\PY{p}{,}\PY{l+m+mi}{23}\PY{p}{,}\PY{l+m+mi}{22}\PY{p}{,}\PY{l+m+mi}{21}\PY{p}{,}\PY{l+m+mi}{20}\PY{p}{,}\PY{l+m+mi}{19}\PY{p}{,}\PY{l+m+mi}{16}\PY{p}{,}\PY{l+m+mi}{7}\PY{p}{,}\PY{l+m+mi}{8}\PY{p}{,}\PY{l+m+mi}{9}\PY{p}{,}\PY{l+m+mi}{5}\PY{p}{,}\PY{l+m+mi}{0}\PY{p}{,}\PY{l+m+mi}{1}\PY{p}{,}
            \PY{l+m+mi}{2}\PY{p}{,}\PY{l+m+mi}{3}\PY{p}{,}\PY{l+m+mi}{4}\PY{p}{,}\PY{l+m+mi}{6}\PY{p}{,}\PY{l+m+mi}{13}\PY{p}{,}\PY{l+m+mi}{14}\PY{p}{,}\PY{l+m+mi}{15}\PY{p}{,}\PY{l+m+mi}{18}\PY{p}{,}\PY{l+m+mi}{27}\PY{p}{,}\PY{l+m+mi}{26}\PY{p}{,}\PY{l+m+mi}{25}\PY{p}{,}\PY{l+m+mi}{29}\PY{p}{,}\PY{l+m+mi}{36}\PY{p}{,}\PY{l+m+mi}{37}\PY{p}{,}\PY{l+m+mi}{38}\PY{p}{,}
            \PY{l+m+mi}{41}\PY{p}{,}\PY{l+m+mi}{50}\PY{p}{,}\PY{l+m+mi}{49}\PY{p}{,}\PY{l+m+mi}{48}\PY{p}{,}\PY{l+m+mi}{47}\PY{p}{,}\PY{l+m+mi}{46}\PY{p}{,}\PY{l+m+mi}{45}\PY{p}{,}\PY{l+m+mi}{44}\PY{p}{,}\PY{l+m+mi}{43}\PY{p}{,}\PY{l+m+mi}{42}\PY{p}{,}\PY{l+m+mi}{39}\PY{p}{,}\PY{l+m+mi}{30}\PY{p}{,}\PY{l+m+mi}{31}\PY{p}{]}
\end{Verbatim}
\end{tcolorbox}

    Now we know how many qubits we have access to, we can create the
repetition code objects for each code that we will run. Note that a code
with \texttt{n} repetitions uses \(n\) code qubits and \(n-1\) link
qubits, and so \(2n-1\) in all.

    \begin{tcolorbox}[ size=fbox, boxrule=1pt, colback=cellbackground, colframe=cellborder]
\prompt{In}{incolor}{31}{\boxspacing}
\begin{Verbatim}[commandchars=\\\{\}]
\PY{n}{n\PYZus{}min} \PY{o}{=} \PY{l+m+mi}{3}
\PY{n}{n\PYZus{}max} \PY{o}{=} \PY{n+nb}{int}\PY{p}{(}\PY{p}{(}\PY{n+nb}{len}\PY{p}{(}\PY{n}{line}\PY{p}{)}\PY{o}{+}\PY{l+m+mi}{1}\PY{p}{)}\PY{o}{/}\PY{l+m+mi}{2}\PY{p}{)}

\PY{n}{code} \PY{o}{=} \PY{p}{\PYZob{}}\PY{p}{\PYZcb{}}

\PY{k}{for} \PY{n}{n} \PY{o+ow}{in} \PY{n+nb}{range}\PY{p}{(}\PY{n}{n\PYZus{}min}\PY{p}{,}\PY{n}{n\PYZus{}max}\PY{o}{+}\PY{l+m+mi}{1}\PY{p}{)}\PY{p}{:}
    \PY{n}{code}\PY{p}{[}\PY{n}{n}\PY{p}{]} \PY{o}{=} \PY{n}{RepetitionCode}\PY{p}{(}\PY{n}{n}\PY{p}{,}\PY{l+m+mi}{1}\PY{p}{)}
\end{Verbatim}
\end{tcolorbox}

    Before running the circuits from these codes, we need to ensure that the
transpiler knows which physical qubits on the device it should use. This
means using the qubit of \texttt{line{[}0{]}} to serve as the first code
qubit, that of \texttt{line{[}1{]}} to be the first link qubit, and so
on. This is done by the following function, which takes a repetition
code object and a \texttt{line}, and creates a Python dictionary to
specify which qubit of the code corresponds to which element of the
line.

    \begin{tcolorbox}[ size=fbox, boxrule=1pt, colback=cellbackground, colframe=cellborder]
\prompt{In}{incolor}{32}{\boxspacing}
\begin{Verbatim}[commandchars=\\\{\}]
\PY{k}{def} \PY{n+nf}{get\PYZus{}initial\PYZus{}layout}\PY{p}{(}\PY{n}{code}\PY{p}{,}\PY{n}{line}\PY{p}{)}\PY{p}{:}
    \PY{n}{initial\PYZus{}layout} \PY{o}{=} \PY{p}{\PYZob{}}\PY{p}{\PYZcb{}}
    \PY{k}{for} \PY{n}{j} \PY{o+ow}{in} \PY{n+nb}{range}\PY{p}{(}\PY{n}{n}\PY{p}{)}\PY{p}{:}
        \PY{n}{initial\PYZus{}layout}\PY{p}{[}\PY{n}{code}\PY{o}{.}\PY{n}{code\PYZus{}qubit}\PY{p}{[}\PY{n}{j}\PY{p}{]}\PY{p}{]} \PY{o}{=} \PY{n}{line}\PY{p}{[}\PY{l+m+mi}{2}\PY{o}{*}\PY{n}{j}\PY{p}{]}
    \PY{k}{for} \PY{n}{j} \PY{o+ow}{in} \PY{n+nb}{range}\PY{p}{(}\PY{n}{n}\PY{o}{\PYZhy{}}\PY{l+m+mi}{1}\PY{p}{)}\PY{p}{:}
        \PY{n}{initial\PYZus{}layout}\PY{p}{[}\PY{n}{code}\PY{o}{.}\PY{n}{link\PYZus{}qubit}\PY{p}{[}\PY{n}{j}\PY{p}{]}\PY{p}{]} \PY{o}{=} \PY{n}{line}\PY{p}{[}\PY{l+m+mi}{2}\PY{o}{*}\PY{n}{j}\PY{o}{+}\PY{l+m+mi}{1}\PY{p}{]}
    \PY{k}{return} \PY{n}{initial\PYZus{}layout}
\end{Verbatim}
\end{tcolorbox}

    Now we can transpile the circuits, to create the circuits that will
actually be run by the device. A check is also made to ensure that the
transpilation indeed has not introduced non-trivial effects by
increasing the number of qubits. Furthermore, the compiled circuits are
collected into a single list, to allow them all to be submitted at once
in the same batch job.

    \begin{tcolorbox}[ size=fbox, boxrule=1pt, colback=cellbackground, colframe=cellborder]
\prompt{In}{incolor}{33}{\boxspacing}
\begin{Verbatim}[commandchars=\\\{\}]
\PY{k}{if} \PY{n}{step\PYZus{}2}\PY{p}{:}
    
    \PY{n}{circuits} \PY{o}{=} \PY{p}{[}\PY{p}{]}
    \PY{k}{for} \PY{n}{n} \PY{o+ow}{in} \PY{n+nb}{range}\PY{p}{(}\PY{n}{n\PYZus{}min}\PY{p}{,}\PY{n}{n\PYZus{}max}\PY{o}{+}\PY{l+m+mi}{1}\PY{p}{)}\PY{p}{:}
        \PY{n}{initial\PYZus{}layout} \PY{o}{=} \PY{n}{get\PYZus{}initial\PYZus{}layout}\PY{p}{(}\PY{n}{code}\PY{p}{[}\PY{n}{n}\PY{p}{]}\PY{p}{,}\PY{n}{line}\PY{p}{)}
        \PY{k}{for} \PY{n}{log} \PY{o+ow}{in} \PY{p}{[}\PY{l+s+s1}{\PYZsq{}}\PY{l+s+s1}{0}\PY{l+s+s1}{\PYZsq{}}\PY{p}{,}\PY{l+s+s1}{\PYZsq{}}\PY{l+s+s1}{1}\PY{l+s+s1}{\PYZsq{}}\PY{p}{]}\PY{p}{:}
            \PY{n}{circuits}\PY{o}{.}\PY{n}{append}\PY{p}{(} \PY{n}{transpile}\PY{p}{(}\PY{n}{code}\PY{p}{[}\PY{n}{n}\PY{p}{]}\PY{o}{.}\PY{n}{circuit}\PY{p}{[}\PY{n}{log}\PY{p}{]}\PY{p}{,}
                                       \PY{n}{backend}\PY{o}{=}\PY{n}{backend}\PY{p}{,}
                                       \PY{n}{initial\PYZus{}layout}\PY{o}{=}\PY{n}{initial\PYZus{}layout}\PY{p}{)} \PY{p}{)}
            \PY{n}{num\PYZus{}cx} \PY{o}{=} \PY{n+nb}{dict}\PY{p}{(}\PY{n}{circuits}\PY{p}{[}\PY{o}{\PYZhy{}}\PY{l+m+mi}{1}\PY{p}{]}\PY{o}{.}\PY{n}{count\PYZus{}ops}\PY{p}{(}\PY{p}{)}\PY{p}{)}\PY{p}{[}\PY{l+s+s1}{\PYZsq{}}\PY{l+s+s1}{cx}\PY{l+s+s1}{\PYZsq{}}\PY{p}{]}
            \PY{k}{assert} \PY{n}{num\PYZus{}cx}\PY{o}{==}\PY{l+m+mi}{2}\PY{o}{*}\PY{p}{(}\PY{n}{n}\PY{o}{\PYZhy{}}\PY{l+m+mi}{1}\PY{p}{)}\PY{p}{,} \PY{n+nb}{str}\PY{p}{(}\PY{n}{num\PYZus{}cx}\PY{p}{)} \PY{o}{+} \PY{l+s+s1}{\PYZsq{}}\PY{l+s+s1}{ instead of }\PY{l+s+s1}{\PYZsq{}} \PY{o}{+} \PY{n+nb}{str}\PY{p}{(}\PY{l+m+mi}{2}\PY{o}{*}\PY{p}{(}\PY{n}{n}\PY{o}{\PYZhy{}}\PY{l+m+mi}{1}\PY{p}{)}\PY{p}{)} \PY{o}{+} \PY{l+s+s1}{\PYZsq{}}\PY{l+s+s1}{ cx gates for n = }\PY{l+s+s1}{\PYZsq{}} \PY{o}{+} \PY{n+nb}{str}\PY{p}{(}\PY{n}{n}\PY{p}{)}
\end{Verbatim}
\end{tcolorbox}

    We are now ready to run the job. As with the simulated jobs considered
already, the results from this are extracted into a dictionary
\texttt{raw\_results}. However, in this case it is extended to hold the
results from different code sizes. This means that
\texttt{raw\_results{[}n{]}} in the following is equivalent to one of
the \texttt{raw\_results} dictionaries used earlier, for a given
\texttt{n}.

    \begin{tcolorbox}[ size=fbox, boxrule=1pt, colback=cellbackground, colframe=cellborder]
\prompt{In}{incolor}{34}{\boxspacing}
\begin{Verbatim}[commandchars=\\\{\}]
\PY{k}{if} \PY{n}{step\PYZus{}2}\PY{p}{:}
    
    \PY{n}{job} \PY{o}{=} \PY{n}{execute}\PY{p}{(}\PY{n}{circuits}\PY{p}{,}\PY{n}{backend}\PY{p}{,}\PY{n}{shots}\PY{o}{=}\PY{l+m+mi}{8192}\PY{p}{)}

    \PY{n}{raw\PYZus{}results} \PY{o}{=} \PY{p}{\PYZob{}}\PY{p}{\PYZcb{}}
    \PY{n}{j} \PY{o}{=} \PY{l+m+mi}{0}
    \PY{k}{for} \PY{n}{n} \PY{o+ow}{in} \PY{n+nb}{range}\PY{p}{(}\PY{n}{n\PYZus{}min}\PY{p}{,}\PY{n}{n\PYZus{}max}\PY{o}{+}\PY{l+m+mi}{1}\PY{p}{)}\PY{p}{:}
        \PY{n}{raw\PYZus{}results}\PY{p}{[}\PY{n}{n}\PY{p}{]} \PY{o}{=} \PY{p}{\PYZob{}}\PY{p}{\PYZcb{}}
        \PY{k}{for} \PY{n}{log} \PY{o+ow}{in} \PY{p}{[}\PY{l+s+s1}{\PYZsq{}}\PY{l+s+s1}{0}\PY{l+s+s1}{\PYZsq{}}\PY{p}{,}\PY{l+s+s1}{\PYZsq{}}\PY{l+s+s1}{1}\PY{l+s+s1}{\PYZsq{}}\PY{p}{]}\PY{p}{:}
            \PY{n}{raw\PYZus{}results}\PY{p}{[}\PY{n}{n}\PY{p}{]}\PY{p}{[}\PY{n}{log}\PY{p}{]} \PY{o}{=} \PY{n}{job}\PY{o}{.}\PY{n}{result}\PY{p}{(}\PY{p}{)}\PY{o}{.}\PY{n}{get\PYZus{}counts}\PY{p}{(}\PY{n}{j}\PY{p}{)}
            \PY{n}{j} \PY{o}{+}\PY{o}{=} \PY{l+m+mi}{1}
\end{Verbatim}
\end{tcolorbox}

    It can be convenient to save the data to file, so that the processing of
step 3 can be done or repeated at a later time.

    \begin{tcolorbox}[ size=fbox, boxrule=1pt, colback=cellbackground, colframe=cellborder]
\prompt{In}{incolor}{35}{\boxspacing}
\begin{Verbatim}[commandchars=\\\{\}]
\PY{k}{if} \PY{n}{step\PYZus{}2}\PY{p}{:} \PY{c+c1}{\PYZsh{} save results}
    \PY{k}{with} \PY{n+nb}{open}\PY{p}{(}\PY{l+s+s1}{\PYZsq{}}\PY{l+s+s1}{results/raw\PYZus{}results\PYZus{}}\PY{l+s+s1}{\PYZsq{}}\PY{o}{+}\PY{n}{device\PYZus{}name}\PY{o}{+}\PY{l+s+s1}{\PYZsq{}}\PY{l+s+s1}{.txt}\PY{l+s+s1}{\PYZsq{}}\PY{p}{,} \PY{l+s+s1}{\PYZsq{}}\PY{l+s+s1}{w}\PY{l+s+s1}{\PYZsq{}}\PY{p}{)} \PY{k}{as} \PY{n}{file}\PY{p}{:}
        \PY{n}{file}\PY{o}{.}\PY{n}{write}\PY{p}{(}\PY{n+nb}{str}\PY{p}{(}\PY{n}{raw\PYZus{}results}\PY{p}{)}\PY{p}{)}
\PY{k}{elif} \PY{n}{step\PYZus{}3}\PY{p}{:} \PY{c+c1}{\PYZsh{} read results}
    \PY{k}{with} \PY{n+nb}{open}\PY{p}{(}\PY{l+s+s1}{\PYZsq{}}\PY{l+s+s1}{results/raw\PYZus{}results\PYZus{}}\PY{l+s+s1}{\PYZsq{}}\PY{o}{+}\PY{n}{device\PYZus{}name}\PY{o}{+}\PY{l+s+s1}{\PYZsq{}}\PY{l+s+s1}{.txt}\PY{l+s+s1}{\PYZsq{}}\PY{p}{,} \PY{l+s+s1}{\PYZsq{}}\PY{l+s+s1}{r}\PY{l+s+s1}{\PYZsq{}}\PY{p}{)} \PY{k}{as} \PY{n}{file}\PY{p}{:}
        \PY{n}{raw\PYZus{}results} \PY{o}{=} \PY{n+nb}{eval}\PY{p}{(}\PY{n}{file}\PY{o}{.}\PY{n}{read}\PY{p}{(}\PY{p}{)}\PY{p}{)}
\end{Verbatim}
\end{tcolorbox}

    As was described previously, some post-processing of the syndromes is required to find the the graph that will be used in the decoding. This is done using the \texttt{process\_results} method of each repetition code object \texttt{code{[}n{]}}. Specifically, we use it to create a results dictionary \texttt{results{[}n{]}} from each \texttt{raw\_results{[}n{]}}.

    \begin{tcolorbox}[ size=fbox, boxrule=1pt, colback=cellbackground, colframe=cellborder]
\prompt{In}{incolor}{36}{\boxspacing}
\begin{Verbatim}[commandchars=\\\{\}]
\PY{k}{if} \PY{n}{step\PYZus{}3}\PY{p}{:}
    \PY{n}{results} \PY{o}{=} \PY{p}{\PYZob{}}\PY{p}{\PYZcb{}}
    \PY{k}{for} \PY{n}{n} \PY{o+ow}{in} \PY{n+nb}{range}\PY{p}{(}\PY{n}{n\PYZus{}min}\PY{p}{,}\PY{n}{n\PYZus{}max}\PY{o}{+}\PY{l+m+mi}{1}\PY{p}{)}\PY{p}{:}
        \PY{n}{results}\PY{p}{[}\PY{n}{n}\PY{p}{]} \PY{o}{=} \PY{n}{code}\PY{p}{[}\PY{n}{n}\PY{p}{]}\PY{o}{.}\PY{n}{process\PYZus{}results}\PY{p}{(} \PY{n}{raw\PYZus{}results}\PY{p}{[}\PY{n}{n}\PY{p}{]} \PY{p}{)}
\end{Verbatim}
\end{tcolorbox}

    The decoding also needs us to set up the \texttt{GraphDecoder} object
for each code. The initialization of these involves the construction of
the graph corresponding to the syndrome, as described in the last
section.

    \begin{tcolorbox}[ size=fbox, boxrule=1pt, colback=cellbackground, colframe=cellborder]
\prompt{In}{incolor}{37}{\boxspacing}
\begin{Verbatim}[commandchars=\\\{\}]
\PY{k}{if} \PY{n}{step\PYZus{}3}\PY{p}{:}
    \PY{n}{dec} \PY{o}{=} \PY{p}{\PYZob{}}\PY{p}{\PYZcb{}}
    \PY{k}{for} \PY{n}{n} \PY{o+ow}{in} \PY{n+nb}{range}\PY{p}{(}\PY{n}{n\PYZus{}min}\PY{p}{,}\PY{n}{n\PYZus{}max}\PY{o}{+}\PY{l+m+mi}{1}\PY{p}{)}\PY{p}{:}
        \PY{n}{dec}\PY{p}{[}\PY{n}{n}\PY{p}{]} \PY{o}{=} \PY{n}{GraphDecoder}\PY{p}{(}\PY{n}{code}\PY{p}{[}\PY{n}{n}\PY{p}{]}\PY{p}{)}
\end{Verbatim}
\end{tcolorbox}

    Finally, the decoder object can be used to process the results. Here the
default algorithm, minimum weight perfect matching, is used. The end
result is a calculation of the logical error probability. This is simply the probability that the decoded output is \texttt{1} when the encoded value was \texttt{0}, or vice-versa.

When running
step 3, the following snippet also saves the logical error
probabilities. Otherwise, it reads in previously saved probabilities.

    \begin{tcolorbox}[ size=fbox, boxrule=1pt, colback=cellbackground, colframe=cellborder]
\prompt{In}{incolor}{38}{\boxspacing}
\begin{Verbatim}[commandchars=\\\{\}]
\PY{k}{if} \PY{n}{step\PYZus{}3}\PY{p}{:}
    
    \PY{n}{logical\PYZus{}prob\PYZus{}match} \PY{o}{=} \PY{p}{\PYZob{}}\PY{p}{\PYZcb{}}
    \PY{k}{for} \PY{n}{n} \PY{o+ow}{in} \PY{n+nb}{range}\PY{p}{(}\PY{n}{n\PYZus{}min}\PY{p}{,}\PY{n}{n\PYZus{}max}\PY{o}{+}\PY{l+m+mi}{1}\PY{p}{)}\PY{p}{:}
        \PY{n}{logical\PYZus{}prob\PYZus{}match}\PY{p}{[}\PY{n}{n}\PY{p}{]} \PY{o}{=} \PY{n}{dec}\PY{p}{[}\PY{n}{n}\PY{p}{]}\PY{o}{.}\PY{n}{get\PYZus{}logical\PYZus{}prob}\PY{p}{(}\PY{n}{results}\PY{p}{[}\PY{n}{n}\PY{p}{]}\PY{p}{)}
        
    \PY{k}{with} \PY{n+nb}{open}\PY{p}{(}\PY{l+s+s1}{\PYZsq{}}\PY{l+s+s1}{results/logical\PYZus{}prob\PYZus{}match\PYZus{}}\PY{l+s+s1}{\PYZsq{}}\PY{o}{+}\PY{n}{device\PYZus{}name}\PY{o}{+}\PY{l+s+s1}{\PYZsq{}}\PY{l+s+s1}{.txt}\PY{l+s+s1}{\PYZsq{}}\PY{p}{,} \PY{l+s+s1}{\PYZsq{}}\PY{l+s+s1}{w}\PY{l+s+s1}{\PYZsq{}}\PY{p}{)} \PY{k}{as} \PY{n}{file}\PY{p}{:}
        \PY{n}{file}\PY{o}{.}\PY{n}{write}\PY{p}{(}\PY{n+nb}{str}\PY{p}{(}\PY{n}{logical\PYZus{}prob\PYZus{}match}\PY{p}{)}\PY{p}{)}
        
\PY{k}{else}\PY{p}{:}
    \PY{k}{with} \PY{n+nb}{open}\PY{p}{(}\PY{l+s+s1}{\PYZsq{}}\PY{l+s+s1}{results/logical\PYZus{}prob\PYZus{}match\PYZus{}}\PY{l+s+s1}{\PYZsq{}}\PY{o}{+}\PY{n}{device\PYZus{}name}\PY{o}{+}\PY{l+s+s1}{\PYZsq{}}\PY{l+s+s1}{.txt}\PY{l+s+s1}{\PYZsq{}}\PY{p}{,} \PY{l+s+s1}{\PYZsq{}}\PY{l+s+s1}{r}\PY{l+s+s1}{\PYZsq{}}\PY{p}{)} \PY{k}{as} \PY{n}{file}\PY{p}{:}
        \PY{n}{logical\PYZus{}prob\PYZus{}match} \PY{o}{=} \PY{n+nb}{eval}\PY{p}{(}\PY{n}{file}\PY{o}{.}\PY{n}{read}\PY{p}{(}\PY{p}{)}\PY{p}{)}
\end{Verbatim}
\end{tcolorbox}

    The resulting logical error probabilities are displayed in the following
graph, which uses a log scale on the y axis. We would expect that
the logical error probability decays exponentially with increasing
\(n\). If this is the case, it is a confirmation that the device is
compatible with this basis test of quantum error correction. If not, it
implies that the qubits and gates are not sufficiently reliable.

Fortunately, the results from this device does show the desired trend. Deviations can be observed, however, which can serve as a starting point for investigations into exactly why the device behaves as it does. However, in this paper we simply seek to show an example of the \texttt{topological\_codes} module in action, and will not perform more in-depth analysis.

    \begin{tcolorbox}[ size=fbox, boxrule=1pt, colback=cellbackground, colframe=cellborder]
\prompt{In}{incolor}{44}{\boxspacing}
\begin{Verbatim}[commandchars=\\\{\}]
\PY{k+kn}{import} \PY{n+nn}{matplotlib}\PY{n+nn}{.}\PY{n+nn}{pyplot} \PY{k}{as} \PY{n+nn}{plt}
\PY{k+kn}{import} \PY{n+nn}{numpy} \PY{k}{as} \PY{n+nn}{np}

\PY{n}{x\PYZus{}axis} \PY{o}{=} \PY{n+nb}{range}\PY{p}{(}\PY{n}{n\PYZus{}min}\PY{p}{,}\PY{n}{n\PYZus{}max}\PY{o}{+}\PY{l+m+mi}{1}\PY{p}{)}
\PY{n}{P} \PY{o}{=} \PY{p}{\PYZob{}} \PY{n}{log}\PY{p}{:} \PY{p}{[}\PY{n}{logical\PYZus{}prob\PYZus{}match}\PY{p}{[}\PY{n}{n}\PY{p}{]}\PY{p}{[}\PY{n}{log}\PY{p}{]} \PY{k}{for} \PY{n}{n} \PY{o+ow}{in} \PY{n}{x\PYZus{}axis}\PY{p}{]} \PY{k}{for} \PY{n}{log} \PY{o+ow}{in} \PY{p}{[}\PY{l+s+s1}{\PYZsq{}}\PY{l+s+s1}{0}\PY{l+s+s1}{\PYZsq{}}\PY{p}{,} \PY{l+s+s1}{\PYZsq{}}\PY{l+s+s1}{1}\PY{l+s+s1}{\PYZsq{}}\PY{p}{]} \PY{p}{\PYZcb{}}

\PY{n}{ax} \PY{o}{=} \PY{n}{plt}\PY{o}{.}\PY{n}{gca}\PY{p}{(}\PY{p}{)}
\PY{n}{plt}\PY{o}{.}\PY{n}{xlabel}\PY{p}{(}\PY{l+s+s1}{\PYZsq{}}\PY{l+s+s1}{Code distance, n}\PY{l+s+s1}{\PYZsq{}}\PY{p}{)}
\PY{n}{plt}\PY{o}{.}\PY{n}{ylabel}\PY{p}{(}\PY{l+s+s1}{\PYZsq{}}\PY{l+s+s1}{Logical error probability}\PY{l+s+s1}{\PYZsq{}}\PY{p}{)}
\PY{n}{ax}\PY{o}{.}\PY{n}{scatter}\PY{p}{(} \PY{n}{x\PYZus{}axis}\PY{p}{,} \PY{n}{P}\PY{p}{[}\PY{l+s+s1}{\PYZsq{}}\PY{l+s+s1}{0}\PY{l+s+s1}{\PYZsq{}}\PY{p}{]}\PY{p}{,} \PY{n}{label}\PY{o}{=}\PY{l+s+s2}{\PYZdq{}}\PY{l+s+s2}{logical 0}\PY{l+s+s2}{\PYZdq{}}\PY{p}{,} \PY{n}{s}\PY{o}{=}\PY{l+m+mi}{512}\PY{p}{)}
\PY{n}{ax}\PY{o}{.}\PY{n}{scatter}\PY{p}{(} \PY{n}{x\PYZus{}axis}\PY{p}{,} \PY{n}{P}\PY{p}{[}\PY{l+s+s1}{\PYZsq{}}\PY{l+s+s1}{1}\PY{l+s+s1}{\PYZsq{}}\PY{p}{]}\PY{p}{,} \PY{n}{label}\PY{o}{=}\PY{l+s+s2}{\PYZdq{}}\PY{l+s+s2}{logical 1}\PY{l+s+s2}{\PYZdq{}}\PY{p}{,} \PY{n}{s}\PY{o}{=}\PY{l+m+mi}{512}\PY{p}{)}
\PY{n}{ax}\PY{o}{.}\PY{n}{set\PYZus{}yscale}\PY{p}{(}\PY{l+s+s1}{\PYZsq{}}\PY{l+s+s1}{log}\PY{l+s+s1}{\PYZsq{}}\PY{p}{)}
\PY{n}{ax}\PY{o}{.}\PY{n}{set\PYZus{}ylim}\PY{p}{(}\PY{n}{ymax}\PY{o}{=}\PY{l+m+mf}{1.5}\PY{o}{*}\PY{n+nb}{max}\PY{p}{(}\PY{n}{P}\PY{p}{[}\PY{l+s+s1}{\PYZsq{}}\PY{l+s+s1}{0}\PY{l+s+s1}{\PYZsq{}}\PY{p}{]}\PY{o}{+}\PY{n}{P}\PY{p}{[}\PY{l+s+s1}{\PYZsq{}}\PY{l+s+s1}{1}\PY{l+s+s1}{\PYZsq{}}\PY{p}{]}\PY{p}{)}\PY{p}{,}
            \PY{n}{ymin}\PY{o}{=}\PY{l+m+mf}{0.75}\PY{o}{*}\PY{n+nb}{min}\PY{p}{(}\PY{p}{[} \PY{n}{p} \PY{k}{for} \PY{n}{p} \PY{o+ow}{in} \PY{n}{P}\PY{p}{[}\PY{l+s+s1}{\PYZsq{}}\PY{l+s+s1}{0}\PY{l+s+s1}{\PYZsq{}}\PY{p}{]}\PY{o}{+}\PY{n}{P}\PY{p}{[}\PY{l+s+s1}{\PYZsq{}}\PY{l+s+s1}{1}\PY{l+s+s1}{\PYZsq{}}\PY{p}{]} \PY{k}{if} \PY{n}{p}\PY{o}{\PYZgt{}}\PY{l+m+mi}{0} \PY{p}{]}\PY{p}{)}\PY{p}{)}
\PY{n}{plt}\PY{o}{.}\PY{n}{legend}\PY{p}{(}\PY{p}{)}

\PY{n}{plt}\PY{o}{.}\PY{n}{show}\PY{p}{(}\PY{p}{)}
\end{Verbatim}
\end{tcolorbox}

    \begin{center}
    \adjustimage{max size={0.667\linewidth}{0.9\paperheight}}{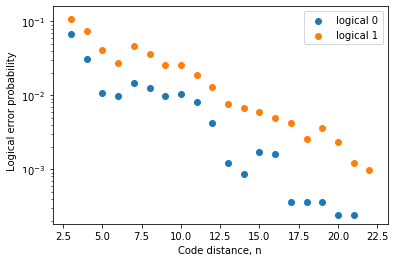}
    \end{center}
     \vspace{0.5cm}

    Another insight we can gain is to use the results to determine how
likely certain error processes are to occur.

To see how this can be done, recall that each node in the syndrome graph corresponds to a particular syndrome measurement being performed at a particular point within the circuit. A pair of nodes are connected by an edge if and only if a single error, occurring on a particular qubit at a particular point within the circuit, can cause the value of both to change. 

For any such pair of adjacent nodes, we will specifically consider the values \(C_{11}\) and \(C_{00}\) , where the former is the number of counts in
\texttt{results{[}n{]}{[}\textquotesingle{}0\textquotesingle{}{]}} corresponding to the syndrome value of both adjacent nodes being
\texttt{1}, and the latter is the same for them both being \texttt{0}.

The most likely cause for each event recorded in \(C_{11}\) is the occurrence of the error corresponding to the edge between these two nodes. Also, it is most likely that this error has not occurred for each event recorded in \(C_{00}\). As such, to first order, we can state that

\[
\frac{p}{1-p} \approx \frac{C_{11}}{C_{00}}
\]

Where \(p\) is the probability of the error corresponding to the edge between these two nodes.

For example, suppose that one of the nodes we are considering corresponds to the syndrome measurement of code qubits $0$ and $1$ in the first round, and the other corresponds to the same for code qubits $1$ and $2$. If both syndrome measurements output \texttt{0}, it is most likely that none of these three qubits suffered an error (during the initial state preparation or during the controlled-NOTs required for the syndrome measurements). If both syndrome measurements output \texttt{1}, it is most likely that a single error occurred on the shared qubit $2$. The probability $p$ in this case would therefore be that for such an error on qubit $2$, either in preparation or during the controlled-NOTs.

The decoder object has a method \texttt{weight\_syndrome\_graph} which
determines these ratios, and assigns each edge the weight
\(-\ln(p/(1-p))\). By employing this method and inspecting the weights,
we can easily retrieve these probabilities.

    \begin{tcolorbox}[size=fbox, boxrule=1pt, colback=cellbackground, colframe=cellborder]
\prompt{In}{incolor}{43}{\boxspacing}
\begin{Verbatim}[commandchars=\\\{\}]
\PY{k+kn}{import} \PY{n+nn}{pandas} \PY{k}{as} \PY{n+nn}{pd}

\PY{k}{if} \PY{n}{step\PYZus{}3}\PY{p}{:}

    \PY{n}{dec}\PY{p}{[}\PY{n}{n\PYZus{}max}\PY{p}{]}\PY{o}{.}\PY{n}{weight\PYZus{}syndrome\PYZus{}graph}\PY{p}{(}\PY{n}{results}\PY{o}{=}\PY{n}{results}\PY{p}{[}\PY{n}{n\PYZus{}max}\PY{p}{]}\PY{p}{)}

    \PY{n}{probs} \PY{o}{=} \PY{p}{[}\PY{p}{]}
    \PY{k}{for} \PY{n}{edge} \PY{o+ow}{in} \PY{n}{dec}\PY{p}{[}\PY{n}{n\PYZus{}max}\PY{p}{]}\PY{o}{.}\PY{n}{S}\PY{o}{.}\PY{n}{edges}\PY{p}{:}
        \PY{n}{ratio} \PY{o}{=} \PY{n}{np}\PY{o}{.}\PY{n}{exp}\PY{p}{(}\PY{o}{\PYZhy{}}\PY{n}{dec}\PY{p}{[}\PY{n}{n\PYZus{}max}\PY{p}{]}\PY{o}{.}\PY{n}{S}\PY{o}{.}\PY{n}{get\PYZus{}edge\PYZus{}data}\PY{p}{(}\PY{n}{edge}\PY{p}{[}\PY{l+m+mi}{0}\PY{p}{]}\PY{p}{,}\PY{n}{edge}\PY{p}{[}\PY{l+m+mi}{1}\PY{p}{]}\PY{p}{)}\PY{p}{[}\PY{l+s+s1}{\PYZsq{}}\PY{l+s+s1}{distance}\PY{l+s+s1}{\PYZsq{}}\PY{p}{]}\PY{p}{)}
        \PY{n}{probs}\PY{o}{.}\PY{n}{append}\PY{p}{(} \PY{n}{ratio}\PY{o}{/}\PY{p}{(}\PY{l+m+mi}{1}\PY{o}{+}\PY{n}{ratio}\PY{p}{)} \PY{p}{)}
        
    \PY{n}{probs\PYZus{}summary} \PY{o}{=} \PY{n}{pd}\PY{o}{.}\PY{n}{Series}\PY{p}{(}\PY{n}{probs}\PY{p}{)}\PY{o}{.}\PY{n}{describe}\PY{p}{(}\PY{p}{)}\PY{o}{.}\PY{n}{to\PYZus{}dict}\PY{p}{(}\PY{p}{)}
        
    \PY{k}{with} \PY{n+nb}{open}\PY{p}{(}\PY{l+s+s1}{\PYZsq{}}\PY{l+s+s1}{results/probs\PYZus{}summary\PYZus{}}\PY{l+s+s1}{\PYZsq{}}\PY{o}{+}\PY{n}{device\PYZus{}name}\PY{o}{+}\PY{l+s+s1}{\PYZsq{}}\PY{l+s+s1}{.txt}\PY{l+s+s1}{\PYZsq{}}\PY{p}{,} \PY{l+s+s1}{\PYZsq{}}\PY{l+s+s1}{w}\PY{l+s+s1}{\PYZsq{}}\PY{p}{)} \PY{k}{as} \PY{n}{file}\PY{p}{:}
        \PY{n}{file}\PY{o}{.}\PY{n}{write}\PY{p}{(}\PY{n+nb}{str}\PY{p}{(}\PY{n}{probs\PYZus{}summary}\PY{p}{)}\PY{p}{)}
        
\PY{k}{else}\PY{p}{:}
    
    \PY{k}{with} \PY{n+nb}{open}\PY{p}{(}\PY{l+s+s1}{\PYZsq{}}\PY{l+s+s1}{results/probs\PYZus{}summary\PYZus{}}\PY{l+s+s1}{\PYZsq{}}\PY{o}{+}\PY{n}{device\PYZus{}name}\PY{o}{+}\PY{l+s+s1}{\PYZsq{}}\PY{l+s+s1}{.txt}\PY{l+s+s1}{\PYZsq{}}\PY{p}{,} \PY{l+s+s1}{\PYZsq{}}\PY{l+s+s1}{r}\PY{l+s+s1}{\PYZsq{}}\PY{p}{)} \PY{k}{as} \PY{n}{file}\PY{p}{:}
        \PY{n}{probs\PYZus{}summary} \PY{o}{=} \PY{n+nb}{eval}\PY{p}{(}\PY{n}{file}\PY{o}{.}\PY{n}{read}\PY{p}{(}\PY{p}{)}\PY{p}{)}

\PY{k}{for} \PY{n}{stat} \PY{o+ow}{in} \PY{n}{probs\PYZus{}summary}\PY{p}{:}
    \PY{n+nb}{print}\PY{p}{(}\PY{n}{stat}\PY{o}{+}\PY{l+s+s1}{\PYZsq{}}\PY{l+s+s1}{:}\PY{l+s+s1}{\PYZsq{}}\PY{p}{,}\PY{n}{probs\PYZus{}summary}\PY{p}{[}\PY{n}{stat}\PY{p}{]}\PY{p}{)}
\end{Verbatim}
\end{tcolorbox}

    \begin{Verbatim}[commandchars=\\\{\}]
count: 85.0
mean: 0.15019428099809715
std: 0.11236995205935389
min: 0.026545002073828285
25\%: 0.05599450360700791
50\%: 0.11155859846959323
75\%: 0.20286006128702758
max: 0.41629711751662973
    \end{Verbatim}
     \vspace{0.25cm}
        
These probabilities come from a novel way of benchmarking the device, that is distinct from currently used methods. Since no other set of available error
probabilities are exactly equivalent, it is not possible to do a direct comparison. Nevertheless, since the results from the standard benchmarking of the device can be easily obtained from the \texttt{backend} object, we will generate a similar summary. Specifically, we will look at the readout errors and controlled-NOT gate errors.

    \begin{tcolorbox}[size=fbox, boxrule=1pt, colback=cellbackground, colframe=cellborder]
\prompt{In}{incolor}{47}{\boxspacing}
\begin{Verbatim}[commandchars=\\\{\}]
\PY{k}{if} \PY{n}{step\PYZus{}3}\PY{p}{:}

    \PY{n}{gate\PYZus{}probs} \PY{o}{=} \PY{p}{[}\PY{p}{]}
    \PY{k}{for} \PY{n}{j}\PY{p}{,}\PY{n}{qubit} \PY{o+ow}{in} \PY{n+nb}{enumerate}\PY{p}{(}\PY{n}{line}\PY{p}{)}\PY{p}{:}
        
        \PY{n}{gate\PYZus{}probs}\PY{o}{.}\PY{n}{append}\PY{p}{(} \PY{n}{backend}\PY{o}{.}\PY{n}{properties}\PY{p}{(}\PY{p}{)}\PY{o}{.}\PY{n}{readout\PYZus{}error}\PY{p}{(}\PY{n}{qubit}\PY{p}{)} \PY{p}{)}
        
        \PY{n}{cx1}\PY{p}{,}\PY{n}{cx2} \PY{o}{=} \PY{l+m+mi}{0}\PY{p}{,}\PY{l+m+mi}{0}
        \PY{k}{if} \PY{n}{j}\PY{o}{\PYZgt{}}\PY{l+m+mi}{0}\PY{p}{:}
            \PY{n}{gate\PYZus{}probs}\PY{o}{.}\PY{n}{append}\PY{p}{(}\PY{n}{backend}\PY{o}{.}\PY{n}{properties}\PY{p}{(}\PY{p}{)}\PY{o}{.}\PY{n}{gate\PYZus{}error}\PY{p}{(}\PY{l+s+s1}{\PYZsq{}}\PY{l+s+s1}{cx}\PY{l+s+s1}{\PYZsq{}}\PY{p}{,}\PY{p}{[}\PY{n}{qubit}\PY{p}{,}\PY{n}{line}\PY{p}{[}\PY{n}{j}\PY{o}{\PYZhy{}}\PY{l+m+mi}{1}\PY{p}{]}\PY{p}{]}\PY{p}{)} \PY{p}{)}
        \PY{k}{if} \PY{n}{j}\PY{o}{\PYZlt{}}\PY{n+nb}{len}\PY{p}{(}\PY{n}{line}\PY{p}{)}\PY{o}{\PYZhy{}}\PY{l+m+mi}{1}\PY{p}{:}
            \PY{n}{gate\PYZus{}probs}\PY{o}{.}\PY{n}{append}\PY{p}{(}\PY{n}{backend}\PY{o}{.}\PY{n}{properties}\PY{p}{(}\PY{p}{)}\PY{o}{.}\PY{n}{gate\PYZus{}error}\PY{p}{(}\PY{l+s+s1}{\PYZsq{}}\PY{l+s+s1}{cx}\PY{l+s+s1}{\PYZsq{}}\PY{p}{,}\PY{p}{[}\PY{n}{qubit}\PY{p}{,}\PY{n}{line}\PY{p}{[}\PY{n}{j}\PY{o}{+}\PY{l+m+mi}{1}\PY{p}{]}\PY{p}{]}\PY{p}{)} \PY{p}{)}
            
    \PY{n}{gate\PYZus{}probs\PYZus{}summary} \PY{o}{=} \PY{n}{pd}\PY{o}{.}\PY{n}{Series}\PY{p}{(}\PY{n}{gate\PYZus{}probs}\PY{p}{)}\PY{o}{.}\PY{n}{describe}\PY{p}{(}\PY{p}{)}\PY{o}{.}\PY{n}{to\PYZus{}dict}\PY{p}{(}\PY{p}{)}
                
    \PY{k}{with} \PY{n+nb}{open}\PY{p}{(}\PY{l+s+s1}{\PYZsq{}}\PY{l+s+s1}{results/gate\PYZus{}probs\PYZus{}summary\PYZus{}}\PY{l+s+s1}{\PYZsq{}}\PY{o}{+}\PY{n}{device\PYZus{}name}\PY{o}{+}\PY{l+s+s1}{\PYZsq{}}\PY{l+s+s1}{.txt}\PY{l+s+s1}{\PYZsq{}}\PY{p}{,} \PY{l+s+s1}{\PYZsq{}}\PY{l+s+s1}{w}\PY{l+s+s1}{\PYZsq{}}\PY{p}{)} \PY{k}{as} \PY{n}{file}\PY{p}{:}
        \PY{n}{file}\PY{o}{.}\PY{n}{write}\PY{p}{(}\PY{n+nb}{str}\PY{p}{(}\PY{n}{gate\PYZus{}probs\PYZus{}summary}\PY{p}{)}\PY{p}{)}
        
\PY{k}{else}\PY{p}{:}
    
    \PY{k}{with} \PY{n+nb}{open}\PY{p}{(}\PY{l+s+s1}{\PYZsq{}}\PY{l+s+s1}{results/gate\PYZus{}probs\PYZus{}summary\PYZus{}}\PY{l+s+s1}{\PYZsq{}}\PY{o}{+}\PY{n}{device\PYZus{}name}\PY{o}{+}\PY{l+s+s1}{\PYZsq{}}\PY{l+s+s1}{.txt}\PY{l+s+s1}{\PYZsq{}}\PY{p}{,} \PY{l+s+s1}{\PYZsq{}}\PY{l+s+s1}{r}\PY{l+s+s1}{\PYZsq{}}\PY{p}{)} \PY{k}{as} \PY{n}{file}\PY{p}{:}
        \PY{n}{gate\PYZus{}probs\PYZus{}summary} \PY{o}{=} \PY{n+nb}{eval}\PY{p}{(}\PY{n}{file}\PY{o}{.}\PY{n}{read}\PY{p}{(}\PY{p}{)}\PY{p}{)}

\PY{k}{for} \PY{n}{stat} \PY{o+ow}{in} \PY{n}{gate\PYZus{}probs\PYZus{}summary}\PY{p}{:}
    \PY{n+nb}{print}\PY{p}{(}\PY{n}{stat}\PY{o}{+}\PY{l+s+s1}{\PYZsq{}}\PY{l+s+s1}{:}\PY{l+s+s1}{\PYZsq{}}\PY{p}{,}\PY{n}{gate\PYZus{}probs\PYZus{}summary}\PY{p}{[}\PY{n}{stat}\PY{p}{]}\PY{p}{)}
\end{Verbatim}
\end{tcolorbox}

    \begin{Verbatim}[commandchars=\\\{\}]
count: 127.0
mean: 0.06850161154431056
std: 0.05891171361248407
min: 0.01937499999999992
25\%: 0.03708774898932393
50\%: 0.046021125148897085
75\%: 0.0857536432141715
max: 0.33624999999999994
    \end{Verbatim}
    \vspace{0.25cm}
        
Note that, despite the different methods used to obtain these values, there is excellent agreement for the minimum and first quartile. However, the values from the repetition code do show more of a skew to larger values in the upper quartiles. This could be due to the fact that repetition code values are obtained from an approximation that is less accurate for higher error probabilities. The possibility of using a better approximation will therefore be investigated in future versions of \texttt{topological\_codes}, as will further study of how the results of this benchmark relates to the standard one.

\section{Conclusions}

In this paper we have provided an introduction to the repetition code and its implementation in Qiskit. It forms part of the \texttt{topological\_codes} module of Qiskit-Ignis. This module has been designed to have a modular form, allowing new codes and decoders to be added with relative ease. It will be expanded to include more codes, such as the 17 qubit variant of the surface code~\cite{svore:14,winkler:17}. It will also be expanded to introduce new decoders, such as Bravyi-Haah \cite{bravyi-haah} and methods devised as part of the 'Decodoku' citizen science project \cite{decodoku}. There are also many other possible ways to expand the module. One aim of this paper is to serve as a starting point for anyone interested in making any contributions to this open-source project.

Results were shown from an example run from a 43 qubit code running on IBM's \emph{Rochester} device. This showed what can be done using just a few lines of code. However, much more could be done to fully probe the working of a device using the repetition code. The most obvious would be to use different possible choices for \texttt{line}, to fully map out where the errors come from and what effect they have. This paper also serves to serve as a starting point for such endeavours, since anyone with an internet connection can now use the tool to probe the 15 qubits of IBM's publicly accessible \emph{Melbourne} device.

\section{Acknowledgements}

This paper was based on resources developed for the Qiskit tutorials and textbook~\cite{qiskit-textbook}. The author would therefore like to thank all members of the Qiskit community.

\bibliography{refs}

\begin{thebibliography}{29}%
\makeatletter
\providecommand \@ifxundefined [1]{%
 \@ifx{#1\undefined}
}%
\providecommand \@ifnum [1]{%
 \ifnum #1\expandafter \@firstoftwo
 \else \expandafter \@secondoftwo
 \fi
}%
\providecommand \@ifx [1]{%
 \ifx #1\expandafter \@firstoftwo
 \else \expandafter \@secondoftwo
 \fi
}%
\providecommand \natexlab [1]{#1}%
\providecommand \enquote  [1]{``#1''}%
\providecommand \bibnamefont  [1]{#1}%
\providecommand \bibfnamefont [1]{#1}%
\providecommand \citenamefont [1]{#1}%
\providecommand \href@noop [0]{\@secondoftwo}%
\providecommand \href [0]{\begingroup \@sanitize@url \@href}%
\providecommand \@href[1]{\@@startlink{#1}\@@href}%
\providecommand \@@href[1]{\endgroup#1\@@endlink}%
\providecommand \@sanitize@url [0]{\catcode `\\12\catcode `\$12\catcode
  `\&12\catcode `\#12\catcode `\^12\catcode `\_12\catcode `\%12\relax}%
\providecommand \@@startlink[1]{}%
\providecommand \@@endlink[0]{}%
\providecommand \url  [0]{\begingroup\@sanitize@url \@url }%
\providecommand \@url [1]{\endgroup\@href {#1}{\urlprefix }}%
\providecommand \urlprefix  [0]{URL }%
\providecommand \Eprint [0]{\href }%
\providecommand \doibase [0]{http://dx.doi.org/}%
\providecommand \selectlanguage [0]{\@gobble}%
\providecommand \bibinfo  [0]{\@secondoftwo}%
\providecommand \bibfield  [0]{\@secondoftwo}%
\providecommand \translation [1]{[#1]}%
\providecommand \BibitemOpen [0]{}%
\providecommand \bibitemStop [0]{}%
\providecommand \bibitemNoStop [0]{.\EOS\space}%
\providecommand \EOS [0]{\spacefactor3000\relax}%
\providecommand \BibitemShut  [1]{\csname bibitem#1\endcsname}%
\let\auto@bib@innerbib\@empty
%</preamble>
\bibitem [{\citenamefont {Temme}\ \emph {et~al.}(2017)\citenamefont {Temme},
  \citenamefont {Bravyi},\ and\ \citenamefont {Gambetta}}]{temme:16}%
  \BibitemOpen
  \bibfield  {author} {\bibinfo {author} {\bibfnamefont {K.}~\bibnamefont
  {Temme}}, \bibinfo {author} {\bibfnamefont {S.}~\bibnamefont {Bravyi}}, \
  and\ \bibinfo {author} {\bibfnamefont {J.~M.}\ \bibnamefont {Gambetta}},\
  }\href {\doibase 10.1103/PhysRevLett.119.180509} {\bibfield  {journal}
  {\bibinfo  {journal} {Phys. Rev. Lett.}\ }\textbf {\bibinfo {volume} {119}},\
  \bibinfo {pages} {180509} (\bibinfo {year} {2017})}\BibitemShut {NoStop}%
\bibitem [{\citenamefont {Endo}\ \emph {et~al.}(2018)\citenamefont {Endo},
  \citenamefont {Benjamin},\ and\ \citenamefont {Li}}]{endo:18}%
  \BibitemOpen
  \bibfield  {author} {\bibinfo {author} {\bibfnamefont {S.}~\bibnamefont
  {Endo}}, \bibinfo {author} {\bibfnamefont {S.~C.}\ \bibnamefont {Benjamin}},
  \ and\ \bibinfo {author} {\bibfnamefont {Y.}~\bibnamefont {Li}},\ }\href
  {\doibase 10.1103/PhysRevX.8.031027} {\bibfield  {journal} {\bibinfo
  {journal} {Phys. Rev. X}\ }\textbf {\bibinfo {volume} {8}},\ \bibinfo {pages}
  {031027} (\bibinfo {year} {2018})}\BibitemShut {NoStop}%
\bibitem [{\citenamefont {Murali}\ \emph {et~al.}(2020)\citenamefont {Murali},
  \citenamefont {McKay}, \citenamefont {Martonosi},\ and\ \citenamefont
  {Javadi-Abhari}}]{murali:19}%
  \BibitemOpen
  \bibfield  {author} {\bibinfo {author} {\bibfnamefont {P.}~\bibnamefont
  {Murali}}, \bibinfo {author} {\bibfnamefont {D.~C.}\ \bibnamefont {McKay}},
  \bibinfo {author} {\bibfnamefont {M.}~\bibnamefont {Martonosi}}, \ and\
  \bibinfo {author} {\bibfnamefont {A.}~\bibnamefont {Javadi-Abhari}},\
  }\href@noop {} {\enquote {\bibinfo {title} {{{Software Mitigation of
  Crosstalk on Noisy Intermediate-Scale Quantum Computers}}},}\ } (\bibinfo
  {year} {2020}),\ \bibinfo {note} {arXiv:2001.02826}\BibitemShut {NoStop}%
\bibitem [{\citenamefont {Lidar}\ and\ \citenamefont {Brun}(2013)}]{lidar:13}%
  \BibitemOpen
  \bibinfo {editor} {\bibfnamefont {D.~A.}\ \bibnamefont {Lidar}}\ and\
  \bibinfo {editor} {\bibfnamefont {T.~A.}\ \bibnamefont {Brun}},\ eds.,\
  \href@noop {} {\emph {\bibinfo {title} {{Quantum Error Correction}}}}\
  (\bibinfo  {publisher} {Cambridge University Press},\ \bibinfo {address}
  {{Cambride, UK}},\ \bibinfo {year} {2013})\BibitemShut {NoStop}%
\bibitem [{\citenamefont {Kelly~{\em et al.}}(2014)}]{kelly:14}%
  \BibitemOpen
  \bibfield  {author} {\bibinfo {author} {\bibfnamefont {J.}~\bibnamefont
  {Kelly~{\em et al.}}},\ }\href {\doibase doi:10.1038/nature14270} {\bibfield
  {journal} {\bibinfo  {journal} {Nature}\ }\textbf {\bibinfo {volume} {519}},\
  \bibinfo {pages} {66} (\bibinfo {year} {2014})}\BibitemShut {NoStop}%
\bibitem [{\citenamefont {Rist{\`e}}\ \emph {et~al.}(2015)\citenamefont
  {Rist{\`e}}, \citenamefont {Poletto}, \citenamefont {Huang}, \citenamefont
  {Bruno}, \citenamefont {Vesterinen}, \citenamefont {Saira},\ and\
  \citenamefont {DiCarlo}}]{riste:15}%
  \BibitemOpen
  \bibfield  {author} {\bibinfo {author} {\bibfnamefont {D.}~\bibnamefont
  {Rist{\`e}}}, \bibinfo {author} {\bibfnamefont {S.}~\bibnamefont {Poletto}},
  \bibinfo {author} {\bibfnamefont {M.~Z.}\ \bibnamefont {Huang}}, \bibinfo
  {author} {\bibfnamefont {A.}~\bibnamefont {Bruno}}, \bibinfo {author}
  {\bibfnamefont {V.}~\bibnamefont {Vesterinen}}, \bibinfo {author}
  {\bibfnamefont {O.~P.}\ \bibnamefont {Saira}}, \ and\ \bibinfo {author}
  {\bibfnamefont {L.}~\bibnamefont {DiCarlo}},\ }\href {\doibase
  10.1038/ncomms7983} {\bibfield  {journal} {\bibinfo  {journal} {Nature
  Communications}\ }\textbf {\bibinfo {volume} {6}},\ \bibinfo {pages} {6983}
  (\bibinfo {year} {2015})}\BibitemShut {NoStop}%
\bibitem [{\citenamefont {Corcoles~{\em et al.}}(2015)}]{corcoles:15}%
  \BibitemOpen
  \bibfield  {author} {\bibinfo {author} {\bibfnamefont {A.~D.}\ \bibnamefont
  {Corcoles~{\em et al.}}},\ }\href {\doibase doi:10.1038/ncomms7979}
  {\bibfield  {journal} {\bibinfo  {journal} {Nature Communications}\ }\textbf
  {\bibinfo {volume} {6}} (\bibinfo {year} {2015}),\
  doi:10.1038/ncomms7979}\BibitemShut {NoStop}%
\bibitem [{\citenamefont {Takita}\ \emph {et~al.}(2016)\citenamefont {Takita},
  \citenamefont {C\'orcoles}, \citenamefont {Magesan}, \citenamefont {Abdo},
  \citenamefont {Brink}, \citenamefont {Cross}, \citenamefont {Chow},\ and\
  \citenamefont {Gambetta}}]{takita:16}%
  \BibitemOpen
  \bibfield  {author} {\bibinfo {author} {\bibfnamefont {M.}~\bibnamefont
  {Takita}}, \bibinfo {author} {\bibfnamefont {A.~D.}\ \bibnamefont
  {C\'orcoles}}, \bibinfo {author} {\bibfnamefont {E.}~\bibnamefont {Magesan}},
  \bibinfo {author} {\bibfnamefont {B.}~\bibnamefont {Abdo}}, \bibinfo {author}
  {\bibfnamefont {M.}~\bibnamefont {Brink}}, \bibinfo {author} {\bibfnamefont
  {A.}~\bibnamefont {Cross}}, \bibinfo {author} {\bibfnamefont {J.~M.}\
  \bibnamefont {Chow}}, \ and\ \bibinfo {author} {\bibfnamefont {J.~M.}\
  \bibnamefont {Gambetta}},\ }\href {\doibase 10.1103/PhysRevLett.117.210505}
  {\bibfield  {journal} {\bibinfo  {journal} {Phys. Rev. Lett.}\ }\textbf
  {\bibinfo {volume} {117}},\ \bibinfo {pages} {210505} (\bibinfo {year}
  {2016})}\BibitemShut {NoStop}%
\bibitem [{\citenamefont {Wootton}(2017{\natexlab{a}})}]{wootton:majorana}%
  \BibitemOpen
  \bibfield  {author} {\bibinfo {author} {\bibfnamefont {J.~R.}\ \bibnamefont
  {Wootton}},\ }\href@noop {} {\bibfield  {journal} {\bibinfo  {journal}
  {Quantum Science and Technology}\ }\textbf {\bibinfo {volume} {2}},\ \bibinfo
  {pages} {015006} (\bibinfo {year} {2017}{\natexlab{a}})}\BibitemShut
  {NoStop}%
\bibitem [{\citenamefont {Takita}\ \emph {et~al.}(2017)\citenamefont {Takita},
  \citenamefont {Cross}, \citenamefont {C\'orcoles}, \citenamefont {Chow},\
  and\ \citenamefont {Gambetta}}]{takita:17}%
  \BibitemOpen
  \bibfield  {author} {\bibinfo {author} {\bibfnamefont {M.}~\bibnamefont
  {Takita}}, \bibinfo {author} {\bibfnamefont {A.~W.}\ \bibnamefont {Cross}},
  \bibinfo {author} {\bibfnamefont {A.~D.}\ \bibnamefont {C\'orcoles}},
  \bibinfo {author} {\bibfnamefont {J.~M.}\ \bibnamefont {Chow}}, \ and\
  \bibinfo {author} {\bibfnamefont {J.~M.}\ \bibnamefont {Gambetta}},\ }\href
  {\doibase 10.1103/PhysRevLett.119.180501} {\bibfield  {journal} {\bibinfo
  {journal} {Phys. Rev. Lett.}\ }\textbf {\bibinfo {volume} {119}},\ \bibinfo
  {pages} {180501} (\bibinfo {year} {2017})}\BibitemShut {NoStop}%
\bibitem [{\citenamefont {Linke}\ \emph {et~al.}(2017)\citenamefont {Linke},
  \citenamefont {Gutierrez}, \citenamefont {Landsman}, \citenamefont {Figgatt},
  \citenamefont {Debnath}, \citenamefont {Brown},\ and\ \citenamefont
  {Monroe}}]{linke:17}%
  \BibitemOpen
  \bibfield  {author} {\bibinfo {author} {\bibfnamefont {N.~M.}\ \bibnamefont
  {Linke}}, \bibinfo {author} {\bibfnamefont {M.}~\bibnamefont {Gutierrez}},
  \bibinfo {author} {\bibfnamefont {K.~A.}\ \bibnamefont {Landsman}}, \bibinfo
  {author} {\bibfnamefont {C.}~\bibnamefont {Figgatt}}, \bibinfo {author}
  {\bibfnamefont {S.}~\bibnamefont {Debnath}}, \bibinfo {author} {\bibfnamefont
  {K.~R.}\ \bibnamefont {Brown}}, \ and\ \bibinfo {author} {\bibfnamefont
  {C.}~\bibnamefont {Monroe}},\ }\href {\doibase 10.1126/sciadv.1701074}
  {\bibfield  {journal} {\bibinfo  {journal} {Science Advances}\ }\textbf
  {\bibinfo {volume} {3}} (\bibinfo {year} {2017}),\
  10.1126/sciadv.1701074}\BibitemShut {NoStop}%
\bibitem [{\citenamefont {Vuillot}(2018)}]{vuillot:18}%
  \BibitemOpen
  \bibfield  {author} {\bibinfo {author} {\bibfnamefont {C.}~\bibnamefont
  {Vuillot}},\ }\href@noop {} {\bibfield  {journal} {\bibinfo  {journal}
  {Quant. Inf. Comp.}\ }\textbf {\bibinfo {volume} {18}},\ \bibinfo {pages}
  {0949} (\bibinfo {year} {2018})}\BibitemShut {NoStop}%
\bibitem [{\citenamefont {Wootton}\ and\ \citenamefont
  {Loss}(2018)}]{wootton:18}%
  \BibitemOpen
  \bibfield  {author} {\bibinfo {author} {\bibfnamefont {J.~R.}\ \bibnamefont
  {Wootton}}\ and\ \bibinfo {author} {\bibfnamefont {D.}~\bibnamefont {Loss}},\
  }\href {\doibase 10.1103/PhysRevA.97.052313} {\bibfield  {journal} {\bibinfo
  {journal} {Phys. Rev. A}\ }\textbf {\bibinfo {volume} {97}},\ \bibinfo
  {pages} {052313} (\bibinfo {year} {2018})}\BibitemShut {NoStop}%
\bibitem [{\citenamefont {Naveh}\ \emph {et~al.}(2018)\citenamefont {Naveh},
  \citenamefont {Kashefi}, \citenamefont {Wootton},\ and\ \citenamefont
  {Bertels}}]{naveh:18}%
  \BibitemOpen
  \bibfield  {author} {\bibinfo {author} {\bibfnamefont {Y.}~\bibnamefont
  {Naveh}}, \bibinfo {author} {\bibfnamefont {E.}~\bibnamefont {Kashefi}},
  \bibinfo {author} {\bibfnamefont {J.~R.}\ \bibnamefont {Wootton}}, \ and\
  \bibinfo {author} {\bibfnamefont {K.}~\bibnamefont {Bertels}},\ }in\
  \href@noop {} {\emph {\bibinfo {booktitle} {Proceedings of the 2018 Design,
  Automation \& Test in Europe (DATE)}}}\ (\bibinfo {year} {2018})\BibitemShut
  {NoStop}%
\bibitem [{\citenamefont {Gong~{\em et al.}}(2019)}]{gong:19}%
  \BibitemOpen
  \bibfield  {author} {\bibinfo {author} {\bibfnamefont {M.}~\bibnamefont
  {Gong~{\em et al.}}},\ }\href@noop {} {\enquote {\bibinfo {title}
  {{{Experimental verification of five-qubit quantum error correction with
  superconducting qubits}}},}\ } (\bibinfo {year} {2019}),\ \bibinfo {note}
  {arXiv:1912.09410}\BibitemShut {NoStop}%
\bibitem [{\citenamefont {Kraglund Andersen~{\em et al.}}(2019)}]{andersen:19}%
  \BibitemOpen
  \bibfield  {author} {\bibinfo {author} {\bibfnamefont {C.}~\bibnamefont
  {Kraglund Andersen~{\em et al.}}},\ }\href@noop {} {\enquote {\bibinfo
  {title} {{{Repeated Quantum Error Detection in a Surface Code}}},}\ }
  (\bibinfo {year} {2019}),\ \bibinfo {note} {arXiv:1907.04507}\BibitemShut
  {NoStop}%
\bibitem [{\citenamefont {Gottesman}(1996)}]{gottesman:96}%
  \BibitemOpen
  \bibfield  {author} {\bibinfo {author} {\bibfnamefont {D.}~\bibnamefont
  {Gottesman}},\ }\href@noop {} {\bibfield  {journal} {\bibinfo  {journal}
  {Phys. Rev. A}\ }\textbf {\bibinfo {volume} {54}},\ \bibinfo {pages} {1862}
  (\bibinfo {year} {1996})}\BibitemShut {NoStop}%
\bibitem [{\citenamefont {Abraham}\ \emph {et~al.}(2019)\citenamefont
  {Abraham}, \citenamefont {Akhalwaya}, \citenamefont {Aleksandrowicz},
  \citenamefont {Alexander}, \citenamefont {Alexandrowics}, \citenamefont
  {Arbel}, \citenamefont {Asfaw}, \citenamefont {Azaustre}, \citenamefont
  {AzizNgoueya}, \citenamefont {Barkoutsos}, \citenamefont {Barron},
  \citenamefont {Bello}, \citenamefont {Ben-Haim}, \citenamefont {Bevenius},
  \citenamefont {Bishop}, \citenamefont {Bosch}, \citenamefont {Bucher},
  \citenamefont {CZ}, \citenamefont {Cabrera}, \citenamefont {Calpin},
  \citenamefont {Capelluto}, \citenamefont {Carballo}, \citenamefont
  {Carrascal}, \citenamefont {Chen}, \citenamefont {Chen}, \citenamefont
  {Chen}, \citenamefont {Chow}, \citenamefont {Claus}, \citenamefont {Clauss},
  \citenamefont {Cross}, \citenamefont {Cross}, \citenamefont {Cross},
  \citenamefont {Cruz-Benito}, \citenamefont {Cryoris}, \citenamefont {Culver},
  \citenamefont {C{\'o}rcoles-Gonzales}, \citenamefont {Dague}, \citenamefont
  {Dartiailh}, \citenamefont {DavideFrr}, \citenamefont {Davila}, \citenamefont
  {Ding}, \citenamefont {Drechsler}, \citenamefont {Drew}, \citenamefont
  {Dumitrescu}, \citenamefont {Dumon}, \citenamefont {Duran}, \citenamefont
  {Eastman}, \citenamefont {Eendebak}, \citenamefont {Egger}, \citenamefont
  {Everitt}, \citenamefont {Fern{\'a}ndez}, \citenamefont {Fern{\'a}ndez},
  \citenamefont {Ferrera}, \citenamefont {Frisch}, \citenamefont {Fuhrer},
  \citenamefont {GEORGE}, \citenamefont {GOULD}, \citenamefont {Gacon},
  \citenamefont {Gadi}, \citenamefont {Gago}, \citenamefont {Gambetta},
  \citenamefont {Garcia}, \citenamefont {Garion}, \citenamefont {Gawel-Kus},
  \citenamefont {Gomez-Mosquera}, \citenamefont {de~la Puente~Gonz{\'a}lez},
  \citenamefont {Greenberg}, \citenamefont {Grinko}, \citenamefont {Guan},
  \citenamefont {Gunnels}, \citenamefont {Haide}, \citenamefont {Hamamura},
  \citenamefont {Havlicek}, \citenamefont {Hellmers}, \citenamefont {Herok},
  \citenamefont {Hillmich}, \citenamefont {Horii}, \citenamefont {Howington},
  \citenamefont {Hu}, \citenamefont {Hu}, \citenamefont {Imai}, \citenamefont
  {Imamichi}, \citenamefont {Ishizaki}, \citenamefont {Iten}, \citenamefont
  {Itoko}, \citenamefont {Javadi-Abhari}, \citenamefont {Jessica},
  \citenamefont {Johns}, \citenamefont {Kanazawa}, \citenamefont {Kang-Bae},
  \citenamefont {Karazeev}, \citenamefont {Kassebaum}, \citenamefont
  {Knabberjoe}, \citenamefont {Kovyrshin}, \citenamefont {Krishnan},
  \citenamefont {Krsulich}, \citenamefont {Kus}, \citenamefont {LaRose},
  \citenamefont {Lambert}, \citenamefont {Latone}, \citenamefont {Lawrence},
  \citenamefont {Liu}, \citenamefont {Liu}, \citenamefont {Mac}, \citenamefont
  {Maeng}, \citenamefont {Malyshev}, \citenamefont {Marecek}, \citenamefont
  {Marques}, \citenamefont {Mathews}, \citenamefont {Matsuo}, \citenamefont
  {McClure}, \citenamefont {McGarry}, \citenamefont {McKay}, \citenamefont
  {Meesala}, \citenamefont {Mezzacapo}, \citenamefont {Midha}, \citenamefont
  {Minev}, \citenamefont {Mooring}, \citenamefont {Morales}, \citenamefont
  {Moran}, \citenamefont {Murali}, \citenamefont {M{\"u}ggenburg},
  \citenamefont {Nadlinger}, \citenamefont {Nannicini}, \citenamefont {Nation},
  \citenamefont {Naveh}, \citenamefont {Nick-Singstock}, \citenamefont
  {Niroula}, \citenamefont {Norlen}, \citenamefont {O'Riordan}, \citenamefont
  {Ogunbayo}, \citenamefont {Ollitrault}, \citenamefont {Oud}, \citenamefont
  {Padilha}, \citenamefont {Paik}, \citenamefont {Perriello}, \citenamefont
  {Phan}, \citenamefont {Pistoia}, \citenamefont {Pozas-iKerstjens},
  \citenamefont {Prutyanov}, \citenamefont {Puzzuoli}, \citenamefont
  {P{\'e}rez}, \citenamefont {Quintiii}, \citenamefont {Raymond}, \citenamefont
  {Redondo}, \citenamefont {Reuter}, \citenamefont {Rodr{\'\i}guez},
  \citenamefont {Ryu}, \citenamefont {SAPV}, \citenamefont {SamFerracin},
  \citenamefont {Sandberg}, \citenamefont {Sathaye}, \citenamefont {Schmitt},
  \citenamefont {Schnabel}, \citenamefont {Scholten}, \citenamefont {Schoute},
  \citenamefont {Sertage}, \citenamefont {Shammah}, \citenamefont {Shi},
  \citenamefont {Silva}, \citenamefont {Siraichi}, \citenamefont {Sitdikov},
  \citenamefont {Sivarajah}, \citenamefont {Smolin}, \citenamefont {Soeken},
  \citenamefont {Steenken}, \citenamefont {Stypulkoski}, \citenamefont
  {Takahashi}, \citenamefont {Taylor}, \citenamefont {Taylour}, \citenamefont
  {Thomas}, \citenamefont {Tillet}, \citenamefont {Tod}, \citenamefont {de~la
  Torre}, \citenamefont {Trabing}, \citenamefont {Treinish}, \citenamefont
  {TrishaPe}, \citenamefont {Turner}, \citenamefont {Vaknin}, \citenamefont
  {Valcarce}, \citenamefont {Varchon}, \citenamefont {Vogt-Lee}, \citenamefont
  {Vuillot}, \citenamefont {Weaver}, \citenamefont {Wieczorek}, \citenamefont
  {Wildstrom}, \citenamefont {Wille}, \citenamefont {Winston}, \citenamefont
  {Woehr}, \citenamefont {Woerner}, \citenamefont {Woo}, \citenamefont {Wood},
  \citenamefont {Wood}, \citenamefont {Wood}, \citenamefont {Wootton},
  \citenamefont {Yeralin}, \citenamefont {Yu}, \citenamefont {Zachow},
  \citenamefont {Zdanski}, \citenamefont {Zoufalc}, \citenamefont {anedumla},
  \citenamefont {azulehner}, \citenamefont {bcamorrison}, \citenamefont
  {brandhsn}, \citenamefont {chlorophyll zz}, \citenamefont {dime10},
  \citenamefont {drholmie}, \citenamefont {elfrocampeador}, \citenamefont
  {faisaldebouni}, \citenamefont {fanizzamarco}, \citenamefont {gruu},
  \citenamefont {kanejess}, \citenamefont {klinvill}, \citenamefont {kurarrr},
  \citenamefont {lerongil}, \citenamefont {ma5x}, \citenamefont {merav
  aharoni}, \citenamefont {mrossinek}, \citenamefont {neupat}, \citenamefont
  {ordmoj}, \citenamefont {sethmerkel}, \citenamefont {strickroman},
  \citenamefont {sumitpuri}, \citenamefont {tigerjack}, \citenamefont {toural},
  \citenamefont {willhbang}, \citenamefont {yang.luh},\ and\ \citenamefont
  {yotamvakninibm}}]{qiskit}%
  \BibitemOpen
  \bibfield  {author} {\bibinfo {author} {\bibfnamefont {H.}~\bibnamefont
  {Abraham}}, \bibinfo {author} {\bibfnamefont {I.~Y.}\ \bibnamefont
  {Akhalwaya}}, \bibinfo {author} {\bibfnamefont {G.}~\bibnamefont
  {Aleksandrowicz}}, \bibinfo {author} {\bibfnamefont {T.}~\bibnamefont
  {Alexander}}, \bibinfo {author} {\bibfnamefont {G.}~\bibnamefont
  {Alexandrowics}}, \bibinfo {author} {\bibfnamefont {E.}~\bibnamefont
  {Arbel}}, \bibinfo {author} {\bibfnamefont {A.}~\bibnamefont {Asfaw}},
  \bibinfo {author} {\bibfnamefont {C.}~\bibnamefont {Azaustre}}, \bibinfo
  {author} {\bibnamefont {AzizNgoueya}}, \bibinfo {author} {\bibfnamefont
  {P.}~\bibnamefont {Barkoutsos}}, \bibinfo {author} {\bibfnamefont
  {G.}~\bibnamefont {Barron}}, \bibinfo {author} {\bibfnamefont
  {L.}~\bibnamefont {Bello}}, \bibinfo {author} {\bibfnamefont
  {Y.}~\bibnamefont {Ben-Haim}}, \bibinfo {author} {\bibfnamefont
  {D.}~\bibnamefont {Bevenius}}, \bibinfo {author} {\bibfnamefont {L.~S.}\
  \bibnamefont {Bishop}}, \bibinfo {author} {\bibfnamefont {S.}~\bibnamefont
  {Bosch}}, \bibinfo {author} {\bibfnamefont {D.}~\bibnamefont {Bucher}},
  \bibinfo {author} {\bibnamefont {CZ}}, \bibinfo {author} {\bibfnamefont
  {F.}~\bibnamefont {Cabrera}}, \bibinfo {author} {\bibfnamefont
  {P.}~\bibnamefont {Calpin}}, \bibinfo {author} {\bibfnamefont
  {L.}~\bibnamefont {Capelluto}}, \bibinfo {author} {\bibfnamefont
  {J.}~\bibnamefont {Carballo}}, \bibinfo {author} {\bibfnamefont
  {G.}~\bibnamefont {Carrascal}}, \bibinfo {author} {\bibfnamefont
  {A.}~\bibnamefont {Chen}}, \bibinfo {author} {\bibfnamefont {C.-F.}\
  \bibnamefont {Chen}}, \bibinfo {author} {\bibfnamefont {R.}~\bibnamefont
  {Chen}}, \bibinfo {author} {\bibfnamefont {J.~M.}\ \bibnamefont {Chow}},
  \bibinfo {author} {\bibfnamefont {C.}~\bibnamefont {Claus}}, \bibinfo
  {author} {\bibfnamefont {C.}~\bibnamefont {Clauss}}, \bibinfo {author}
  {\bibfnamefont {A.~J.}\ \bibnamefont {Cross}}, \bibinfo {author}
  {\bibfnamefont {A.~W.}\ \bibnamefont {Cross}}, \bibinfo {author}
  {\bibfnamefont {S.}~\bibnamefont {Cross}}, \bibinfo {author} {\bibfnamefont
  {J.}~\bibnamefont {Cruz-Benito}}, \bibinfo {author} {\bibnamefont {Cryoris}},
  \bibinfo {author} {\bibfnamefont {C.}~\bibnamefont {Culver}}, \bibinfo
  {author} {\bibfnamefont {A.~D.}\ \bibnamefont {C{\'o}rcoles-Gonzales}},
  \bibinfo {author} {\bibfnamefont {S.}~\bibnamefont {Dague}}, \bibinfo
  {author} {\bibfnamefont {M.}~\bibnamefont {Dartiailh}}, \bibinfo {author}
  {\bibnamefont {DavideFrr}}, \bibinfo {author} {\bibfnamefont {A.~R.}\
  \bibnamefont {Davila}}, \bibinfo {author} {\bibfnamefont {D.}~\bibnamefont
  {Ding}}, \bibinfo {author} {\bibfnamefont {E.}~\bibnamefont {Drechsler}},
  \bibinfo {author} {\bibnamefont {Drew}}, \bibinfo {author} {\bibfnamefont
  {E.}~\bibnamefont {Dumitrescu}}, \bibinfo {author} {\bibfnamefont
  {K.}~\bibnamefont {Dumon}}, \bibinfo {author} {\bibfnamefont
  {I.}~\bibnamefont {Duran}}, \bibinfo {author} {\bibfnamefont
  {E.}~\bibnamefont {Eastman}}, \bibinfo {author} {\bibfnamefont
  {P.}~\bibnamefont {Eendebak}}, \bibinfo {author} {\bibfnamefont
  {D.}~\bibnamefont {Egger}}, \bibinfo {author} {\bibfnamefont
  {M.}~\bibnamefont {Everitt}}, \bibinfo {author} {\bibfnamefont {P.~M.}\
  \bibnamefont {Fern{\'a}ndez}}, \bibinfo {author} {\bibfnamefont {P.~M.}\
  \bibnamefont {Fern{\'a}ndez}}, \bibinfo {author} {\bibfnamefont {A.~H.}\
  \bibnamefont {Ferrera}}, \bibinfo {author} {\bibfnamefont {A.}~\bibnamefont
  {Frisch}}, \bibinfo {author} {\bibfnamefont {A.}~\bibnamefont {Fuhrer}},
  \bibinfo {author} {\bibfnamefont {M.}~\bibnamefont {GEORGE}}, \bibinfo
  {author} {\bibfnamefont {I.}~\bibnamefont {GOULD}}, \bibinfo {author}
  {\bibfnamefont {J.}~\bibnamefont {Gacon}}, \bibinfo {author} {\bibnamefont
  {Gadi}}, \bibinfo {author} {\bibfnamefont {B.~G.}\ \bibnamefont {Gago}},
  \bibinfo {author} {\bibfnamefont {J.~M.}\ \bibnamefont {Gambetta}}, \bibinfo
  {author} {\bibfnamefont {L.}~\bibnamefont {Garcia}}, \bibinfo {author}
  {\bibfnamefont {S.}~\bibnamefont {Garion}}, \bibinfo {author} {\bibnamefont
  {Gawel-Kus}}, \bibinfo {author} {\bibfnamefont {J.}~\bibnamefont
  {Gomez-Mosquera}}, \bibinfo {author} {\bibfnamefont {S.}~\bibnamefont {de~la
  Puente~Gonz{\'a}lez}}, \bibinfo {author} {\bibfnamefont {D.}~\bibnamefont
  {Greenberg}}, \bibinfo {author} {\bibfnamefont {D.}~\bibnamefont {Grinko}},
  \bibinfo {author} {\bibfnamefont {W.}~\bibnamefont {Guan}}, \bibinfo {author}
  {\bibfnamefont {J.~A.}\ \bibnamefont {Gunnels}}, \bibinfo {author}
  {\bibfnamefont {I.}~\bibnamefont {Haide}}, \bibinfo {author} {\bibfnamefont
  {I.}~\bibnamefont {Hamamura}}, \bibinfo {author} {\bibfnamefont
  {V.}~\bibnamefont {Havlicek}}, \bibinfo {author} {\bibfnamefont
  {J.}~\bibnamefont {Hellmers}}, \bibinfo {author} {\bibfnamefont
  {{\L}.}~\bibnamefont {Herok}}, \bibinfo {author} {\bibfnamefont
  {S.}~\bibnamefont {Hillmich}}, \bibinfo {author} {\bibfnamefont
  {H.}~\bibnamefont {Horii}}, \bibinfo {author} {\bibfnamefont
  {C.}~\bibnamefont {Howington}}, \bibinfo {author} {\bibfnamefont
  {S.}~\bibnamefont {Hu}}, \bibinfo {author} {\bibfnamefont {W.}~\bibnamefont
  {Hu}}, \bibinfo {author} {\bibfnamefont {H.}~\bibnamefont {Imai}}, \bibinfo
  {author} {\bibfnamefont {T.}~\bibnamefont {Imamichi}}, \bibinfo {author}
  {\bibfnamefont {K.}~\bibnamefont {Ishizaki}}, \bibinfo {author}
  {\bibfnamefont {R.}~\bibnamefont {Iten}}, \bibinfo {author} {\bibfnamefont
  {T.}~\bibnamefont {Itoko}}, \bibinfo {author} {\bibfnamefont
  {A.}~\bibnamefont {Javadi-Abhari}}, \bibinfo {author} {\bibnamefont
  {Jessica}}, \bibinfo {author} {\bibfnamefont {K.}~\bibnamefont {Johns}},
  \bibinfo {author} {\bibfnamefont {N.}~\bibnamefont {Kanazawa}}, \bibinfo
  {author} {\bibnamefont {Kang-Bae}}, \bibinfo {author} {\bibfnamefont
  {A.}~\bibnamefont {Karazeev}}, \bibinfo {author} {\bibfnamefont
  {P.}~\bibnamefont {Kassebaum}}, \bibinfo {author} {\bibnamefont
  {Knabberjoe}}, \bibinfo {author} {\bibfnamefont {A.}~\bibnamefont
  {Kovyrshin}}, \bibinfo {author} {\bibfnamefont {V.}~\bibnamefont {Krishnan}},
  \bibinfo {author} {\bibfnamefont {K.}~\bibnamefont {Krsulich}}, \bibinfo
  {author} {\bibfnamefont {G.}~\bibnamefont {Kus}}, \bibinfo {author}
  {\bibfnamefont {R.}~\bibnamefont {LaRose}}, \bibinfo {author} {\bibfnamefont
  {R.}~\bibnamefont {Lambert}}, \bibinfo {author} {\bibfnamefont
  {J.}~\bibnamefont {Latone}}, \bibinfo {author} {\bibfnamefont
  {S.}~\bibnamefont {Lawrence}}, \bibinfo {author} {\bibfnamefont
  {D.}~\bibnamefont {Liu}}, \bibinfo {author} {\bibfnamefont {P.}~\bibnamefont
  {Liu}}, \bibinfo {author} {\bibfnamefont {P.~B.~Z.}\ \bibnamefont {Mac}},
  \bibinfo {author} {\bibfnamefont {Y.}~\bibnamefont {Maeng}}, \bibinfo
  {author} {\bibfnamefont {A.}~\bibnamefont {Malyshev}}, \bibinfo {author}
  {\bibfnamefont {J.}~\bibnamefont {Marecek}}, \bibinfo {author} {\bibfnamefont
  {M.}~\bibnamefont {Marques}}, \bibinfo {author} {\bibfnamefont
  {D.}~\bibnamefont {Mathews}}, \bibinfo {author} {\bibfnamefont
  {A.}~\bibnamefont {Matsuo}}, \bibinfo {author} {\bibfnamefont {D.~T.}\
  \bibnamefont {McClure}}, \bibinfo {author} {\bibfnamefont {C.}~\bibnamefont
  {McGarry}}, \bibinfo {author} {\bibfnamefont {D.}~\bibnamefont {McKay}},
  \bibinfo {author} {\bibfnamefont {S.}~\bibnamefont {Meesala}}, \bibinfo
  {author} {\bibfnamefont {A.}~\bibnamefont {Mezzacapo}}, \bibinfo {author}
  {\bibfnamefont {R.}~\bibnamefont {Midha}}, \bibinfo {author} {\bibfnamefont
  {Z.}~\bibnamefont {Minev}}, \bibinfo {author} {\bibfnamefont {M.~D.}\
  \bibnamefont {Mooring}}, \bibinfo {author} {\bibfnamefont {R.}~\bibnamefont
  {Morales}}, \bibinfo {author} {\bibfnamefont {N.}~\bibnamefont {Moran}},
  \bibinfo {author} {\bibfnamefont {P.}~\bibnamefont {Murali}}, \bibinfo
  {author} {\bibfnamefont {J.}~\bibnamefont {M{\"u}ggenburg}}, \bibinfo
  {author} {\bibfnamefont {D.}~\bibnamefont {Nadlinger}}, \bibinfo {author}
  {\bibfnamefont {G.}~\bibnamefont {Nannicini}}, \bibinfo {author}
  {\bibfnamefont {P.}~\bibnamefont {Nation}}, \bibinfo {author} {\bibfnamefont
  {Y.}~\bibnamefont {Naveh}}, \bibinfo {author} {\bibnamefont
  {Nick-Singstock}}, \bibinfo {author} {\bibfnamefont {P.}~\bibnamefont
  {Niroula}}, \bibinfo {author} {\bibfnamefont {H.}~\bibnamefont {Norlen}},
  \bibinfo {author} {\bibfnamefont {L.~J.}\ \bibnamefont {O'Riordan}}, \bibinfo
  {author} {\bibfnamefont {O.}~\bibnamefont {Ogunbayo}}, \bibinfo {author}
  {\bibfnamefont {P.}~\bibnamefont {Ollitrault}}, \bibinfo {author}
  {\bibfnamefont {S.}~\bibnamefont {Oud}}, \bibinfo {author} {\bibfnamefont
  {D.}~\bibnamefont {Padilha}}, \bibinfo {author} {\bibfnamefont
  {H.}~\bibnamefont {Paik}}, \bibinfo {author} {\bibfnamefont {S.}~\bibnamefont
  {Perriello}}, \bibinfo {author} {\bibfnamefont {A.}~\bibnamefont {Phan}},
  \bibinfo {author} {\bibfnamefont {M.}~\bibnamefont {Pistoia}}, \bibinfo
  {author} {\bibfnamefont {A.}~\bibnamefont {Pozas-iKerstjens}}, \bibinfo
  {author} {\bibfnamefont {V.}~\bibnamefont {Prutyanov}}, \bibinfo {author}
  {\bibfnamefont {D.}~\bibnamefont {Puzzuoli}}, \bibinfo {author}
  {\bibfnamefont {J.}~\bibnamefont {P{\'e}rez}}, \bibinfo {author}
  {\bibnamefont {Quintiii}}, \bibinfo {author} {\bibfnamefont {R.}~\bibnamefont
  {Raymond}}, \bibinfo {author} {\bibfnamefont {R.~M.-C.}\ \bibnamefont
  {Redondo}}, \bibinfo {author} {\bibfnamefont {M.}~\bibnamefont {Reuter}},
  \bibinfo {author} {\bibfnamefont {D.~M.}\ \bibnamefont {Rodr{\'\i}guez}},
  \bibinfo {author} {\bibfnamefont {M.}~\bibnamefont {Ryu}}, \bibinfo {author}
  {\bibfnamefont {T.}~\bibnamefont {SAPV}}, \bibinfo {author} {\bibnamefont
  {SamFerracin}}, \bibinfo {author} {\bibfnamefont {M.}~\bibnamefont
  {Sandberg}}, \bibinfo {author} {\bibfnamefont {N.}~\bibnamefont {Sathaye}},
  \bibinfo {author} {\bibfnamefont {B.}~\bibnamefont {Schmitt}}, \bibinfo
  {author} {\bibfnamefont {C.}~\bibnamefont {Schnabel}}, \bibinfo {author}
  {\bibfnamefont {T.~L.}\ \bibnamefont {Scholten}}, \bibinfo {author}
  {\bibfnamefont {E.}~\bibnamefont {Schoute}}, \bibinfo {author} {\bibfnamefont
  {I.~F.}\ \bibnamefont {Sertage}}, \bibinfo {author} {\bibfnamefont
  {N.}~\bibnamefont {Shammah}}, \bibinfo {author} {\bibfnamefont
  {Y.}~\bibnamefont {Shi}}, \bibinfo {author} {\bibfnamefont {A.}~\bibnamefont
  {Silva}}, \bibinfo {author} {\bibfnamefont {Y.}~\bibnamefont {Siraichi}},
  \bibinfo {author} {\bibfnamefont {I.}~\bibnamefont {Sitdikov}}, \bibinfo
  {author} {\bibfnamefont {S.}~\bibnamefont {Sivarajah}}, \bibinfo {author}
  {\bibfnamefont {J.~A.}\ \bibnamefont {Smolin}}, \bibinfo {author}
  {\bibfnamefont {M.}~\bibnamefont {Soeken}}, \bibinfo {author} {\bibfnamefont
  {D.}~\bibnamefont {Steenken}}, \bibinfo {author} {\bibfnamefont
  {M.}~\bibnamefont {Stypulkoski}}, \bibinfo {author} {\bibfnamefont
  {H.}~\bibnamefont {Takahashi}}, \bibinfo {author} {\bibfnamefont
  {C.}~\bibnamefont {Taylor}}, \bibinfo {author} {\bibfnamefont
  {P.}~\bibnamefont {Taylour}}, \bibinfo {author} {\bibfnamefont
  {S.}~\bibnamefont {Thomas}}, \bibinfo {author} {\bibfnamefont
  {M.}~\bibnamefont {Tillet}}, \bibinfo {author} {\bibfnamefont
  {M.}~\bibnamefont {Tod}}, \bibinfo {author} {\bibfnamefont {E.}~\bibnamefont
  {de~la Torre}}, \bibinfo {author} {\bibfnamefont {K.}~\bibnamefont
  {Trabing}}, \bibinfo {author} {\bibfnamefont {M.}~\bibnamefont {Treinish}},
  \bibinfo {author} {\bibnamefont {TrishaPe}}, \bibinfo {author} {\bibfnamefont
  {W.}~\bibnamefont {Turner}}, \bibinfo {author} {\bibfnamefont
  {Y.}~\bibnamefont {Vaknin}}, \bibinfo {author} {\bibfnamefont {C.~R.}\
  \bibnamefont {Valcarce}}, \bibinfo {author} {\bibfnamefont {F.}~\bibnamefont
  {Varchon}}, \bibinfo {author} {\bibfnamefont {D.}~\bibnamefont {Vogt-Lee}},
  \bibinfo {author} {\bibfnamefont {C.}~\bibnamefont {Vuillot}}, \bibinfo
  {author} {\bibfnamefont {J.}~\bibnamefont {Weaver}}, \bibinfo {author}
  {\bibfnamefont {R.}~\bibnamefont {Wieczorek}}, \bibinfo {author}
  {\bibfnamefont {J.~A.}\ \bibnamefont {Wildstrom}}, \bibinfo {author}
  {\bibfnamefont {R.}~\bibnamefont {Wille}}, \bibinfo {author} {\bibfnamefont
  {E.}~\bibnamefont {Winston}}, \bibinfo {author} {\bibfnamefont {J.~J.}\
  \bibnamefont {Woehr}}, \bibinfo {author} {\bibfnamefont {S.}~\bibnamefont
  {Woerner}}, \bibinfo {author} {\bibfnamefont {R.}~\bibnamefont {Woo}},
  \bibinfo {author} {\bibfnamefont {C.~J.}\ \bibnamefont {Wood}}, \bibinfo
  {author} {\bibfnamefont {R.}~\bibnamefont {Wood}}, \bibinfo {author}
  {\bibfnamefont {S.}~\bibnamefont {Wood}}, \bibinfo {author} {\bibfnamefont
  {J.}~\bibnamefont {Wootton}}, \bibinfo {author} {\bibfnamefont
  {D.}~\bibnamefont {Yeralin}}, \bibinfo {author} {\bibfnamefont
  {J.}~\bibnamefont {Yu}}, \bibinfo {author} {\bibfnamefont {C.}~\bibnamefont
  {Zachow}}, \bibinfo {author} {\bibfnamefont {L.}~\bibnamefont {Zdanski}},
  \bibinfo {author} {\bibnamefont {Zoufalc}}, \bibinfo {author} {\bibnamefont
  {anedumla}}, \bibinfo {author} {\bibnamefont {azulehner}}, \bibinfo {author}
  {\bibnamefont {bcamorrison}}, \bibinfo {author} {\bibnamefont {brandhsn}},
  \bibinfo {author} {\bibnamefont {chlorophyll zz}}, \bibinfo {author}
  {\bibnamefont {dime10}}, \bibinfo {author} {\bibnamefont {drholmie}},
  \bibinfo {author} {\bibnamefont {elfrocampeador}}, \bibinfo {author}
  {\bibnamefont {faisaldebouni}}, \bibinfo {author} {\bibnamefont
  {fanizzamarco}}, \bibinfo {author} {\bibnamefont {gruu}}, \bibinfo {author}
  {\bibnamefont {kanejess}}, \bibinfo {author} {\bibnamefont {klinvill}},
  \bibinfo {author} {\bibnamefont {kurarrr}}, \bibinfo {author} {\bibnamefont
  {lerongil}}, \bibinfo {author} {\bibnamefont {ma5x}}, \bibinfo {author}
  {\bibnamefont {merav aharoni}}, \bibinfo {author} {\bibnamefont {mrossinek}},
  \bibinfo {author} {\bibnamefont {neupat}}, \bibinfo {author} {\bibnamefont
  {ordmoj}}, \bibinfo {author} {\bibnamefont {sethmerkel}}, \bibinfo {author}
  {\bibnamefont {strickroman}}, \bibinfo {author} {\bibnamefont {sumitpuri}},
  \bibinfo {author} {\bibnamefont {tigerjack}}, \bibinfo {author} {\bibnamefont
  {toural}}, \bibinfo {author} {\bibnamefont {willhbang}}, \bibinfo {author}
  {\bibnamefont {yang.luh}}, \ and\ \bibinfo {author} {\bibnamefont
  {yotamvakninibm}},\ }\href {\doibase 10.5281/zenodo.2562110} {\enquote
  {\bibinfo {title} {Qiskit: An open-source framework for quantum computing},}\
  } (\bibinfo {year} {2019})\BibitemShut {NoStop}%
\bibitem [{\citenamefont {Kitaev}(2003)}]{double}%
  \BibitemOpen
  \bibfield  {author} {\bibinfo {author} {\bibfnamefont {A.~Y.}\ \bibnamefont
  {Kitaev}},\ }\href@noop {} {\bibfield  {journal} {\bibinfo  {journal} {Ann.
  Phys.}\ }\textbf {\bibinfo {volume} {303}},\ \bibinfo {pages} {2} (\bibinfo
  {year} {2003})}\BibitemShut {NoStop}%
\bibitem [{\citenamefont {Dennis}\ \emph {et~al.}(2002)\citenamefont {Dennis},
  \citenamefont {Kitaev}, \citenamefont {Landahl},\ and\ \citenamefont
  {Preskill}}]{dennis}%
  \BibitemOpen
  \bibfield  {author} {\bibinfo {author} {\bibfnamefont {E.}~\bibnamefont
  {Dennis}}, \bibinfo {author} {\bibfnamefont {A.}~\bibnamefont {Kitaev}},
  \bibinfo {author} {\bibfnamefont {A.}~\bibnamefont {Landahl}}, \ and\
  \bibinfo {author} {\bibfnamefont {J.}~\bibnamefont {Preskill}},\ }\href@noop
  {} {\bibfield  {journal} {\bibinfo  {journal} {J. Math. Phys.}\ }\textbf
  {\bibinfo {volume} {43}},\ \bibinfo {pages} {4452} (\bibinfo {year}
  {2002})}\BibitemShut {NoStop}%
\bibitem [{\citenamefont {Bombin}\ and\ \citenamefont
  {Martin-Delgado}(2006)}]{bombin:06}%
  \BibitemOpen
  \bibfield  {author} {\bibinfo {author} {\bibfnamefont {H.}~\bibnamefont
  {Bombin}}\ and\ \bibinfo {author} {\bibfnamefont {M.~A.}\ \bibnamefont
  {Martin-Delgado}},\ }\href {\doibase 10.1103/PhysRevLett.97.180501}
  {\bibfield  {journal} {\bibinfo  {journal} {Phys. Rev. Lett.}\ }\textbf
  {\bibinfo {volume} {97}},\ \bibinfo {pages} {180501} (\bibinfo {year}
  {2006})}\BibitemShut {NoStop}%
\bibitem [{\citenamefont {Delfosse}(2014)}]{delfosse:14}%
  \BibitemOpen
  \bibfield  {author} {\bibinfo {author} {\bibfnamefont {N.}~\bibnamefont
  {Delfosse}},\ }\href {\doibase 10.1103/PhysRevA.89.012317} {\bibfield
  {journal} {\bibinfo  {journal} {Phys. Rev. A}\ }\textbf {\bibinfo {volume}
  {89}},\ \bibinfo {pages} {012317} (\bibinfo {year} {2014})}\BibitemShut
  {NoStop}%
\bibitem [{\citenamefont {Wootton}(2015)}]{wootton:15}%
  \BibitemOpen
  \bibfield  {author} {\bibinfo {author} {\bibfnamefont {J.~R.}\ \bibnamefont
  {Wootton}},\ }\href {\doibase 10.1088/1751-8113/48/21/215302} {\bibfield
  {journal} {\bibinfo  {journal} {Journal of Physics A: Mathematical and
  Theoretical}\ }\textbf {\bibinfo {volume} {48}},\ \bibinfo {pages} {215302}
  (\bibinfo {year} {2015})}\BibitemShut {NoStop}%
\bibitem [{\citenamefont {Fowler}(2012)}]{fowler:12}%
  \BibitemOpen
  \bibfield  {author} {\bibinfo {author} {\bibfnamefont {A.~G.}\ \bibnamefont
  {Fowler}},\ }\href {\doibase 10.1103/PhysRevLett.109.180502} {\bibfield
  {journal} {\bibinfo  {journal} {Phys. Rev. Lett.}\ }\textbf {\bibinfo
  {volume} {109}},\ \bibinfo {pages} {180502} (\bibinfo {year}
  {2012})}\BibitemShut {NoStop}%
\bibitem [{\citenamefont {Tomita}\ and\ \citenamefont
  {Svore}(2014)}]{svore:14}%
  \BibitemOpen
  \bibfield  {author} {\bibinfo {author} {\bibfnamefont {Y.}~\bibnamefont
  {Tomita}}\ and\ \bibinfo {author} {\bibfnamefont {K.~M.}\ \bibnamefont
  {Svore}},\ }\href@noop {} {\bibfield  {journal} {\bibinfo  {journal} {Phys.
  Rev. A}\ }\textbf {\bibinfo {volume} {90}},\ \bibinfo {pages} {062320}
  (\bibinfo {year} {2014})}\BibitemShut {NoStop}%
\bibitem [{\citenamefont {Wootton}\ \emph {et~al.}(2017)\citenamefont
  {Wootton}, \citenamefont {Peter}, \citenamefont {Winkler},\ and\
  \citenamefont {Loss}}]{winkler:17}%
  \BibitemOpen
  \bibfield  {author} {\bibinfo {author} {\bibfnamefont {J.~R.}\ \bibnamefont
  {Wootton}}, \bibinfo {author} {\bibfnamefont {A.}~\bibnamefont {Peter}},
  \bibinfo {author} {\bibfnamefont {J.~R.}\ \bibnamefont {Winkler}}, \ and\
  \bibinfo {author} {\bibfnamefont {D.}~\bibnamefont {Loss}},\ }\href {\doibase
  10.1103/PhysRevA.96.032338} {\bibfield  {journal} {\bibinfo  {journal} {Phys.
  Rev. A}\ }\textbf {\bibinfo {volume} {96}},\ \bibinfo {pages} {032338}
  (\bibinfo {year} {2017})}\BibitemShut {NoStop}%
\bibitem [{\citenamefont {Bravyi}\ and\ \citenamefont
  {Haah}(2013)}]{bravyi-haah}%
  \BibitemOpen
  \bibfield  {author} {\bibinfo {author} {\bibfnamefont {S.}~\bibnamefont
  {Bravyi}}\ and\ \bibinfo {author} {\bibfnamefont {J.}~\bibnamefont {Haah}},\
  }\href {\doibase 10.1103/PhysRevLett.111.200501} {\bibfield  {journal}
  {\bibinfo  {journal} {Phys. Rev. Lett.}\ }\textbf {\bibinfo {volume} {111}},\
  \bibinfo {pages} {200501} (\bibinfo {year} {2013})}\BibitemShut {NoStop}%
\bibitem [{\citenamefont {Wootton}(2017{\natexlab{b}})}]{decodoku}%
  \BibitemOpen
  \bibfield  {author} {\bibinfo {author} {\bibfnamefont {J.~R.}\ \bibnamefont
  {Wootton}},\ }\href@noop {} {\enquote {\bibinfo {title} {{{Getting the public
  involved in Quantum Error Correction}}},}\ } (\bibinfo {year}
  {2017}{\natexlab{b}}),\ \bibinfo {note} {arXiv:1712.09649}\BibitemShut
  {NoStop}%
\bibitem [{\citenamefont {Asfaw}\ \emph {et~al.}(2020)\citenamefont {Asfaw},
  \citenamefont {Bello}, \citenamefont {Ben-Haim}, \citenamefont {Bravyi},
  \citenamefont {Capelluto}, \citenamefont {Vazquez}, \citenamefont {Ceroni},
  \citenamefont {Gambetta}, \citenamefont {Garion}, \citenamefont {Gil},
  \citenamefont {Gonzalez}, \citenamefont {McKay}, \citenamefont {Minev},
  \citenamefont {Nation}, \citenamefont {Phan}, \citenamefont {Rattew},
  \citenamefont {Schaefer}, \citenamefont {Shabani}, \citenamefont {Smolin},
  \citenamefont {Temme}, \citenamefont {Tod},\ and\ \citenamefont
  {Wootton.}}]{qiskit-textbook}%
  \BibitemOpen
  \bibfield  {author} {\bibinfo {author} {\bibfnamefont {A.}~\bibnamefont
  {Asfaw}}, \bibinfo {author} {\bibfnamefont {L.}~\bibnamefont {Bello}},
  \bibinfo {author} {\bibfnamefont {Y.}~\bibnamefont {Ben-Haim}}, \bibinfo
  {author} {\bibfnamefont {S.}~\bibnamefont {Bravyi}}, \bibinfo {author}
  {\bibfnamefont {L.}~\bibnamefont {Capelluto}}, \bibinfo {author}
  {\bibfnamefont {A.~C.}\ \bibnamefont {Vazquez}}, \bibinfo {author}
  {\bibfnamefont {J.}~\bibnamefont {Ceroni}}, \bibinfo {author} {\bibfnamefont
  {J.}~\bibnamefont {Gambetta}}, \bibinfo {author} {\bibfnamefont
  {S.}~\bibnamefont {Garion}}, \bibinfo {author} {\bibfnamefont
  {L.}~\bibnamefont {Gil}}, \bibinfo {author} {\bibfnamefont {S.~D. L.~P.}\
  \bibnamefont {Gonzalez}}, \bibinfo {author} {\bibfnamefont {D.}~\bibnamefont
  {McKay}}, \bibinfo {author} {\bibfnamefont {Z.}~\bibnamefont {Minev}},
  \bibinfo {author} {\bibfnamefont {P.}~\bibnamefont {Nation}}, \bibinfo
  {author} {\bibfnamefont {A.}~\bibnamefont {Phan}}, \bibinfo {author}
  {\bibfnamefont {A.}~\bibnamefont {Rattew}}, \bibinfo {author} {\bibfnamefont
  {J.}~\bibnamefont {Schaefer}}, \bibinfo {author} {\bibfnamefont
  {J.}~\bibnamefont {Shabani}}, \bibinfo {author} {\bibfnamefont
  {J.}~\bibnamefont {Smolin}}, \bibinfo {author} {\bibfnamefont
  {K.}~\bibnamefont {Temme}}, \bibinfo {author} {\bibfnamefont
  {M.}~\bibnamefont {Tod}}, \ and\ \bibinfo {author} {\bibfnamefont
  {J.}~\bibnamefont {Wootton.}},\ }\href {http://community.qiskit.org/textbook}
  {\enquote {\bibinfo {title} {Learn quantum computation using qiskit},}\ }
  (\bibinfo {year} {2020})\BibitemShut {NoStop}%
\end{thebibliography}%

\end{document}